\documentclass[preprint,11pt,review]{elsarticle}

\usepackage{amssymb,amsmath,mathrsfs,bm,caption,graphicx,multirow,lineno,booktabs,threeparttable,subfig,makecell,hyperref}

\begin{document}

\begin{frontmatter}

\title{EXFormer: A Multi-Scale Trend-Aware Transformer with Dynamic Variable Selection for Foreign Exchange Returns Prediction}

\author[inst1,inst4]{Dinggao Liu}
\ead{dinggao.liu@fafu.edu.cn}

\author[inst2]{Robert Ślepaczuk}
\ead{rslepaczuk@wne.uw.edu.pl}

\author[inst3]{Zhenpeng Tang\corref{cor1}}
\ead{zhenpt@126.com}

\cortext[cor1]{Corresponding author.}

\affiliation[inst1]{organization={College of Forestry},
            addressline={Fujian Agriculture and Forestry University}, 
            city={Fuzhou 350002},
            country={China}}

\affiliation[inst3]{organization={College of Economics and Management},
            addressline={Fujian Agriculture and Forestry University}, 
            city={Fuzhou 350002},
            country={China}}

\affiliation[inst2]{organization={Quantitative Finance Research Group, Department of Quantitative Finance and Machine Learning, Faculty of Economic Sciences},
            addressline={University of Warsaw}, 
            city={Warsaw Ul. Długa 44/50, 00-241},
            country={Poland}}

\affiliation[inst4]{organization={Faculty of Forestry},
            addressline={University of British Columbia}, 
            city={Vancouver, BC V6T 1Z4},
            country={Canada}}


\begin{abstract}
Accurately forecasting daily exchange rate returns represents a longstanding challenge in international finance, as the exchange rate returns are driven by a multitude of correlated market factors and exhibit high-frequency fluctuations. This paper proposes EXFormer, a novel Transformer-based architecture specifically designed for forecasting the daily exchange rate returns. We introduce a multi-scale trend-aware self-attention mechanism that employs parallel convolutional branches with differing receptive fields to align observations on the basis of local slopes, preserving long-range dependencies while remaining sensitive to regime shifts. A dynamic variable selector assigns time-varying importance weights to 28 exogenous covariates related to exchange rate returns, providing pre-hoc interpretability. An embedded squeeze-and-excitation block recalibrates channel responses to emphasize informative features and depress noise in the forecasting. Using the daily data for EUR/USD, USD/JPY, and GBP/USD, we conduct out-of-sample evaluations across five different sliding windows. EXFormer consistently outperforms the random walk and other baselines, improving directional accuracy by a statistically significant margin of up to 8.5--22.8\%. In nearly one year of trading backtests, the model converts these gains into cumulative returns of 18\%, 25\%, and 18\% for the three pairs, with Sharpe ratios exceeding 1.8. When conservative transaction costs and slippage are accounted for, EXFormer retains cumulative returns of 7\%, 19\%, and 9\%, while other baselines achieve negative. The robustness checks further confirm the model's superiority under high-volatility and bear-market regimes. EXFormer furnishes both economically valuable forecasts and transparent, time-varying insights into the drivers of exchange rate dynamics for international investors, corporations, and central bank practitioners.
\end{abstract}



\end{frontmatter}

\section{Introduction}
\label{sec:sample1}
Exchange rates lie at the core of global finance, shaping trade, capital flows, monetary policy transmission, portfolio allocation, and corporate risk management \citep{fang202430}. Accurate day-ahead forecasts support hedging by multinational firms, guide central bank interventions, and inform systematic trading strategies used by hedge funds \citep{rime2010exchange}. 
Yet despite decades of research, forecasting daily exchange rate returns remains one of the most persistent challenges in international finance \citep{faust2003exchange, rossi2013exchange}.
\citet{meese1983empirical} demonstrated that structural and time-series models routinely fail to outperform a simple random walk in out-of-sample tests. The Meese--Rogoff puzzle reflects two fundamental difficulties. First, daily returns combine high-frequency noise---microstructure effects, news-driven volatility, liquidity shocks---with slower patterns such as medium-term cycles from interest differentials and carry-trade activity \citep{hasanov2024structural,mueller2017exchange,narayan2022understanding,mao2024unveiling}.
Second, it is necessary to model the influence and cross-market interactions of diverse exogenous drivers, such as other currency pairs, stock and commodity indices, and scheduled macroeconomic announcements on the daily returns behavior \citep{stavrakeva2024fundamental}. 
Traditional econometric models, especially those designed for lower-frequency data, often produce unstable daily forecasts and are prone to overfitting \citep{darvas2025forecasting,iyke2022exchange}.
Portfolio-balance theory further highlights the role of cross-asset capital flows and time-varying risk preferences, which conventional models typically overlook  \citep{engel2019uncovered}.
The core research gap is thus how to jointly model multi-scale temporal dependencies and multi-factor cross-market linkages. Although uncovered interest parity (UIP) implies unpredictability of exchange rate changes, its persistent empirical rejection indicates the presence of time-varying risk premia that more advanced forecasting models may exploit \citep{fang202430,ayitey2023forex}. 

Recent work increasingly applies machine learning and deep learning to forecasting daily exchange rate returns \citep{plakandaras2015forecasting,sun2020new,neghab2025explaining}. 
Classical methods such as SVMs and random forests offer limited gains, as they ignore temporal structure \citep{sadeghi2021combined,gyamerah2020long}.
RNN-based models including LSTM and GRU networks, better capture sequential dependencies \citep{zhou2025forecasting,cao2020deep} and have been widely tested in financial forecasting (\citet{fischer2018deep}; \citet{gu2020empirical}), yet still struggle with long-range patterns and with modeling complex non-local interactions across economic drivers \citep{ullah2025forecasting}. 
Methodologically, many studies emphasize technical improvements without rigorous comparison to the random walk benchmark \citep{abedin2025deep}, and often report only statistical metrics. Transaction-cost-adjusted evaluations remain scarce, leaving open whether model forecasts yield economically meaningful gains \citep{galeshchuk2016neural,ayitey2023forex}.
Determining whether any approach can outperform the random walk both statistically and economically remains a core question in daily exchange rate forecasting.

The emergence of Transformer-based architectures offers a promising alternative \citep{fischer2024fx}, but their direct application to financial time series faces three critical limitations. First, the standard pointwise self-attention mechanism, designed for discrete tokens, fails to capture the continuous, trend-like features of financial data \citep{zhang2025forecasting,wu2021autoformer}. Second, standard Transformers lack mechanisms to differentiate influential market drivers from redundant noise among correlated external factors \citep{wang2024timexer,mishra2024volatility}. Third, these models remain black boxes. While post-hoc techniques like SHAP provide retrospective explanations, they fail to offer the forward-looking, time-varying interpretability required for adaptive investment decisions \citep{bybee2023narrative,shi2025econometrics}.

To address these challenges and gaps, we propose EXFormer, a novel Transformer-based architecture specifically engineered for forecasting the daily exchange rate returns. EXFormer introduces several critical innovations. 
At its core is a novel multi-scale trend-aware attention mechanism, which is designed to capture features across different time horizons while being sensitive to local trends. This is complemented by a squeeze-and-excitation (SE) block that learns to amplify the most informative feature channels while suppressing redundancy. 
To manage the complex web of external impact factors, we integrate a dynamic variable selector that learns daily importance weights for each covariate, allowing the model to focus on the most relevant market information and adapt as market conditions evolve. 
The encoder's rich output representation is then passed to a decoder, which utilizes a GRU and a linear layer to generate the final returns forecast. 
Moreover, for interpretability, the dynamic variable selector provides a built-in pre-hoc explanation, assigning a specific weight to each covariate before the prediction is made, which allows for both global and time-varying interpretability, offering a granular view of how each variable's importance evolves at every time step.

We evaluate EXFormer using daily data for three major currency pairs (EUR/USD, USD/JPY, GBP/USD) alongside 28 macroeconomic and financial covariates. Our validation strategy is threefold: (1) we assess statistical accuracy using the Mean Squared Forecast Error (MSFE) ratio relative to the random walk, confirmed by the Clark--West test \citep{clark2007approximately}; (2) we evaluate Directional Accuracy (DA) to verify the prediction of market turning points\citet{blaskowitz2014testing}; and (3) we conduct a realistic trading backtest to bridge the gap between statistical gains and economic value, explicitly accounting for transaction costs.

This study makes the following contributions to the exchange rate forecasting literature. 
First, we propose a specialized deep learning architecture (EXFormer) that effectively addresses the Meese--Rogoff puzzle by jointly modeling local trends and long-range dependencies. Second, we advance model interpretability by embedding a time-varying variable selection mechanism, offering a transparent alternative to black box models and post-hoc explanations. Third, we demonstrate through extensive backtesting and regime analysis that these statistical improvements translate into economically meaningful trading profits, providing a theoretically grounded tool for international investors and policy institutions.
Moreover, by uncovering systematic deviations from UIP through persistent returns, and by dynamically weighting covariates such as commodity indices and equity benchmarks in line with portfolio balance theory, the model provides a theoretically grounded explanation for the sources of predictability in daily exchange rate returns.

The remainder of this paper is organized as follows. Section 2 reviews the literature on the exchange rate returns forecasting. Section 3 details the architecture of EXFormer. Section 4 describes the data, variable construction, and experimental design. Section 5 presents the main empirical results and robustness checks. 
Section 6 discusses the findings and compares our work with prior literature. Section 7 concludes and suggests directions for future research.

\section{Literature review}
\label{sec:sample2}
This section surveys the literature on exchange rate forecasting, structuring the discussion into three main streams that motivate our proposed EXFormer model.

\subsection{The econometric models and the high-frequency challenge}
The academic pursuit of exchange rate predictability is historically rooted in structural macroeconomic \citep{forbes2018shocks}. Grounded in economic theory, models such as the monetary model and its sticky-price variant posit that exchange rate movements are driven by fundamentals like money supply, output levels, and inflation differentials \citep{cheung2019exchange}. The subsequent research extended these to incorporate interest rate differentials, as formalized in uncovered interest parity \citep{candian2025imperfect}, and portfolio balance effects \citep{camanho2022global}. While these models offer theoretical narratives, their empirical performance has been downgrade at daily horizons \citep{bulut2018google}.

The work of \citet{meese1983empirical} crystallized this failure into what is now famously known as the "Meese--Rogoff puzzle". Their out-of-sample forecasting exercise demonstrated that a random walk model consistently outperformed a suite of structural models for major currencies at horizons of up to one year. 
A comprehensive review by \citet{cheung2005empirical} re-affirmed the puzzle, concluding that no single model consistently and significantly dominates the random walk. 
The decades of following research have sought to resolve the puzzle. For example, \citet{mark2001nominal} found evidence of long-horizon predictability using monetary fundamentals, but this finding was later challenged for its reliance on in-sample fitting and specific time periods \citep{hong2024twofold}. Other research has explored nonlinearities, suggesting that transaction costs or policy interventions create thresholds beyond which fundamentals reassert their influence \citep{rossi2013exchange,engel2019uncovered,flores2022forecasting,candian2025imperfect}.

The macroeconomic variables central to these models are typically measured at low frequencies (weekly or monthly) \citep{lilley2022exchange}, making them inefficient to explain or predict the high-frequency, noisy dynamics of daily returns. Applying these models to the daily returns data often requires temporal disaggregation or interpolation techniques that introduce measurement error and model instabilitye \citep{cheung2019exchange}. 
This leaves a research gap: the need for a framework that can integrate information across different time scales---from high-frequency market noise to slower-moving fundamental trends---without being constrained by the low-frequency nature of traditional economic data. Our EXFormer model addresses this directly through its multi-scale trend-aware attention mechanism, which is designed to analyze temporal patterns at various frequencies within a single, unified architecture.

\subsection{The rise of machine learning and hybrid models}
The limitations of linear econometric models prompted a shift towards more flexible, data-driven approaches from machine learning \citep{mallqui2019predicting}. The early applications included SVMs and RFs, which demonstrated some success in capturing nonlinear relationships between predictors and exchange rate returns \citep{fu2019evolutionary,gyamerah2020long}. However, they have treated financial data as a collection of independent and identically distributed feature vectors, largely ignore the temporal ordering of observations, failing to model path dependency, volatility clustering, and other sequential effects that are symbols of exchange rate time series \citep{abedin2025deep}.
Furthermore, the RNNs and their more advanced variants, LSTM and GRU networks incorporating feedback loops and memory cells, designed to capture sequential dependencies \citep{fischer2018deep}. A growing body of literature has applied these architectures to exchange rates returns forecasting, with some studies reporting improved forecast accuracy over both linear models and machine learning methods \citep{zhou2025forecasting,cao2020deep}. The study by \citep{gu2020empirical} also demonstrated the power of deep learning in the broader context of asset pricing, showing they could identify risk premia more effectively than traditional methods.

The classical deep learning architectures face persistent challenges when it comes to exchange rate returns forecasting \citep{wei2018hybrid}. First, while LSTMs mitigate the vanishing gradient problem of simple RNNs, they can still struggle to capture very long-range dependencies, which are relevant for modeling multi-month cycles in currency markets \citep{yildirim2021forecasting,chi2025enhancing}. Second, their strictly sequential processing mechanism is a bottleneck \citep{mao2024unveiling}. They process information one time step at a time, making it difficult to model complex, non-adjacent interactions \citep{han2024lest}. For example, a monetary policy announcement from 60 days prior might have a more direct impact on today's return than a minor commodity price fluctuation from yesterday \citep{shi2025econometrics}.  
Third, much of the existing literature suffers from a statistical versus economic significance gap \citep{windsor2022improving,abedin2025deep}. Many studies report improvements in statistical error metrics without conducting rigorous backtesting \citep{liu2024new}. As pointed out by \citet{rossi2013exchange}, statistical outperformance does not guarantee economic value \citep{darvas2025forecasting}. 
EXFormer is designed to overcome these specific limitations. Its Transformer-based core replaces sequential processing with a self-attention mechanism, enabling the model to directly assess relationships across all points in the input sequence simultaneously. Furthermore, we bridge the economic significance gap by conducting a trading backtests to validate the practical value of our forecasts, a step missing from much of the prior deep learning literature in this domain.

Meanwhile, some of the literature concerns how exchange rate forecasts are evaluated. Much of the empirical studies of applying machine learning and deep learning have assessed performance through point forecasting error measures such as RMSE, MAE, and MSFE ratios \citep{rossi2013exchange,darvas2025forecasting}. While informative about average accuracy, these symmetric, scale-dependent metrics are skeptical of sign, downside risk, and payoff asymmetries that matter for trading and risk management. Directional accuracy and associated inference have been used less systematically, and even when reported, they are not always linked to trading simulations or to risk-adjusted performance measures (e.g., Sharpe, Sortino, drawdowns), leaving a gap between statistical and economic significance.
Recent work in artificial intelligence-driven finance has argued for objective functions that better reflect investment goals by penalizing costly errors more heavily. For example, the magnitude-aware directional losses \citep{michankow2024mean,michankow2025alternative} weight sign mistakes by their return amplitudes, aligning optimization with realized trading outcomes rather than with squared-error fit alone. Related proposals incorporate asymmetry, tail sensitivity, or downside aversion into training objectives and evaluation metrics, improving the congruence between model learning and portfolio utility \citep{windsor2022improving,abedin2025deep}. A further limitation of prior studies is the frequent absence of realistic backtesting (transaction costs, slippage, and liquidity), rolling re-estimation, and regime-aware analysis. When these ingredients are missing, reported gains can be overstated and fragile out of sample \citep{rossi2013exchange}.
Based on this background, we position our contribution along both MSFE-based measures and DA with robust inference, and we translate forecasts into implementable long--short strategies evaluated with risk-adjusted statistics and trading frictions. The evaluation perspective complements our architectural advances and narrows the gap between predictive accuracy and actionable performance in daily exchange rate markets.

Furthermore, a parallel strand of the literature employs hybrid models for exchange rate returns forecasting. On the one hand is decomposition-ensemble frameworks \citep{hao2023bi,henriquez2019combined,sun2018clustering,sun2019new}. The exchange rate prices or returns are first decomposed into trend, cyclical and residual components via empirical mode decomposition or wavelet transforms, then each component is forecast with a specialized network and recombined \citep{wei2025wavelet}. These methods can isolate scale-specific patterns, yet they amplify model complexity, require ad-hoc frequency splits \citep{zhou2025forecasting}. Furthermore, decomposition-ensemble approaches do not inherently address the dynamic relevance of external covariates spreads, volatility indices, commodity prices across time \citep{ullah2025forecasting}.

On the other hand, hybrid models that combine different econometric models, machine learning, and deep learning approaches have also emerged \citep{neghab2025explaining,petrosino2025garch,khashei2021kalman}. These typically involve using one model to generate features or signals that are then fed into another model for final prediction \citep{plakandaras2015forecasting}. 
The rationale is that econometric models, such as GARCH-type volatility processes, cointegration frameworks, or error-correction mechanisms, embed structural domain knowledge and well-established statistical properties, while deep networks provide flexibility to capture nonlinearities and long-range dependencies \citep{petrosino2025garch,khashei2021kalman,neghab2025explaining}. For example, ARIMA-LSTM or GARCH-RNN hybrids have been shown to improve short-horizon volatility and return forecasts by jointly modeling conditional mean and variance dynamics \citep{plakandaras2015forecasting,wei2018hybrid}. While the methods can leverage the strengths of both paradigms, it often results in a loss of interpretability and does not fully exploit the temporal structure of the data \citep{shi2025econometrics}.
Furthermore, these methods have proven valuable in high-frequency daily data environments, where volatility clustering, regime switches, and heavy tails challenge either paradigm in isolation. However, hybrid frameworks often face trade-offs in interpretability, computational efficiency, and robustness out of sample \citep{shi2025econometrics,ullah2025forecasting}. Our proposed EXFormer shares the hybrid motivation but differs in design. Instead of ad-hoc model stacking or decomposition, it embeds econometric intuition (trend extraction, multi-scale dependence) directly within the attention mechanism and feature selection modules, offering an integrated architecture that remains both scalable and interpretable for daily exchange rate forecasting.

\subsection{Transformer models frontier}
The Transformer architecture, first introduced by \citet{vaswani2017attention} for natural language processing, has recently emerged as a powerful alternative for time series forecasting. Its core innovation, the self-attention mechanism, allows the model to weigh the influence of all other time steps when processing a given point in a sequence, addressing the long-range dependency and sequential bottleneck issues. Early applications in finance have shown promise \citep{liu2024interpretable,mishra2024volatility}, and have been developed to improve efficiency for very long sequence forecasting \citep{wang2024timexer,wu2021autoformer}.
However, traditional self-attention treats each time stamp as an exchangeable token and computes pointwise query-key similarity \citep{zhang2025forecasting}. For noisy financial series, where whether an observation is an outlier, a break, or part of a pattern depends on its local trend, pointwise matching can misalign relevant contexts \citep{hong2024twofold,han2024lest}. Recent decomposition or frequency domain Transformers partially mitigate this by separating trend, seasonality or filtering in the spectral domain, yet they do not explicitly encode local slope information into the attention kernel \citep{fischer2024fx}. Moreover, exchange rate returns are influenced by a broad, time-varying set of exogenous drivers \citep{stavrakeva2024fundamental,candian2025imperfect,iyke2022exchange}. The standard multi-head attention distributes capacity across heads but does not by itself distinguish influential covariates from redundant ones \citep{mao2024unveiling}. The variable selection layers as in \citet{petrosino2025garch} are a step forward, yet most implementations provide post-hoc importance or static gating, offering limited transparency about which inputs are used at prediction time. 
EXFormer is purpose-built to fill the gaps. The study aims to advance the state-of-the-art in the daily exchange rate returns forecasting.

\section{Methodology}
\label{sec:sample3}

\begin{figure}[htbp]
    \centering
    \setlength{\leftskip}{-110pt}
    \includegraphics[width=200mm]{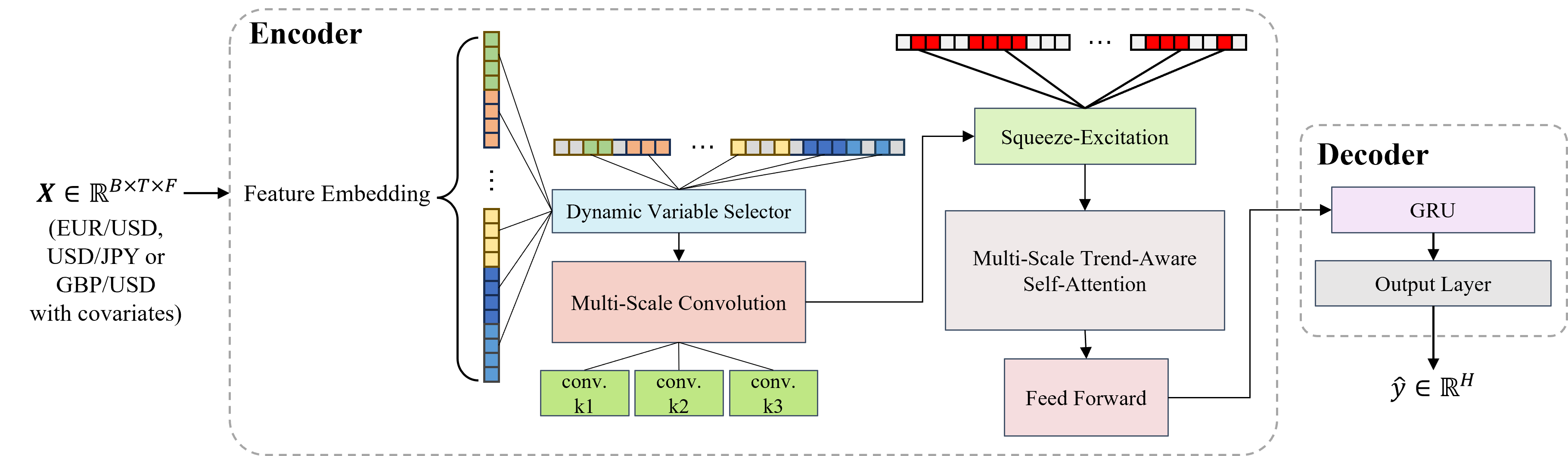}
    \caption{The architecture of EXFormer. The model consists of an Encoder and a Decoder. The Encoder includes a dynamic variable selector, multi-scale convolution and squeeze-and-excitation block, and multi-scale trend-aware self-attention. The Decoder comprises a position-wise feed-forward layer and a GRU followed by a linear output layer.}
    \label{fig:1}
\end{figure}

In this section, we present the EXFormer architecture in detail. 
The proposed EXFormer architecture is conceptually inspired by the trend-aware self-attention model \citep{li2019enhancing,guo2021learning}. We have redesigned this foundation into a novel multi-scale trend-aware self-attention mechanism. The motivation for this multi-scale approach is to effectively capture the complex temporal patterns in the daily exchange rate of high-frequency fluctuations, medium-term cycles, and long-run drifts. 
Recognizing that the daily exchange rates are driven by numerous external factors, we have developed a dynamic variable selector that not only reweights variables dynamically but also imparts pre-hoc interpretability to the architecture. Meanwhile, to contend with the issue of information redundancy in financial time series forecasting, the model employs a squeeze-and-excitation (SE) block. The above is among the Encoder. For the Decoder, we are inspired by \citet{shu2025deformtime} to introduce GRU and linear mapping.
Figure \ref{fig:1} provides a schematic overview of the entire model.

\subsection{Problem formulation}
\label{sec:notation}

Let $\{r_t\}_{t=1}^T$ denote the sequence of daily log-returns for one of the three target currency pairs (EUR/USD, USD/JPY, GBP/USD).  At each time $t$, we observe an $F$-dimensional vector of exogenous covariates
\begin{equation}
x_t = \bigl(x_{1,t},\,x_{2,t},\dots,x_{F,t}\bigr)^\top \in \mathbb R^F,
\end{equation}
which includes exchange rates, stock and commodity indices, and macroeconomic releases. We use a sliding window of length $T$, forming
\begin{equation}
X_t = [\,x_{t-T+1},\dots,x_t\,]\in\mathbb R^{T\times F},
\end{equation}
and in minibatch form
\begin{equation}
\mathcal X \in \mathbb R^{B\times T\times F},
\end{equation}
where $B$ is the batch size.  All covariates and returns are standardised using training-set statistics. Our goal is to learn a mapping
\begin{equation}
\mathcal M:\;\{x_{t-T+1},\dots,x_t\} \;\mapsto\; \hat r_{t+1},
\end{equation}
where $\hat r_{t+1}$ is the predicted return for the next day, $t+1$.  The model is trained by minimizing the mean squared error (MSE) loss:
minimising the forecast error $\varepsilon_{t+1}=r_{t+1}-\hat r_{t+1}$.
To mitigate overfitting and enhance out-of-sample generalization, we incorporate standard regularization techniques, including dropout applied to intermediate layers and an early stopping criterion based on validation loss.

\subsection{Dynamic variable selector}
\label{sec:dvs}

Given an input tensor $X\in\mathbb R^{B\times T\times F}$, we first embed each series
\begin{equation}
E_i = X_{:,:,i}\,W^{(e)}_i + b^{(e)}_i
\;\in\; \mathbb R^{B\times T\times D},
\end{equation}
for $i=1,\dots,F$.  A shared linear layer then computes raw scores
\begin{equation}
s_{i,t} = E_{i,t,:}\,w^{(s)} + b^{(s)},
\quad
s\in\mathbb R^{B\times T\times1}.
\end{equation}
These are normalised across $i$ by softmax:
\begin{equation}
\omega_{i,t}
= \frac{\exp\bigl(s_{i,t}\bigr)}{\sum_{j=1}^F \exp\bigl(s_{j,t}\bigr)},
\quad
\sum_{i=1}^F \omega_{i,t}=1.
\end{equation}
The weighted embeddings $\omega_{i,t} E_{i,t,:}$ are concatenated into
$\tilde E\in\mathbb R^{B\times T\times (F\,D)}$.  The map $\{\omega_{i,t}\}$ serves as a pre-hoc explanation of feature relevance.

\subsection{Multi-scale convolution and squeeze-and-excitation}
\label{sec:msc_se}

We reshape $\tilde E$ to $\mathbb R^{B\times (F\,D)\times T}$ and apply $K$ same-padded 1D convolutions of kernel sizes $\{k_1,\dots,k_K\}$:
\begin{equation}
H^{(k)} = \mathrm{ReLU}\bigl(\mathrm{Conv1D}_{k}(\tilde E)\bigr),
\quad
H^{(k)}\in\mathbb R^{B\times D\times T}.
\end{equation}
The concatenation across kernels yields $H\in\mathbb R^{B\times(KD)\times T}$, followed by dropout and a linear projection
\begin{equation}
H_{\mathrm{proj}}
= \bigl[H^{(k_1)};\dots;H^{(k_K)}\bigr]^\top W^{(c)} + b^{(c)}
\;\in\;\mathbb R^{B\times T\times D}.
\end{equation}

In line with prior studies on multi-scale convolution for financial and economic time series \citep{liang2022airformer,pmlr-v162-lan22a,guo2021learning}, we adopt 3 convolutional branches to capture short-, medium-, and long-horizon return dynamics. The kernel sizes are initialized as $\{3,5,7\}$ to reflect these scales, while their exact values are subsequently tuned automatically during hyperparameter optimization.

Then, for squeeze-and-excitation block, we compute channel-wise statistics via temporal pooling
\begin{equation}
z = \frac1T \sum_{t=1}^T H_{\mathrm{proj},t,:}
\in\mathbb R^D,
\end{equation}
and pass $z$ through a bottleneck MLP:
\begin{equation}
u = \sigma\bigl(W_2\,\mathrm{ReLU}(W_1\,z)\bigr)
\in[0,1]^D,
\end{equation}
where $W_1\in\mathbb R^{\frac D r\times D}$, $W_2\in\mathbb R^{D\times\frac D r}$, $r$ is the reduction ratio, and $\sigma$ is the sigmoid.  The recalibrated features are
\begin{equation}
\hat H_{t,d} = u_d\;H_{\mathrm{proj},t,d}.
\end{equation}

\subsection{Multi-scale trend-aware self-attention}
\label{sec:ta}
\begin{figure}[htbp]
    \centering
    \setlength{\leftskip}{-90pt}
    \includegraphics[width=190mm]{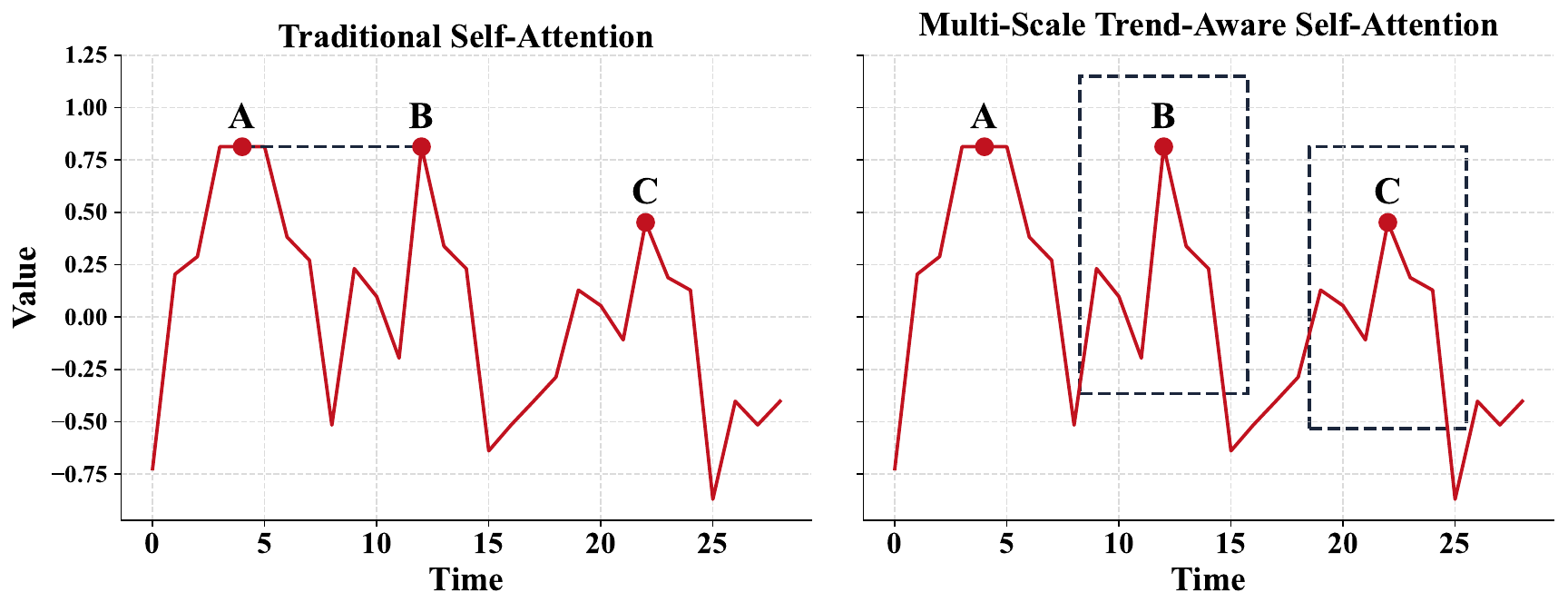}
    \caption{The comparison of the standard self-attention versus our multi-scale trend-aware self-attention in a daily exchange rate forecasting task. The standard self-attention mistakenly aligns observation A with B simply because their instantaneous values coincide. In contrast, our multi-scale trend-aware self-attention incorporates local trend dynamics and correctly associates B with C, the two most relevant time points.}
    \label{fig:2}
\end{figure}

Due to the financial time-series patterns can shift dramatically as a result of market-driving events, whether an observation represents an anomaly, a regime change, or simply part of an existing pattern depends on its local trend. The vanilla Transformer self-attention computes query-key affinities pointwise, based on individual feature values, and thus fails to exploit rich information from neighbouring time steps. When an observation lies at an outlier, a change point, or within a broader pattern, the trend irrelevant matching can confuse the self-attention module and hinder model optimization.
Figure \ref{fig:2} contrasts the traditional self-attention mechanism with the multi-scale trend-aware attention we propose in EXFormer.

As suggested in related work on trend-aware attention mechanisms for time series \citep{liang2022airformer,pmlr-v162-lan22a,guo2021learning}, we design three parallel convolutional branches, each operating with a distinct kernel size to represent short-, medium-, and long-term dependencies. The convolution kernel sizes are initialized at $\{3,5,7\}$ following common practice, while the actual receptive fields are subject to automatic selection in the hyperparameter tuning process.

The standard self-attention computes
\begin{equation}
\mathrm{Attn}(Q,K,V)
=\mathrm{softmax}\!\Bigl(\frac{QK^\top}{\sqrt d}\Bigr)\,V,
\end{equation}
where $Q,K,V\in\mathbb R^{B\times T\times D}$ are the query, key, and value matrices, respectively, and $d$ is the dimension of the model.  The attention weights are computed as a scaled dot product of queries and keys, followed by a softmax operation.  The output is a weighted sum of values, where the weights depend on the similarity between queries and keys.  This approach is effective for many tasks but can struggle with time series data, especially when local trends or anomalies are present. This pointwise matching ignores local trend, so anomalies or regime shifts can confuse the attention weights.

We replace the linear projections for $Q$ and $K$ with $K$ parallel 2D convolutions of kernel sizes $\{k_1,\dots,k_K\}$ over the (head, time) dimensions.  Let $H_{\mathrm{SE}}\in\mathbb R^{B\times T\times D}$ and reshape to $\mathbb R^{B\times (H d)\times 1\times T}$, where $D=H\,d$.  For each branch $k$:
\begin{equation}
Q^{(k)} = \mathrm{Conv2D}_k(H_{\mathrm{SE}}),\quad
K^{(k)} = \mathrm{Conv2D}_k(H_{\mathrm{SE}}),
\end{equation}
where $\mathrm{Conv2D}_k$ is a 2D convolution with kernel size $k\times 1$ and stride $1\times 1$.  The output is reshaped to $\mathbb R^{B\times T\times d}$, where $d$ is the dimension of each attention head.  The value projection $V$ is a single linear layer, which is shared across branches:
\begin{equation}
V = H_{\mathrm{SE}} W^{(v)} + b^{(v)}.
\end{equation}

Each branch attends within its local context window:
\begin{equation}
A^{(k)} = \mathrm{softmax}\!\Bigl(\frac{Q^{(k)}(K^{(k)})^\top}{\sqrt d}\Bigr)\,V
\;\in\;\mathbb R^{B\times T\times D}.
\end{equation}
Concatenating $\{A^{(k)}\}$ and fusing:
\begin{equation}
H_{\mathrm{attn}}
= \bigl[A^{(1)};\dots;A^{(K)}\bigr]\,W^{(f)} + b^{(f)}
\;\in\;\mathbb R^{B\times T\times D}.
\end{equation}

\subsection{Decoder}
The attention‐enriched sequence $H_{\mathrm{attn}}\in\mathbb R^{B\times T\times D}$ is passed through a position‐wise feed‐forward layer and into a single‐layer GRU:
\begin{align}
h_t &= \mathrm{GRU}\bigl(h_{t-1},\,H_{\mathrm{attn},t}\bigr),
\quad
h_t\in\mathbb R^{B\times D}.
\end{align}
We take the final hidden state $h_T$ and apply a linear output layer:
\begin{align}
\hat r_{t+1} = h_T\,W^{(o)} + b^{(o)},
\quad
W^{(o)}\in\mathbb R^{D\times1},\; b^{(o)}\in\mathbb R.
\end{align}

\section{Data and experimental design}
\label{sec:sample4}
This section outlines the datasets, variable construction, and empirical design adopted in this study. We first describe the daily exchange rates for the three most liquid currency pairs and the computation of returns, followed by a detailed summary of their statistical properties. We then present the broad set of exogenous covariates, capturing key information from global foreign exchange, equity, fixed-income, commodity, and macroeconomic markets. Finally, we specify the experimental setup, including the forecasting horizon, sliding window lengths, benchmark models, hyperparameter tuning procedures, and the evaluation metrics used to assess both statistical and economic forecasting performance.

\subsection{Data description}
The empirical evaluation employs daily spot exchange rates for the three most liquid currency pairs: Euro to US Dollar (EUR/USD), US Dollar to Japanese Yen (USD/JPY), and British Pound to US Dollar (GBP/USD) for the period May 7, 2010 through August 29, 2024.
Let $S_{i,t}$ denote the closing spot rate of pair $i$ trading day $t$. We compute the continuously compounded return in percent as
\begin{equation}
r_{i,t} = \ln\bigl(S_{i,t}\bigr) - \ln\bigl(S_{i,t-1}\bigr).
\end{equation}

\begin{figure}[htbp]
    \centering
    \setlength{\leftskip}{-80pt}
    \includegraphics[width=190mm]{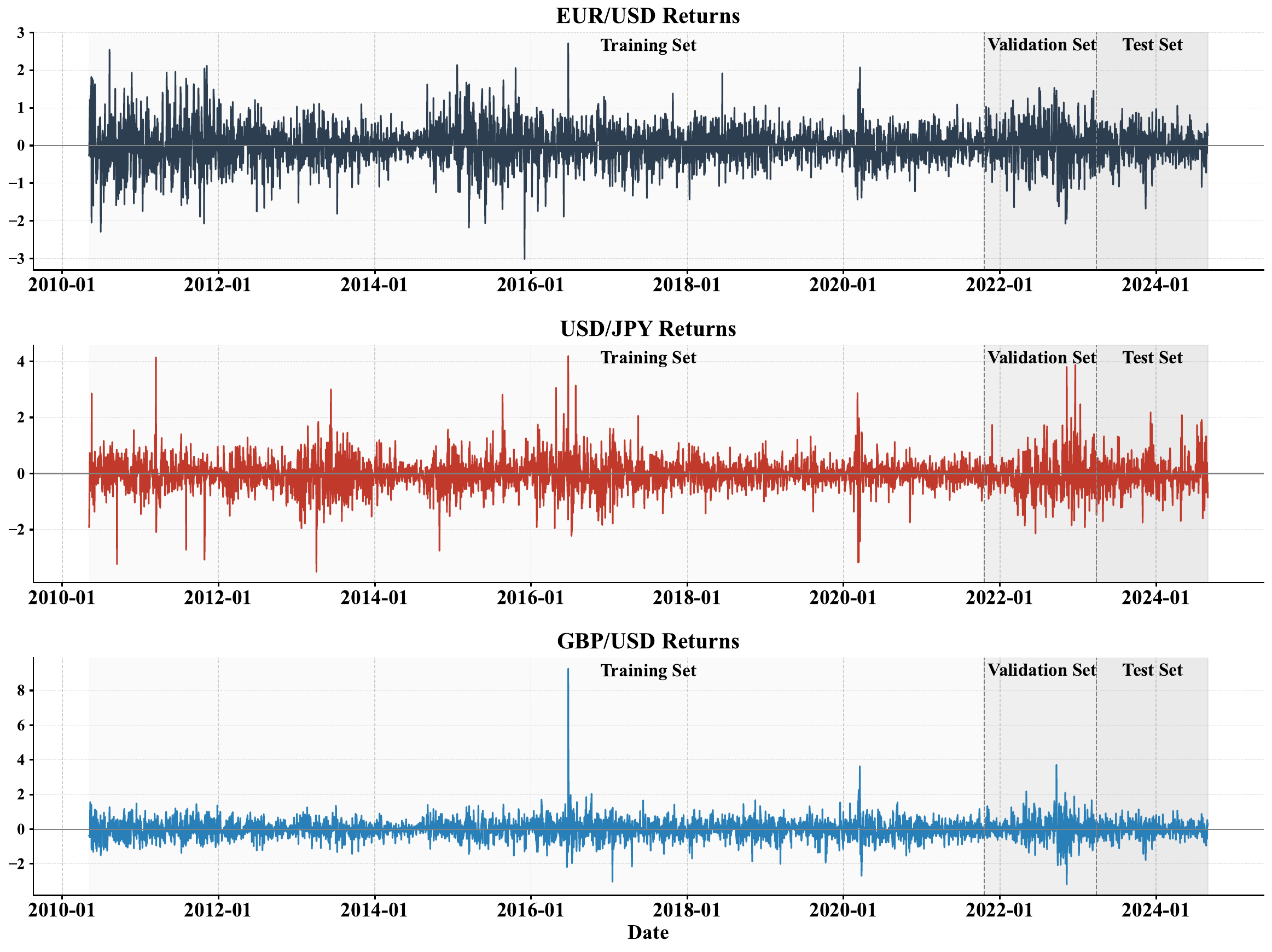}
    \caption{The daily returns of EUR/USD, USD/JPY, and GBP/USD from May 7, 2010 to August 29, 2024. The shaded areas indicate the training (first 80\%), validation (next 10\%), and test (final 10\%) periods.}
    \label{fig:3}
\end{figure}

Figure~\ref{fig:3} plots the three standardized return series on a common time axis. The shaded regions delineate the training (first 80\%), validation (next 10\%) and test (final 10\%) splits, ensuring that each model is estimated, tuned and evaluated in a strictly out-of-sample practice.

For the descriptive statistics, Table~\ref{table:1} provides a summary for the return series of EUR/USD, USD/JPY, and GBP/USD, broken down by the overall sample and the training, validation, and test subsets. Several key characteristics of financial returns are evident.
The distributions of returns are not normal. The skewness statistics indicate a slight asymmetry in the distributions. Furthermore, the kurtosis values for USD/JPY and GBP/USD indicate leptokurtic distributions, or fat tails, implying that extreme price movements are more frequent than a normal distribution would suggest. The Jarque-Bera test statistic, which formally tests the null hypothesis of normality, is highly significant (*** denotes $p<0.001$) for nearly all series and subsets, providing strong statistical evidence to reject normality.
Moreover, the Ljung-Box Q(5) statistic, which tests for autocorrelation up to a lag of 5, is generally insignificant regarding the time series properties. This suggests that the daily returns themselves exhibit little to no linear serial correlation. Conversely, the Augmented Dickey-Fuller (ADF) test yields highly significant negative test statistics for all series and subsets, leading to a rejection of the null hypothesis of a unit root, confirming that the returns are stationary.

Because cointegration is a property of levels rather than returns, we diagnose long-run co-movements among the three spot exchange rates. The Johansen trace statistics (26.56 for $r\leq0$, 13.97 for $r\leq1$, and 2.94 for $r\leq2$) are all below their respective 5\% critical values (35.01, 18.40, and 3.84), and the pairwise Engle-Granger tests similarly yield non-rejections of the null (e.g., $t=-2.75$, $p=0.184$ for EUR/USD--USD/JPY). The results indicate that no strong cointegrating relation is present among the three series. Nevertheless, our predictive targets are daily returns, which are stationary by construction and confirmed by the ADF tests above, so the forecasting setup does not rely on any particular long-run relation.

In addition, we assess volatility clustering in the daily returns using ARCH-LM tests. For all three currency pairs, the null hypothesis of no ARCH effects is rejected at conventional significance levels across multiple lags (i.e., LM-statistics exceeding 100 with $p<0.001$). The detailed test statistics are provided in the~\ref{table:a0}. This diagnostic addresses the concern that conditional heteroskedasticity may bias inference. In our empirical strategy, it is further mitigated by using heteroskedasticity and autocorrelation consistent (HAC) \citep{newey1986simple} standard errors in forecast evaluation tests and by relying on risk-adjusted criteria in backtesting.

\begin{table}[htbp]
  \centering
  \setlength{\leftskip}{-90pt}
  \caption{The descriptive statistics of daily exchange rate returns for EUR/USD, USD/JPY, and GBP/USD.}
  \resizebox{190mm}{!}{%
\begin{tabular}{llllllllllll}
\toprule
Type & Subset & Mean & Median & Std & Min & Max & Kurtosis & Skewness & Jarque-Bera & Q(5) Ljung-Box & ADF \\
\midrule
EUR/USD & Overall & 0.004 & -0.004 & 0.524 & -3.016 & 2.716 & 1.975 & 0.023 & 604.537*** & 5.917 & -61.934*** \\
 & Training & 0.003 & -0.004 & 0.530 & -3.016 & 2.716 & 2.138 & 0.063 & 568.090*** & 8.217 & -55.563*** \\
 & Validation & 0.020 & -0.005 & 0.590 & -2.076 & 1.537 & 0.530 & -0.187 & 6.162* & 1.818 & -18.939*** \\
 & Test & -0.006 & -0.008 & 0.389 & -1.678 & 1.061 & 0.793 & -0.191 & 11.407** & 6.006 & -8.238*** \\ \cline{2-12} 
USD/JPY & Overall & -0.013 & -0.013 & 0.573 & -3.506 & 4.197 & 5.914 & 0.292 & 5479.363*** & 2.447 & -61.235*** \\
 & Training & -0.007 & 0.002 & 0.552 & -3.506 & 4.197 & 6.608 & 0.086 & 5418.411*** & 3.053 & -55.550*** \\
 & Validation & -0.038 & -0.078 & 0.711 & -2.141 & 3.877 & 4.794 & 1.004 & 407.486*** & 5.952 & -14.763*** \\
 & Test & -0.028 & -0.081 & 0.583 & -1.752 & 2.187 & 1.742 & 0.470 & 58.823*** & 7.147 & -17.610*** \\ \cline{2-12} 
GBP/USD & Overall & 0.003 & 0.000 & 0.561 & -3.190 & 9.261 & 22.051 & 1.266 & 76452.267*** & 6.815 & -43.630*** \\
 & Training & 0.002 & 0.002 & 0.555 & -3.026 & 9.261 & 27.566 & 1.593 & 95536.649*** & 3.231 & -53.629*** \\
 & Validation & 0.032 & 0.020 & 0.716 & -3.190 & 3.716 & 3.052 & 0.116 & 140.216*** & 7.166 & -14.631*** \\
 & Test & -0.018 & -0.019 & 0.419 & -1.787 & 1.152 & 0.923 & -0.086 & 12.939** & 3.313 & -19.738*** \\
 \bottomrule
\end{tabular}%
}
      \begin{tablenotes}
    \small
    \item \textbf{Notes:} Significance levels: $^{*}p<0.05$, $^{**}p<0.01$, $^{***}p<0.001$.
  \end{tablenotes}
  \label{table:1}
\end{table}

We incorporate a broad set of external variables that capture information from various segments of the global financial and economic situation. Table~\ref{table:2} lists the set of covariates employed in this study, which are categorized into five groups: global foreign exchange, stock indices, fixed-income yields, commodity indices, and macroeconomic variables. These variables are chosen to provide a rich informational context, reflecting investor sentiment, monetary policy expectations, commodity price dynamics, and the overall health of the macroeconomy \citep{candian2025imperfect,fang202430,mueller2017exchange,rime2010exchange}.
The set includes other major exchange rates, key US stock indices like the S\&P 500 and NASDAQ, a range of US Treasury yields from 3-months to 10-years, broad commodity indices, and macroeconomic indicators such as the economic policy uncertainty (EPU) index, CPI, and GDP growth. When forecasting one of the target exchange rates returns, the other two are included as covariates to capture the dynamic relationships within the foreign exchange market. For example, in the task of predicting EUR/USD returns, the daily returns of USD/JPY and GBP/USD are used as input features. To handle the different frequencies of the data, lower-frequency variables were forward-filled to align with the daily timescale, ensuring that no future information is leaked into the training process.
All series spot exchange rates, equity indices, treasury yields, commodity indices, and macroeconomic indicators were retrieved from the Wind Database.

\begin{table}[htbp]
  \centering
  \setlength{\leftskip}{-50pt}
  \caption{The list of exogenous covariates used in the study.}
  \resizebox{160mm}{!}{%
\begin{tabular}{lll}
\toprule
Type & Abbreviation & Full Name \\
\midrule
Global Foreign Exchange & EUR to USD (EUR/USD) & Euro to US Dollar Exchange Rate \\
 & USD to JPY (USD/JPY) & US Dollar to Japanese Yen Exchange Rate \\
 & GBP to USD (GBP/USD) & British Pound to US Dollar Exchange Rate \\
 & USD to CHF & US Dollar to Swiss Franc Exchange Rate \\
 & USD to CAD & US Dollar to Canadian Dollar Exchange Rate \\
 & AUD to USD & Australian Dollar to US Dollar Exchange Rate \\
 & NZD to USD & New Zealand Dollar to US Dollar Exchange Rate \\
 & USD to HKD & US Dollar to Hong Kong Dollar Exchange Rate \\
 & USD to SGD & US Dollar to Singapore Dollar Exchange Rate \\ 
 \midrule
Stock Indices & NASDAQ & NASDAQ Composite Index \\
 & Dow Jones & Dow Jones Industrial Average \\
 & S\&P & S\&P 500 Index \\
  \midrule
Fixed-income yields & 3-mo & 3-Month US Treasury Bill Yield \\
 & 6-mo & 6-Month US Treasury Bill Yield \\
 & 1-yr & 1-Year US Treasury Note Yield \\
 & 5-yr & 5-Year US Treasury Note Yield \\
 & 10-yr & 10-Year US Treasury Bond Yield \\
 & Federal Fund & Federal Funds Effective Rate \\
  \midrule
Commodity Indices & Dow Jones Commodity & Dow Jones Commodity Index \\
 & Bloomberg Commodity & Bloomberg Commodity Index \\
 & S\&P GSCI Commodity & S\&P GSCI Commodity Index \\
 & TR\_CC CRB Excess Return & Thomson Reuters/Constant Maturity CRB Excess Return Index \\
  \midrule
Macroeconomic Variables & EPU & Economic Policy Uncertainty Index \\
 & Unemploy & Monthly US unemployment rate \\
 & CPI & Monthly US CPI for all-urban-consumer group \\
 & GDP & Quarterly US real GDP growth rate (\%) \\
 & PPI & Monthly US Producer Price Index \\
 & M2 & Monthly US M2 money supply (in billions USD) \\
 \bottomrule
\end{tabular}%
}
  \label{table:2}
\end{table}

\subsection{Experimental design and evaluation criteria}
This study focus on 1 day-ahead forecasting of the exchange rate returns, the most challenging and practically relevant horizon for exchange rate market participants \citep{darvas2025forecasting,engel2019uncovered,byrne2018sources}, and assess performance across five look-back window lengths $T \in \{5, 10, 15, 20, 30\}$ trading days. The choice of these window lengths is motivated by the need to balance the amount of historical information used for prediction with the model's ability to adapt to changing market conditions. The shorter windows capture recent trends and volatility, while the longer windows provide a broader context of historical behavior.

We compare EXFormer against popular deep learning architectures and a classical econometric model.
The vanilla Transformer with identical embedding dimension and number of layers, to isolate the impact of our multi-scale trend-aware modifications.
Temporal Convolutional Network (TCN) is a purely convolutional sequence model that achieves long receptive fields via dilated convolutions, representing an alternative multi-scale feature extractor.
GRU and LSTM are two recurrent architectures widely used in financial time-series forecasting, capable of capturing sequential dependencies but lacking multi-scale or interpretability modules.
Multilayer Perceptron (MLP) is a dense feed-forward network on lagged inputs, serving as a simple non-sequential benchmark.
ARIMA is a univariate autoregressive integrated moving-average model, chosen for its frequent use in time-series analysis.
Beyond these standard baselines, we also include: Informer, an attention-based architecture designed for long-sequence time-series forecasting with a probabilistic sparse self-attention mechanism that reduces computational complexity while preserving predictive power.
LSTM-ARIMA \citep{kashif2025lstm}, a hybrid approach that integrates linear ARIMA components with nonlinear sequence modeling from LSTM, allowing the capture of both autoregressive structures and nonlinear dependencies; 
and the large language model (LLM) Chronos (chronos-t5-base) \citep{ansari2024chronos}, a transformer-based foundation model recently adapted for time-series forecasting, which serves as a cutting-edge benchmark reflecting the capabilities of pretrained LLMs applied to financial prediction.

All deep learning models (EXFormer, Transformer, TCN, GRU, LSTM, MLP, Informer, LSTM-ARIMA, Chronos) are implemented in PyTorch 2.5.1 and trained on an ``NVIDIA RTX 3090 GPU'' with an ``Intel Xeon Gold 6330 CPU''. 
We employ Optuna to conduct automated hyperparameter tuning. The full search ranges appear in Appendix Table~\ref{table:a1}, and the optimal hyperparameter settings for each sliding window length are reported in Appendix Tables \ref{table:a2}, \ref{table:a3}, \ref{table:a4}. For ARIMA, we use $(p,d,q)=(1,1,0)$.

We assess predictive accuracy along three dimensions.
First, following \citep{darvas2025forecasting,rossi2013exchange}, we compute the mean squared forecast error (MSFE) of each model and express it as a percentage relative to that of the random walk. 
For model $m$, the MSFE is defined as
\begin{equation}
\text{MSFE}_m = \frac{1}{H}\sum_{h=1}^{H} \bigl(r_{t+h} - \hat{r}_{t+h|t}^m \bigr)^2,
\end{equation}
where $r_{t+h}$ denotes the realized return at horizon $h$, and $\hat{r}_{t+h|t}^m$ is the $h$-step-ahead forecast from model $m$. 
We then report the MSFE ratio
\begin{equation}
\text{MSFE Ratio}_m = \frac{\text{MSFE}_m}{\text{MSFE}_{\text{RW}}} \times 100,
\end{equation}
with values below $100$ indicating an improvement over the random walk (RW). 
To assess statistical significance in nested comparisons, we apply the Clark--West test \citep{clark2007approximately}, estimating HAC standard errors via \citet{newey1986simple} estimator with Bartlett kernel. Clark--West's correction accounts for the noise introduced by parameter estimation in the larger model.

Second, we measure directional accuracy (DA), defined as the proportion of days on which the model correctly predicts the sign of the return:
\begin{equation}
       \text{DA} = \frac{1}{N}\sum_{i=1}^{N}a(t),\quad a(t)\begin{cases}
        1,&\hat{y}_iy_i\geq 0,\\
        0,&\mathrm{otherwise}. \end{cases}
\end{equation}
Here, $N$ is the number of forecasting instances, $\hat{y}_i$ is the predicted return, and $y_i$ is the true return. 
The values of DA above 0.5 indicate better-than-chance performance. 
To account for serial correlation in the indicator series when using multi-day horizons, we follow \citet{blaskowitz2014testing} and estimate the dynamic regression using Newey--West HAC standard errors to obtain a robust $t$-statistic and $p$-value for the null that the model's directional accuracy does not exceed that of the random walk.

Third, to assess the economic value of daily exchange rate returns forecasting, we convert predicted returns into daily trading signals and simulate a series of long-short strategies across rolling estimation windows.
We construct the following strategies: 
At each day $t$, we generate a signal
\begin{equation}
    \mathrm{Signal}_{m,t} = \mathrm{sign}\bigl(\hat r_{m,t}\bigr)\in\{-1,0,1\},
\end{equation}
where \(\hat r_{m,t}\) is the model’s one‐day forecast.  The resulting strategy return is
\begin{equation}
      R_{m,t} = \mathrm{Signal}_{m,t}\,\times\,r_t,
\end{equation}
    with \(r_t\) the actual daily log‐return.  Cumulative return over \(T\) days is
\begin{equation}
      \mathrm{CR}_{m,T} = \prod_{t=1}^T\bigl(1 + R_{m,t}\bigr) - 1.
\end{equation}

A random walk (RW) benchmark uses $\mathrm{Signal}_{\mathrm{RW},t}=\mathrm{sign}(r_{t-1})$.
Buy and Hold (B\&H) strategy uses $\mathrm{Signal}_{\mathrm{BH},t}=1$ for all $t$.
Moving‐Average Crossover (MA) define a synthetic price series $\tilde S_t = 100\prod_{i=1}^t(1+r_i)$.  Compute $\mathrm{MA}_{20,t}$ and $\mathrm{MA}_{50,t}$ and set $\mathrm{Signal}_{\mathrm{MA},t} =
      \begin{cases}
        +1, & \mathrm{MA}_{20,t} > \mathrm{MA}_{50,t},\\
        -1, & \mathrm{MA}_{20,t} < \mathrm{MA}_{50,t},\\
        0, & \text{otherwise}.
      \end{cases}
    $
For each strategy, we compute the following performance metrics on the test set period, including mean, min, max daily return (\%), final cumulative return (\%), max drawdown (\%), Sharpe ratio, and Sortino ratio.

\section{Results}
\label{sec:sample5}
The results are organized into four core evaluations. First, in the forecasting results, we compare EXFormer against baselines using MSFE ratios and directional accuracy tests. Second, the trading backtests convert predictions into long--short signals. Third, the interpretability analysis traces the shifting importance of external drivers over time. Finally, ablation and robustness checks confirm the necessity of each architectural component and stability across market regimes.
\subsection{Forecasting results}

Table~\ref{table:3} reports the out-of-sample MSFE ratios (in percent) for EXFormer and all benchmarks, with the Clark--West $t$-statistics in parentheses. By construction, a ratio below 100 indicates improvement over the random walk benchmark.

Across all three currency pairs and every sliding window length, EXFormer achieves MSFE ratios well below 100, confirming consistent gains. For EUR/USD, EXFormer reduces MSFE to roughly 67--76 for 5--20 day windows (the Clark--West $t$-stat $>$ 7.8, $p <$ 0.001). Its superior performance deteriorates only in the 30-day window (98.8), where longer-horizon noise dominates.
For USD/JPY, EXFormer delivers MSFE ratios between 54 and 61 across windows, outstripping the random walk with $t$-statistics around 6.3--6.5 ($p < 0.001$). 
GBP/USD results are similar: EXFormer attains 49--58 MSFE ratios, outperforming Chronos, GRU, LSTM, TCN, LSTM-ARIMA and ARIMA models in nearly every configuration.
In every case, the Clark--West tests strongly reject the null of equal predictive accuracy at the 5\% level, demonstrating that the gains of EXFormer over both statistical and machine learning benchmarks are both economically meaningful and statistically robust.

The results also highlight the critical influence of the sliding window hyperparameter, as no single look-back period is universally optimal across all models and datasets. The performance of most models, including EXFormer, varies with the window length, reinforcing the necessity of the subsequent analysis to determine which configuration translates predictive accuracy into the most profitable trading strategy.
\begin{table}[htbp]
 \centering
 \vspace*{-30mm} 
 \setlength{\leftskip}{-70pt}
 \caption{The out-of-sample Mean Squared Forecast Error (MSFE) ratios with the Clark--West $t$-statistics for EXFormer and all benchmarks.}
 \resizebox{180mm}{!}{%
\begin{tabular}{lllllll}
\toprule
 &  &  &  & Sliding Window &  &  \\ \cline{3-7}
Type & Model & 5 & 10 & 15 & 20 & 30 \\
\midrule
EUR/USD & EXFormer & 76.042 (7.850***) & 76.835 (7.847***) & 71.463 (8.417***) & 67.053 (8.124***) & 98.800 (5.915***) \\
 & Chronos & 67.427 (10.569***) & 60.616 (10.429***) & 55.933 (10.482***) & 53.571 (10.505***) & 52.101 (9.961***) \\
 & Informer & 48.544 (10.566***) & 49.388 (10.334***) & 48.282 (10.236***) & 49.179 (10.109***) & 49.049 (9.827***) \\
 & Transformer & 48.386 (10.587***) & 48.977 (10.371***) & 48.281 (10.236***) & 48.499 (10.15***) & 49.138 (9.820***) \\
 & TCN & 64.166 (9.174***) & 73.745 (7.839***) & 52.472 (9.740***) & 53.655 (9.934***) & 56.250 (9.451***) \\
 & GRU & 52.527 (10.090***) & 59.901 (9.432***) & 52.494 (10.173***) & 61.250 (8.352***) & 52.547 (9.660***) \\
 & LSTM & 54.186 (9.717***) & 71.322 (8.241***) & 82.260 (6.934***) & 61.824 (8.518***) & 93.472 (6.402***) \\
 & MLP & 48.378 (10.578***) & 49.025 (10.335***) & 48.281 (10.235***) & 48.418 (10.149***) & 69.907 (8.891***) \\
 & ARIMA & 99.728 (8.777***) & 86.723 (9.766***) & 82.783 (9.891***) & 80.291 (9.753***) & 79.980 (9.490***) \\
 & LSTM-ARIMA & 66.271 (8.57***) & 52.380 (10.036***) & 48.362 (10.223***) & 50.898 (9.876***) & 50.800 (9.609***) \\
 \midrule
USD/JPY & EXFormer & 61.104 (6.333***) & 55.122 (6.512***) & 55.017 (6.397***) & 56.216 (6.340***) & 54.336 (6.186***) \\
 & Chronos & 72.016 (6.133***) & 64.877 (6.221***) & 60.378 (6.282***) & 79.760 (5.000***) & 57.999 (6.093***) \\
 & Informer & 56.816 (6.387***) & 55.493 (6.441***) & 55.581 (6.35***) & 55.453 (6.348***) & 56.764 (6.003***) \\
 & Transformer & 55.500 (6.477***) & 55.923 (6.389***) & 55.785 (6.329***) & 55.493 (6.338***) & 54.808 (6.126***) \\
 & TCN & 57.167 (6.413***) & 61.529 (6.010***) & 59.016 (6.215***) & 58.826 (6.101***) & 72.897 (5.086***) \\
 & GRU & 58.903 (6.432***) & 61.667 (5.974***) & 66.821 (5.826***) & 56.896 (6.119***) & 56.655 (6.058***) \\
 & LSTM & 59.389 (6.432***) & 59.324 (6.101***) & 143.701 (1.906***) & 110.725 (3.065***) & 59.264 (5.939***) \\
 & MLP & 55.545 (6.500***) & 83.748 (4.859***) & 62.121 (6.036***) & 56.052 (6.321***) & 54.365 (6.167***) \\
 & ARIMA & 108.593 (4.951***) & 94.084 (5.544***) & 89.121 (5.634***) & 85.070 (5.692***) & 83.032 (5.524***) \\
 & LSTM-ARIMA & 61.153 (6.284***) & 61.834 (5.884***) & 305.135 (-2.994***) & 67.707 (5.525***) & 79.017 (4.88***) \\
 \midrule
GBP/USD & EXFormer & 57.962 (9.876***) & 49.202 (9.471***) & 48.737 (9.350***) & 67.487 (7.797***) & 50.043 (8.972***) \\
 & Chronos & 68.425 (9.752***) & 58.779 (9.494***) & 56.869 (9.313***) & 54.177 (9.284***) & 51.367 (9.023***) \\
 & Informer & 74.020 (10.442***) & 50.261 (9.403***) & 49.070 (9.336***) & 49.556 (9.239***) & 51.107 (8.993***) \\
 & Transformer & 48.966 (9.571***) & 49.474 (9.448***) & 49.084 (9.335***) & 49.475 (9.251***) & 49.538 (8.960***) \\
 & TCN & 56.932 (9.239***) & 59.319 (8.941***) & 50.589 (9.261***) & 62.589 (8.408***) & 55.818 (8.751***) \\
 & GRU & 77.841 (7.535***) & 79.276 (6.905***) & 102.916 (5.206***) & 74.232 (7.191***) & 77.202 (6.938***) \\
 & LSTM & 86.978 (6.686***) & 53.842 (9.091***) & 49.077 (9.336*) & 49.021 (9.277***) & 121.463 (4.328***) \\
 & ARIMA & 111.392 (9.152***) & 90.759 (9.451***) & 84.115 (9.451***) & 81.946 (9.408***) & 80.082 (9.145***) \\
 & MLP & 49.282 (9.544***) & 53.706 (9.235***) & 49.303 (9.318***) & 49.042 (9.278***) & 49.626 (8.953***) \\
 & LSTM-ARIMA & 95.466 (5.869***) & 113.049 (3.819***) & 49.028 (9.34***) & 109.206 (4.551***) & 203.795 (-0.135) \\

\bottomrule
\end{tabular}%
}
  \begin{tablenotes}
  \small
  \item \textbf{Notes:} A value below 100 indicates that the model outperforms the random walk. The Clark--West $t$-statistics are shown in parentheses. $^{*}p<0.10$, $^{**}p<0.05$, $^{***}p<0.01$.
 \end{tablenotes}
 \label{table:3}
\end{table}

Across all three currency pairs, the DA results in Table~\ref{table:4} corroborate the MSFE findings by showing that EXFormer not only forecasts returns more precisely but also calls the sign of the move more often than the random walk. For EUR/USD, the random walk hit rate hovers around 0.499. EXFormer rises to 0.525--0.556 for four of the five window lengths, and the Blaskowitz--Herwartz $t$-statistics reach 1.65 in the 15-day window, significant at the 10\% level. While remaining positive and near unity elsewhere, competing deep learning models seldom exceed 0.510, and ARIMA stays below 0.470.

The directional gains are even more pronounced for USD/JPY. EXFormer attains DA values between 0.570 and 0.598, demonstrating a vivid improvement over the random walk with $t$-statistics between 2.37 and 3.21 ($p<0.05$ to $p<0.01$). 
Only isolated baseline configurations (e.g., Transformer at the 20-day window and MLP at 10-day) approach similar accuracy, whereas most others fall below 0.56 or even underperform the random walk. 

For GBP/USD, EXFormer again outperforms in four of five windows, achieving hit rates of 0.537--0.560. The statistical significance is weaker here (maximum $t=1.44$), yet the model remains directionally superior to both the random walk and the majority of baselines. 
In every currency pair, ARIMA exhibits hit rates below 0.50 and negative $t$-values, confirming its inferiority in sign prediction. 
These results indicate that EXFormer's improvements are not confined to the MSFE ratio but translate into meaningful increases in the probability of correctly anticipating the direction of daily exchange rate returns changes.

\begin{table}[htbp]
 \centering
\vspace*{-40mm} 
 \setlength{\leftskip}{-70pt}
 \caption{The out-of-sample directional accuracy (DA) with the Blaskowitz--Herwartz $t$-statistics for EXFormer and all benchmarks in 1 day-ahead.}
 \resizebox{180mm}{!}{%
\begin{tabular}{lllllll}
    \toprule
 &  &  &  & Sliding Window &  &  \\ \cline{3-7} 
Type & Model & 5 & 10 & 15 & 20 & 30 \\
\midrule
EUR/USD & Random Walk & 0.499 & 0.499 & 0.499 & 0.499 & 0.499 \\ \cline{2-7} 
 & EXFormer & \textbf{0.530 (0.974)} & \textbf{0.510 (0.445)} & \textbf{0.556 (1.647*)} & \textbf{0.538 (1.162)} & \textbf{0.525 (0.560)} \\
 & Chronos & 0.478 (-0.806) & 0.482 (-0.709) & 0.469 (-1.151) & 0.462 (-1.446) & 0.478 (-0.842) \\
 & Informer & 0.505 (0.407) & 0.496 (0.074) & 0.503 (0.248) & 0.499 (0.074) & 0.507 (0.175) \\
 & Transformer & 0.500 (0.157) & 0.499 (0.153) & 0.503 (0.248) & 0.499 (0.074) & 0.510 (0.335) \\
 & TCN & 0.497 (0.075) & 0.488 (-0.239) & 0.489 (-0.210) & 0.521 (0.594) & 0.501 (0.004) \\
 & GRU & 0.516 (0.603) & 0.504 (0.243) & 0.475 (-0.567) & 0.516 (0.482) & 0.496 (-0.07) \\
 & LSTM & 0.500 (0.173) & 0.510 (0.373) & 0.480 (-0.430) & 0.513 (0.427) & 0.510 (0.299) \\
 & MLP & 0.503 (0.327) & 0.504 (0.248) & 0.503 (0.248) & 0.501 (0.093) & 0.513 (0.362) \\
 & ARIMA & 0.470 (-0.706) & 0.466 (-0.906) & 0.466 (-0.942) & 0.462 (-1.173) & 0.458 (-1.501) \\
 & LSTM-ARIMA & 0.519 (0.600) & 0.463 (-0.826) & 0.503 (0.248) & 0.521 (0.632) & 0.516 (0.358) \\
 \midrule
USD/JPY & Random Walk & 0.487 & 0.487 & 0.487 & 0.487 & 0.487 \\ \cline{2-7} 
 & EXFormer & \textbf{0.582 (2.633***)} & 0.570 (2.365**) & \textbf{0.598 (3.206***)} & \textbf{0.586 (2.763***)} & \textbf{0.589 (2.788***)} \\
 & Chronos & 0.489 (-0.146) & 0.466 (-1.197) & 0.525 (1.366) & 0.541 (1.408) & 0.531 (1.932*) \\
 & Informer & 0.413 (-2.024**) & 0.416 (-1.900*) & 0.413 (-1.902*) & 0.569 (2.526**) & 0.411 (-1.941*) \\
 & Transformer & 0.424 (-1.728**) & 0.408 (-2.042**) & 0.413 (-1.902*) & 0.579 (2.626***) & 0.420 (-1.674*) \\
 & TCN & 0.563 (2.023**) & 0.466 (-0.599) & 0.556 (1.628) & 0.504 (0.354) & 0.431 (-1.368) \\
 & GRU & 0.563 (2.018**) & 0.435 (-1.306) & 0.422 (-1.745*) & 0.465 (-0.593) & 0.528 (0.923) \\
 & LSTM & 0.573 (2.294**) & 0.482 (-0.197) & 0.455 (-0.808) & 0.453 (-0.932) & 0.539 (1.230) \\
 & MLP & 0.413 (-2.024**) & \textbf{0.590 (2.755***)} & 0.573 (2.185**) & 0.484 (-0.147) & 0.545 (1.427) \\
 & ARIMA & 0.497 (0.205) & 0.501 (0.544) & 0.480 (-0.319) & 0.510 (0.650) & 0.499 (0.425) \\
 & LSTM-ARIMA & 0.574 (2.328**) & 0.540 (1.305) & 0.427 (-1.641) & 0.419 (-1.837) & 0.51 (0.534*) \\
 \midrule
GBP/USD & Random Walk & 0.507 & 0.507 & 0.507 & 0.507 & 0.507 \\ \cline{2-7} 
 & EXFormer & \textbf{0.560 (1.441)} & \textbf{0.537 (0.902)} & \textbf{0.550 (1.256)} & 0.516 (0.167) & \textbf{0.542 (1.090)} \\
 & Chronos & 0.481 (-1.536) & 0.474 (-1.691) & 0.489 (-0.827) & 0.504 (-0.137) & 0.507 (0.122) \\
 & Informer & 0.489 (-1.218) & 0.527 (0.802) & 0.536 (0.902) & 0.465 (-1.059) & 0.531 (0.891) \\
 & Transformer & 0.533 (0.727) & 0.460 (-1.125) & 0.461 (-1.057) & 0.535 (0.751) & 0.469 (-0.849) \\
 & TCN & 0.511 (0.091) & 0.468 (-0.923) & 0.511 (0.238) & 0.493 (-0.480) & 0.525 (0.502) \\
 & GRU & 0.516 (0.145) & 0.471 (-0.870) & 0.517 (0.393) & 0.510 (0.076) & 0.539 (0.927) \\
 & LSTM & 0.505 (-0.149) & 0.466 (-1.019) & 0.536 (0.902) & \textbf{0.538 (0.823)} & 0.510 (0.163) \\
 & MLP & 0.465 (-1.196) & 0.477 (-0.922) & 0.461 (-1.057) & 0.535 (0.751) & 0.469 (-0.849) \\
 & ARIMA & 0.489 (-0.760) & 0.485 (-0.759) & 0.500 (-0.227) & 0.482 (-1.055) & 0.487 (-0.744) \\
 & LSTM-ARIMA & 0.516 (0.233) & 0.468 (-0.936) & 0.536 (0.902) & 0.516 (0.219) & 0.531 (0.648) \\
    \bottomrule
\end{tabular}%
}
 \begin{tablenotes}
  \small
  \item \textbf{Notes:} The Blaskowitz--Herwartz $t$-statistics (in parentheses) test whether DA exceeds that of the random walk. Significance levels: $^{*}p<0.10$, $^{**}p<0.05$, $^{***}p<0.01$.
 \end{tablenotes}
 \label{table:4}
\end{table}

\begin{figure}[htbp]
  \centering
  \setlength{\leftskip}{-80pt}
   \vspace*{-38mm} 
  \includegraphics[width=185mm]{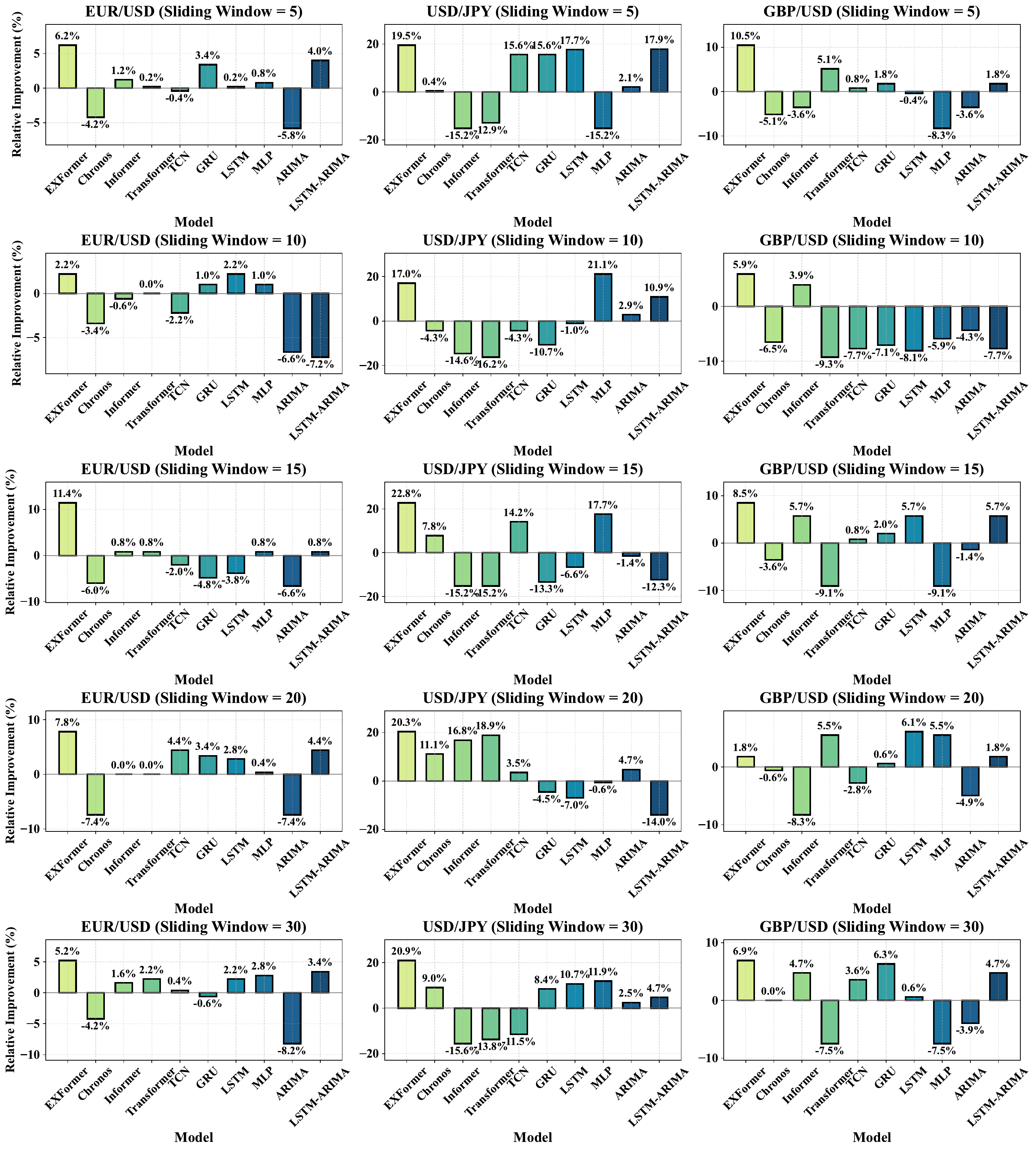}
  \caption{The directional accuracy (DA) improvement of EXFormer and all benchmarks relative to the random walk for EUR/USD, USD/JPY, and GBP/USD in 1 day-ahead. The bars show the percentage point lift for each model and sliding window size $T\in\{5,10,15,20,30\}$. Positive values indicate that the model predicts the correct sign more often than the random walk.}
  \label{fig:4}
\end{figure}

Moreover, Figure~\ref{fig:4} presents each model's DA improvement relative to the random walk benchmark, across sliding windows \(T\in\{5,10,15,20,30\}\) and the three currency pairs. 
EXFormer uniformly delivers the largest positive lifts, peaking at 22.8\% for USD/JPY with \(T=15\), 11.4\% for EUR/USD at \(T=15\), and 10.5\% for GBP/USD at \(T=5\), and maintains gains above 5\% even at the rest of sliding windows (except for GBP/USD \(T=20\) and EUR/USD \(T=10\)). 
In contrast, the vanilla Transformer yields only modest improvements, but a lot of time not outperform the random walk. TCN show the same improvement of EXFormer in the EUR/USD at \(T=20\)) but fails at others. 
Moreover, GRU and LSTM occasionally posts gains (up to 14\% for USD/JPY at \(T=5\)) but also suffers negative performance in several configurations. MLP exhibits marginal improvements under most settings and ARIMA consistently underperforms the random walk. 
The recently proposed Chronos LLM displays moderate gains in some of slinding windows of USD/JPY, typically in the 7.8--11.1\% range, but hard to supress the random walk, highlighting its potential to capture broad semantic-temporal patterns but with less specialization for exchange rate returns.
The Informer, designed for efficient long-sequence forecasting, shows sporadic improvements (notably for USD/JPY with $T=20$), but lacks consistency across pairs, underscoring the need for more tailored inductive biases in high-frequency financial data.
The LSTM-ARIMA hybrid achieves modest gains in some short-horizon windows (e.g., GBP/USD at $T=15$), but overall its improvements remain small and unstable, reflecting the difficulty of integrating linear ARIMA structure with nonlinear dynamics at the daily frequency.
The results confirming that EXFormer's multi-scale trend-aware attention and pre-hoc variable selection translate into substantively larger and more stable directional-forecasting gains than both modern deep-learning and traditional time-series benchmarks.

\subsection{Backtesting}
To assess the practical applicability and profitability of our proposed model, a rigorous backtesting simulation was performed. This step is critical in financial forecasting, as it translates predictive accuracy into tangible economic value by simulating a trading strategy based on the model's signals. We evaluated the performance of all models across various sliding window lengths (5, 10, 15, 20, and 30 days) to identify the optimal look-back period, a key hyperparameter that significantly influences model performance. Our analysis revealed that a sliding window of 15 days consistently yielded the most robust and profitable results for the EXFormer model.

Table \ref{table:5} presents the backtesting outcomes for the optimal 15-day sliding window across three major currency pairs (EUR/USD, USD/JPY, and GBP/USD). For the EUR/USD pair, EXFormer demonstrated superior performance, achieving a mean daily return of 0.050\% and a cumulative return of 18.572\%. This performance starkly contrasts with all baseline models, most of which yielded negligible or negative returns. For example, LSTM and ARIMA posted cumulative returns of -6.640\% and -3.942\%, respectively. Furthermore, EXFormer also exhibited superior risk management, as evidenced by the lowest maximum drawdown of 4.068\%. The combination of high returns and controlled risk is outlined in its Sharpe ratio (2.087) and Sortino ratio (3.237), indicating a highly favorable risk-adjusted return profile that no other model approached.

The pattern of the proposed model's dominance was replicated across the other currency pairs. For USD/JPY, EXFormer achieved the highest cumulative return of 25.782\% and a mean return of 0.068\%, again coupled with an Sharpe ratio of 1.891. While some models like MLP and TCN managed positive returns, they were less profitable and, in many cases, incurred greater risk, as seen in the high drawdowns for models like Informer and Transformer (17.703\%). Similarly, for the GBP/USD pair, EXFormer secured a cumulative return of 18.373\% and a Sharpe ratio of 1.903. This performance was better than the next best models, TCN and GRU, and highlighted the unprofitability of strategies based on Transformer, MLP, and Buy \& Hold, which all resulted in cumulative losses of nearly -4\%.

\begin{table}[htbp]
 \centering
 \vspace*{-20mm} 
 \setlength{\leftskip}{-100pt}
 \caption{The backtesting results for EXFormer and all benchmarks using a 15-day sliding window ($T = 15$) across EUR/USD, USD/JPY, and GBP/USD.}
 \resizebox{200mm}{!}{%
\begin{tabular}{ccccccccc}
\toprule
Type & Model & Mean Return(\%) & Minimum Return(\%) & Maximum Return(\%) & Max Drawdown(\%) & Sharpe Ratio & Sortino Ratio & Cumulative Return(\%) \\
\midrule
EUR/USD & Random Walk & 0.008 & -1.101 & 1.678 & 7.003 & 0.323 & 0.545 & 2.469 \\ \cline{2-9}
 & \textbf{EXFormer} & \textbf{0.050} & \textbf{-1.101} & \textbf{1.678} & \textbf{4.068} & \textbf{2.087} & \textbf{3.237} & \textbf{18.572} \\
 & Chronos & -0.009 & -1.101 & 1.678 & 8.629 & -0.384 & -0.633 & -3.401 \\
 & Informer & 0.001 & -1.061 & 1.678 & 6.906 & 0.047 & 0.076 & 0.140 \\
 & Transformer & 0.001 & -1.061 & 1.678 & 6.906 & 0.047 & 0.076 & 0.140 \\
 & TCN & -0.002 & -1.678 & 1.096 & 5.549 & -0.102 & -0.160 & -1.101 \\
 & GRU & -0.001 & -1.678 & 1.101 & 8.297 & -0.028 & -0.042 & -0.485 \\
 & LSTM & -0.019 & -1.678 & 1.061 & 9.334 & -0.793 & -1.159 & -6.640 \\
 & MLP & 0.001 & -1.061 & 1.678 & 6.906 & 0.047 & 0.076 & 0.140 \\
 & ARIMA & -0.011 & -1.101 & 1.678 & 7.579 & -0.451 & -0.772 & -3.942 \\
 & LSTM-ARIMA & 0.001 & -1.061 & 1.678 & 6.906 & 0.047 & 0.076 & 0.140 \\
 & Buy \& Hold & -0.001 & -1.678 & 1.061 & 6.645 & -0.047 & -0.070 & -0.648 \\
 & Moving-Average & -0.018 & -1.061 & 1.678 & 10.233 & -0.800 & -1.257 & -6.047 \\
 \midrule
USD/JPY & Random Walk & 0.040 & -2.187 & 1.916 & 8.211 & 1.089 & 1.643 & 13.923 \\ \cline{2-9}
 & \textbf{EXFormer} & \textbf{0.068} & \textbf{-2.187} & \textbf{1.847} & \textbf{6.778} & \textbf{1.891} & \textbf{2.529} & \textbf{25.782} \\
 & Chronos & 0.016 & -2.090 & 2.187 & 9.551 & 0.431 & 0.602 & 4.912 \\
 & Informer & -0.023 & -1.752 & 2.187 & 17.246 & -0.630 & -1.084 & -8.108 \\
 & Transformer & -0.023 & -1.752 & 2.187 & 17.246 & -0.630 & -1.084 & -8.108 \\
 & TCN & 0.018 & -2.090 & 2.187 & 8.374 & 0.488 & 0.682 & 5.692 \\
 & GRU & -0.031 & -2.090 & 2.187 & 17.050 & -0.839 & -1.312 & -10.478 \\
 & LSTM & -0.021 & -2.090 & 2.187 & 12.467 & -0.575 & -0.883 & -7.471 \\
 & MLP & 0.033 & -2.090 & 2.187 & 7.860 & 0.895 & 1.223 & 11.202 \\
 & ARIMA & 0.001 & -2.187 & 2.090 & 16.747 & 0.039 & 0.059 & -0.089 \\
 & LSTM-ARIMA & -0.019 & -1.752 & 2.187 & 16.165 & -0.514 & -0.883 & -6.762 \\
 & Buy \& Hold & -0.023 & -1.752 & 2.187 & 17.246 & -0.630 & -1.084 & -8.108 \\
 & Moving-Average & 0.028 & -2.090 & 2.187 & 5.223 & 0.825 & 1.156 & 9.571 \\
 \midrule
GBP/USD & Random Walk & 0.009 & -1.152 & 1.787 & 8.934 & 0.326 & 0.532 & 2.700 \\ \cline{2-9}
 & \textbf{EXFormer} & \textbf{0.050} & \textbf{-1.097} & \textbf{1.787} & \textbf{3.526} & \textbf{1.903} & \textbf{3.607} & \textbf{18.373} \\
 & Chronos & -0.004 & -1.787 & 1.432 & 7.149 & -0.166 & -0.253 & -1.795 \\
 & Informer & 0.011 & -1.152 & 1.787 & 8.120 & 0.435 & 0.680 & 3.716 \\
 & Transformer & -0.011 & -1.787 & 1.152 & 9.133 & -0.435 & -0.660 & -4.163 \\
 & TCN & 0.017 & -1.787 & 1.432 & 7.586 & 0.636 & 1.013 & 5.629 \\
 & GRU & 0.015 & -1.152 & 1.787 & 9.430 & 0.551 & 0.873 & 4.818 \\
 & LSTM & 0.011 & -1.152 & 1.787 & 8.120 & 0.435 & 0.680 & 3.716 \\
 & MLP & -0.011 & -1.787 & 1.152 & 9.133 & -0.435 & -0.660 & -4.163 \\
 & ARIMA & -0.001 & -1.152 & 1.787 & 8.596 & -0.022 & -0.036 & -0.502 \\
 & LSTM-ARIMA & 0.011 & -1.152 & 1.787 & 8.120 & 0.435 & 0.680 & 3.716 \\
 & Buy \& Hold & -0.011 & -1.787 & 1.152 & 9.133 & -0.435 & -0.660 & -4.163 \\
 & Moving-Average & 0.005 & -1.787 & 1.172 & 4.957 & 0.199 & 0.262 & 1.398 \\
\bottomrule
\end{tabular}%
}
 \label{table:5}
\end{table}

The sensitivity of model performance to the sliding window hyperparameter is detailed in Tables \ref{table:6}, \ref{table:7}, and \ref{table:8} for EUR/USD, USD/JPY, and GBP/USD, respectively. These results emphasize the rationale for selecting the 15-day window. As shown in Table \ref{table:6} (EUR/USD), while EXFormer maintained positive returns across most window sizes, its performance metrics, including mean daily return, Sharpe ratio, and final cumulative return, peaked decisively at the 15-day length. Other models displayed random performance. For example, TCN and GRU performed well with a 20-day window but were suboptimal at other lengths, demonstrating a lack of robustness. A similar trend is evident in Tables \ref{table:7} (USD/JPY) and \ref{table:8} (GBP/USD). Across all currency pairs, the 15-day window consistently emerged as a ``sweet spot'' for EXFormer, maximizing its predictive leverage. In contrast, the baseline models either failed to show a clear optimal window or performed poorly across all tested configurations, reinforcing the conclusion that EXFormer's architecture is not only more powerful but also more stable in a practical trading simulation.

The results for alternative sliding windows (Tables \ref{table:6}, \ref{table:7}, and \ref{table:8}) confirm the robustness of these findings. For EUR/USD the 5-day window already yields a positive Sharpe of 0.736, but performance peaks at 15 days before tapering off as longer windows dilute recent information. A similar pattern emerges for USD/JPY, where EXFormer's Sharpe climbs from 0.270 (5 days) to 1.891 (15 days) and declines thereafter, yet remains above every comparator. GBP/USD exhibits its best Sharpe at 15 days as well, with shorter windows offering smaller gains and longer windows dominated by market noise. In all currencies, traditional models or strategies (ARIMA, Buy \& Hold, Moving-Average) and most neural baselines either underperform the random walk or achieve positive returns only at the cost of larger drawdowns and lower risk-adjusted metrics. The backtest results demonstrate that EXFormer's forecasting edge translates into consistent, superior trading performance and that a 15-day estimation window provides the optimal trade-off between adaptability and signal strength.

\begin{table}[htbp]
 \centering
 \setlength{\leftskip}{-70pt}
 \vspace*{-45mm} 
 \caption{The backtesting results for EXFormer and all benchmarks using sliding window size $T\in\{5,10,20,30\}$ across EUR/USD.}
 \resizebox{180mm}{!}{%
\begin{tabular}{ccccccccc}
\toprule
Sliding   Window & Model & Mean Return(\%) & Minimum Return(\%) & Maximum Return(\%) & Max Drawdown(\%) & Sharpe Ratio & Sortino Ratio & Cumulative Return(\%) \\
\midrule
5 & Random Walk & 0.008 & -1.101 & 1.678 & 7.003 & 0.323 & 0.545 & 2.469 \\ \cline{2-9}
 & EXFormer & 0.018 & -1.678 & 1.096 & 4.273 & 0.736 & 1.100 & 6.068 \\
 & Chronos & 0.019 & -1.101 & 1.678 & 6.293 & 0.794 & 1.393 & 6.578 \\
 & Informer & 0.001 & -1.061 & 1.678 & 6.906 & 0.052 & 0.084 & 0.182 \\
 & Transformer & -0.005 & -1.101 & 1.678 & 5.572 & -0.192 & -0.296 & -1.841 \\
 & TCN & 0.000 & -1.678 & 1.101 & 6.566 & -0.002 & -0.003 & -0.273 \\
 & GRU & 0.003 & -1.096 & 1.678 & 8.782 & 0.136 & 0.213 & 0.886 \\
 & LSTM & 0.005 & -1.096 & 1.678 & 5.358 & 0.221 & 0.361 & 1.607 \\
 & MLP & 0.001 & -1.061 & 1.678 & 6.906 & 0.047 & 0.076 & 0.140 \\
 & ARIMA & -0.022 & -1.101 & 1.678 & 9.554 & -0.899 & -1.448 & -7.459 \\
 & LSTM-ARIMA & -0.002 & -1.678 & 1.061 & 5.325 & -0.082 & -0.126 & -0.940 \\
 & Buy \& Hold & -0.001 & -1.678 & 1.061 & 6.645 & -0.047 & -0.070 & -0.648 \\
 & Moving-Average & -0.018 & -1.061 & 1.678 & 10.233 & -0.800 & -1.257 & -6.047 \\
 \midrule
10 & Random Walk & 0.008 & -1.101 & 1.678 & 7.003 & 0.323 & 0.545 & 2.469 \\ \cline{2-9}
 & EXFormer & 0.008 & -1.678 & 1.072 & 8.090 & 0.334 & 0.502 & 2.565 \\
 & Chronos & 0.000 & -1.101 & 1.678 & 7.844 & 0.019 & 0.031 & -0.095 \\
 & Informer & -0.001 & -1.678 & 1.061 & 6.645 & -0.047 & -0.070 & -0.648 \\
 & Transformer & -0.011 & -1.678 & 1.072 & 11.090 & -0.453 & -0.655 & -3.959 \\
 & TCN & -0.009 & -1.678 & 1.101 & 6.072 & -0.376 & -0.586 & -3.340 \\
 & GRU & 0.003 & -1.072 & 1.678 & 5.533 & 0.128 & 0.204 & 0.815 \\
 & LSTM & -0.005 & -1.101 & 1.678 & 7.530 & -0.225 & -0.353 & -2.110 \\
 & MLP & 0.001 & -1.061 & 1.678 & 6.906 & 0.047 & 0.076 & 0.140 \\
 & ARIMA & -0.011 & -1.101 & 1.678 & 8.955 & -0.457 & -0.790 & -3.992 \\
 & LSTM-ARIMA & -0.036 & -1.678 & 1.061 & 11.901 & -1.490 & -2.174 & -11.883 \\
 & Buy \& Hold & -0.001 & -1.678 & 1.061 & 6.645 & -0.047 & -0.070 & -0.648 \\
 & Moving-Average & -0.018 & -1.061 & 1.678 & 10.233 & -0.800 & -1.257 & -6.047 \\
 \midrule
20 & Random Walk & 0.008 & -1.101 & 1.678 & 7.003 & 0.323 & 0.545 & 2.469 \\ \cline{2-9}
 & EXFormer & 0.014 & -1.678 & 1.072 & 5.282 & 0.557 & 0.831 & 4.492 \\
 & Chronos & -0.012 & -1.101 & 1.678 & 11.437 & -0.480 & -0.768 & -4.176 \\
 & Informer & -0.001 & -1.678 & 1.061 & 6.645 & -0.047 & -0.070 & -0.648 \\
 & Transformer & -0.001 & -1.678 & 1.061 & 6.645 & -0.047 & -0.070 & -0.648 \\
 & TCN & 0.022 & -1.678 & 1.101 & 5.032 & 0.925 & 1.446 & 7.748 \\
 & GRU & 0.028 & -1.678 & 1.101 & 5.883 & 1.159 & 1.798 & 9.861 \\
 & LSTM & 0.021 & -1.101 & 1.678 & 4.987 & 0.866 & 1.490 & 7.220 \\
 & MLP & 0.001 & -1.061 & 1.678 & 6.906 & 0.047 & 0.076 & 0.140 \\
 & ARIMA & -0.014 & -1.101 & 1.678 & 7.911 & -0.579 & -0.996 & -4.965 \\
 & LSTM-ARIMA & 0.013 & -1.096 & 1.678 & 7.091 & 0.525 & 0.795 & 4.212 \\
 & Buy \& Hold & -0.001 & -1.678 & 1.061 & 6.645 & -0.047 & -0.070 & -0.648 \\
 & Moving-Average & -0.018 & -1.061 & 1.678 & 10.233 & -0.800 & -1.257 & -6.047 \\
 \midrule
30 & Random Walk & 0.008 & -1.101 & 1.678 & 7.003 & 0.323 & 0.545 & 2.469 \\ \cline{2-9}
 & EXFormer & 0.018 & -1.101 & 1.678 & 6.644 & 0.747 & 1.192 & 6.161 \\
 & Chronos & 0.001 & -1.101 & 1.678 & 7.594 & 0.026 & 0.043 & -0.037 \\
 & Informer & 0.004 & -1.061 & 1.678 & 6.216 & 0.156 & 0.251 & 1.056 \\
 & Transformer & 0.008 & -1.061 & 1.678 & 6.906 & 0.318 & 0.511 & 2.427 \\
 & TCN & 0.017 & -1.678 & 1.101 & 5.459 & 0.689 & 1.080 & 5.653 \\
 & GRU & 0.003 & -1.678 & 1.061 & 6.711 & 0.130 & 0.194 & 0.833 \\
 & LSTM & -0.001 & -1.101 & 1.678 & 7.932 & -0.055 & -0.087 & -0.708 \\
 & MLP & -0.007 & -1.678 & 1.101 & 7.836 & -0.279 & -0.392 & -2.556 \\
 & ARIMA & -0.022 & -1.101 & 1.678 & 10.378 & -0.906 & -1.546 & -7.511 \\
 & LSTM-ARIMA & 0.002 & -1.101 & 1.678 & 6.083 & 0.072 & 0.113 & 0.349 \\
 & Buy \& Hold & -0.001 & -1.678 & 1.061 & 6.645 & -0.047 & -0.070 & -0.648 \\
 & Moving-Average & -0.018 & -1.061 & 1.678 & 10.233 & -0.800 & -1.257 & -6.047 \\
\bottomrule
\end{tabular}%
}
 \label{table:6}
\end{table}

\begin{table}[htbp]
 \centering
 \setlength{\leftskip}{-70pt}
 \vspace*{-45mm} 
 \caption{The backtesting results for EXFormer and all benchmarks using sliding window size $T\in\{5,10,20,30\}$ across USD/JPY.}
 \resizebox{180mm}{!}{%
\begin{tabular}{ccccccccc}
    \toprule
Sliding Window & Model & Mean Return(\%) & Minimum Return(\%) & Maximum Return(\%) & Max Drawdown(\%) & Sharpe Ratio & Sortino Ratio & Cumulative Return(\%) \\
\midrule
5 & Random Walk & 0.040 & -2.187 & 1.916 & 8.211 & 1.089 & 1.643 & 13.923 \\ \cline{2-9}
 & EXFormer & 0.010 & -2.187 & 1.752 & 10.914 & 0.270 & 0.342 & 2.852 \\
 & Chronos & -0.003 & -2.187 & 1.916 & 8.599 & -0.081 & -0.110 & -1.570 \\
 & Informer & -0.023 & -1.752 & 2.187 & 17.246 & -0.630 & -1.084 & -8.108 \\
 & Transformer & -0.011 & -1.752 & 2.187 & 15.604 & -0.300 & -0.534 & -4.229 \\
 & TCN & 0.043 & -2.090 & 2.187 & 6.027 & 1.176 & 1.588 & 15.162 \\
 & GRU & 0.020 & -2.187 & 1.701 & 8.446 & 0.542 & 0.693 & 6.408 \\
 & LSTM & 0.011 & -2.187 & 1.752 & 11.004 & 0.299 & 0.382 & 3.225 \\
 & MLP & -0.023 & -1.752 & 2.187 & 17.246 & -0.630 & -1.084 & -8.108 \\
 & ARIMA & 0.009 & -1.775 & 2.187 & 11.482 & 0.254 & 0.397 & 2.648 \\
 & LSTM-ARIMA & 0.012 & -2.187 & 1.752 & 11.095 & 0.338 & 0.427 & 3.725 \\
 & Buy \& Hold & -0.023 & -1.752 & 2.187 & 17.246 & -0.630 & -1.084 & -8.108 \\
 & Moving-Average & 0.028 & -2.090 & 2.187 & 5.223 & 0.825 & 1.156 & 9.571 \\
 \midrule
10 & Random Walk & 0.040 & -2.187 & 1.916 & 8.211 & 1.089 & 1.643 & 13.923 \\ \cline{2-9}
 & EXFormer & 0.015 & -2.090 & 2.187 & 13.229 & 0.399 & 0.526 & 4.525 \\
 & Chronos & 0.014 & -2.090 & 1.916 & 12.964 & 0.388 & 0.617 & 4.282 \\
 & Informer & -0.023 & -1.752 & 2.187 & 17.246 & -0.630 & -1.084 & -8.108 \\
 & Transformer & -0.029 & -1.752 & 2.187 & 19.060 & -0.789 & -1.365 & -9.907 \\
 & TCN & 0.025 & -1.916 & 2.187 & 5.435 & 0.691 & 1.131 & 8.412 \\
 & GRU & -0.020 & -1.916 & 2.187 & 10.712 & -0.538 & -0.905 & -7.041 \\
 & LSTM & -0.023 & -1.916 & 2.187 & 11.974 & -0.646 & -0.967 & -8.288 \\
 & MLP & 0.002 & -2.187 & 1.752 & 11.957 & 0.067 & 0.085 & 0.260 \\
 & ARIMA & 0.014 & -2.187 & 2.090 & 13.018 & 0.396 & 0.586 & 4.480 \\
 & LSTM-ARIMA & 0.028 & -1.752 & 2.187 & 11.480 & 0.773 & 1.185 & 9.524 \\
 & Buy \& Hold & -0.023 & -1.752 & 2.187 & 17.246 & -0.630 & -1.084 & -8.108 \\
 & Moving-Average & 0.028 & -2.090 & 2.187 & 5.223 & 0.825 & 1.156 & 9.571 \\
 \midrule
20 & Random Walk & 0.040 & -2.187 & 1.916 & 8.211 & 1.089 & 1.643 & 13.923 \\ \cline{2-9}
 & EXFormer & 0.023 & -2.187 & 1.752 & 13.814 & 0.625 & 0.774 & 7.514 \\
 & Chronos & 0.021 & -2.187 & 1.775 & 12.435 & 0.566 & 0.773 & 6.679 \\
 & Informer & 0.023 & -2.187 & 1.752 & 11.095 & 0.630 & 0.801 & 7.584 \\
 & Transformer & 0.023 & -2.187 & 1.752 & 11.095 & 0.630 & 0.801 & 7.584 \\
 & TCN & 0.006 & -2.090 & 2.187 & 9.857 & 0.161 & 0.236 & 1.455 \\
 & GRU & 0.004 & -1.916 & 2.187 & 7.014 & 0.124 & 0.207 & 0.980 \\
 & LSTM & -0.002 & -1.916 & 2.187 & 13.912 & -0.068 & -0.113 & -1.415 \\
 & MLP & 0.001 & -2.090 & 2.187 & 8.024 & 0.028 & 0.042 & -0.219 \\
 & ARIMA & 0.012 & -2.187 & 2.090 & 12.801 & 0.331 & 0.492 & 3.635 \\
 & LSTM-ARIMA & -0.050 & -1.916 & 2.187 & 21.670 & -1.382 & -2.164 & -16.318 \\
 & Buy \& Hold & -0.023 & -1.752 & 2.187 & 17.246 & -0.630 & -1.084 & -8.108 \\
 & Moving-Average & 0.028 & -2.090 & 2.187 & 5.223 & 0.825 & 1.156 & 9.571 \\
 \midrule
30 & Random Walk & 0.040 & -2.187 & 1.916 & 8.211 & 1.089 & 1.643 & 13.923 \\ \cline{2-9}
 & EXFormer & 0.030 & -2.187 & 1.775 & 11.095 & 0.834 & 1.090 & 10.351 \\
 & Chronos & 0.020 & -2.187 & 1.916 & 9.702 & 0.546 & 0.762 & 6.458 \\
 & Informer & -0.023 & -1.752 & 2.187 & 17.246 & -0.630 & -1.084 & -8.108 \\
 & Transformer & -0.013 & -1.752 & 2.187 & 15.188 & -0.361 & -0.623 & -4.960 \\
 & TCN & -0.011 & -1.701 & 2.187 & 12.180 & -0.300 & -0.475 & -4.236 \\
 & GRU & 0.012 & -2.090 & 2.187 & 10.327 & 0.333 & 0.473 & 3.657 \\
 & LSTM & -0.008 & -2.187 & 2.090 & 10.914 & -0.217 & -0.300 & -3.234 \\
 & MLP & 0.031 & -2.090 & 2.187 & 5.394 & 0.854 & 1.219 & 10.638 \\
 & ARIMA & 0.017 & -2.187 & 2.090 & 14.610 & 0.460 & 0.693 & 5.322 \\
 & LSTM-ARIMA & 0.039 & -1.916 & 2.187 & 7.120 & 1.082 & 1.710 & 13.816 \\
 & Buy \& Hold & -0.023 & -1.752 & 2.187 & 17.246 & -0.630 & -1.084 & -8.108 \\
 & Moving-Average & 0.028 & -2.090 & 2.187 & 5.223 & 0.825 & 1.156 & 9.571 \\
 \bottomrule
\end{tabular}%
}
 \label{table:7}
\end{table}

\begin{table}[htbp]
 \centering
 \setlength{\leftskip}{-70pt}
 \vspace*{-45mm} 
 \caption{The backtesting results for EXFormer and all benchmarks using sliding window size $T\in\{5,10,20,30\}$ across GBP/USD.}
 \resizebox{180mm}{!}{%
\begin{tabular}{ccccccccc}
    \toprule
Sliding Window & Model & Mean Return(\%) & Minimum Return(\%) & Maximum Return(\%) & Max Drawdown(\%) & Sharpe Ratio & Sortino Ratio & Cumulative Return(\%) \\
\midrule
5 & Random Walk & 0.009 & -1.152 & 1.787 & 8.934 & 0.326 & 0.532 & 2.700 \\ \cline{2-9}
 & EXFormer & 0.022 & -1.152 & 1.787 & 7.421 & 0.824 & 1.272 & 7.445 \\
 & Chronos & 0.011 & -1.432 & 1.787 & 6.732 & 0.425 & 0.696 & 3.619 \\
 & Informer & 0.010 & -1.152 & 1.787 & 10.234 & 0.369 & 0.611 & 3.095 \\
 & Transformer & 0.011 & -1.152 & 1.787 & 8.120 & 0.435 & 0.680 & 3.716 \\
 & TCN & 0.013 & -1.787 & 1.172 & 5.058 & 0.477 & 0.679 & 4.114 \\
 & GRU & 0.010 & -1.787 & 1.152 & 5.981 & 0.369 & 0.552 & 3.102 \\
 & LSTM & 0.003 & -1.787 & 1.152 & 7.578 & 0.105 & 0.147 & 0.655 \\
 & MLP & -0.011 & -1.787 & 1.152 & 9.133 & -0.435 & -0.660 & -4.163 \\
 & ARIMA & -0.010 & -1.152 & 1.787 & 9.997 & -0.392 & -0.625 & -3.789 \\
 & LSTM-ARIMA & 0.016 & -1.787 & 1.432 & 7.570 & 0.624 & 0.927 & 5.508 \\
 & Buy \& Hold & -0.011 & -1.787 & 1.152 & 9.133 & -0.435 & -0.660 & -4.163 \\
 & Moving-Average & 0.005 & -1.787 & 1.172 & 4.957 & 0.199 & 0.262 & 1.398 \\
 \midrule
10 & Random Walk & 0.009 & -1.152 & 1.787 & 8.934 & 0.326 & 0.532 & 2.700 \\ \cline{2-9}
 & EXFormer & 0.011 & -1.152 & 1.787 & 8.120 & 0.435 & 0.680 & 3.716 \\
 & Chronos & -0.006 & -1.787 & 1.432 & 12.275 & -0.211 & -0.334 & -2.196 \\
 & Informer & 0.011 & -1.152 & 1.787 & 8.120 & 0.435 & 0.680 & 3.716 \\
 & Transformer & -0.011 & -1.787 & 1.152 & 9.133 & -0.435 & -0.660 & -4.163 \\
 & TCN & -0.006 & -1.787 & 1.172 & 9.563 & -0.241 & -0.362 & -2.459 \\
 & GRU & 0.003 & -1.432 & 1.787 & 6.878 & 0.126 & 0.194 & 0.845 \\
 & LSTM & -0.025 & -1.787 & 1.172 & 11.793 & -0.963 & -1.412 & -8.643 \\
 & MLP & -0.022 & -1.787 & 1.152 & 8.976 & -0.830 & -1.206 & -7.538 \\
 & ARIMA & -0.010 & -1.152 & 1.787 & 12.079 & -0.392 & -0.630 & -3.791 \\
 & LSTM-ARIMA & -0.013 & -1.432 & 1.787 & 7.837 & -0.478 & -0.776 & -4.534 \\
 & Buy \& Hold & -0.011 & -1.787 & 1.152 & 9.133 & -0.435 & -0.660 & -4.163 \\
 & Moving-Average & 0.005 & -1.787 & 1.172 & 4.957 & 0.199 & 0.262 & 1.398 \\
    \midrule
20 & Random Walk & 0.009 & -1.152 & 1.787 & 8.934 & 0.326 & 0.532 & 2.700 \\ \cline{2-9}
 & EXFormer & 0.025 & -1.152 & 1.787 & 9.428 & 0.941 & 1.610 & 8.583 \\
 & Chronos & 0.005 & -1.152 & 1.787 & 6.320 & 0.194 & 0.329 & 1.476 \\
 & Informer & -0.011 & -1.787 & 1.152 & 9.133 & -0.435 & -0.660 & -4.163 \\
 & Transformer & 0.011 & -1.152 & 1.787 & 8.120 & 0.435 & 0.680 & 3.716 \\
 & TCN & 0.006 & -1.152 & 1.787 & 7.824 & 0.244 & 0.413 & 1.932 \\
 & GRU & -0.004 & -1.432 & 1.787 & 9.187 & -0.149 & -0.217 & -1.639 \\
 & LSTM & 0.013 & -1.152 & 1.787 & 8.120 & 0.486 & 0.757 & 4.195 \\
 & MLP & 0.011 & -1.152 & 1.787 & 8.120 & 0.435 & 0.680 & 3.716 \\
 & ARIMA & -0.011 & -1.152 & 1.787 & 10.568 & -0.428 & -0.680 & -4.106 \\
 & LSTM-ARIMA & 0.005 & -1.152 & 1.787 & 9.220 & 0.199 & 0.319 & 1.517 \\
 & Buy \& Hold & -0.011 & -1.787 & 1.152 & 9.133 & -0.435 & -0.660 & -4.163 \\
 & Moving-Average & 0.005 & -1.787 & 1.172 & 4.957 & 0.199 & 0.262 & 1.398 \\
 \midrule
30 & Random Walk & 0.009 & -1.152 & 1.787 & 8.934 & 0.326 & 0.532 & 2.700 \\ \cline{2-9}
 & EXFormer & 0.005 & -1.787 & 1.432 & 5.437 & 0.204 & 0.288 & 1.571 \\
 & Chronos & -0.001 & -1.152 & 1.787 & 8.966 & -0.043 & -0.069 & -0.692 \\
 & Informer & 0.011 & -1.152 & 1.787 & 7.226 & 0.429 & 0.675 & 3.652 \\
 & Transformer & -0.013 & -1.787 & 1.152 & 9.133 & -0.481 & -0.734 & -4.574 \\
 & TCN & -0.012 & -1.787 & 1.432 & 7.478 & -0.470 & -0.689 & -4.473 \\
 & GRU & 0.025 & -1.152 & 1.787 & 7.138 & 0.951 & 1.509 & 8.700 \\
 & LSTM & 0.020 & -1.432 & 1.787 & 5.747 & 0.740 & 1.180 & 6.638 \\
 & MLP & -0.013 & -1.787 & 1.152 & 9.133 & -0.481 & -0.734 & -4.574 \\
 & ARIMA & -0.004 & -1.152 & 1.787 & 8.897 & -0.168 & -0.268 & -1.815 \\
 & LSTM-ARIMA & 0.005 & -1.152 & 1.787 & 8.658 & 0.195 & 0.301 & 1.487 \\
 & Buy \& Hold & -0.013 & -1.787 & 1.152 & 9.133 & -0.481 & -0.734 & -4.574 \\
 & Moving-Average & 0.005 & -1.787 & 1.172 & 4.957 & 0.199 & 0.262 & 1.398 \\
\bottomrule
\end{tabular}%
}
 \label{table:8}
\end{table}

\begin{figure}[htbp]
  \centering
  \setlength{\leftskip}{-80pt}
  \includegraphics[width=180mm]{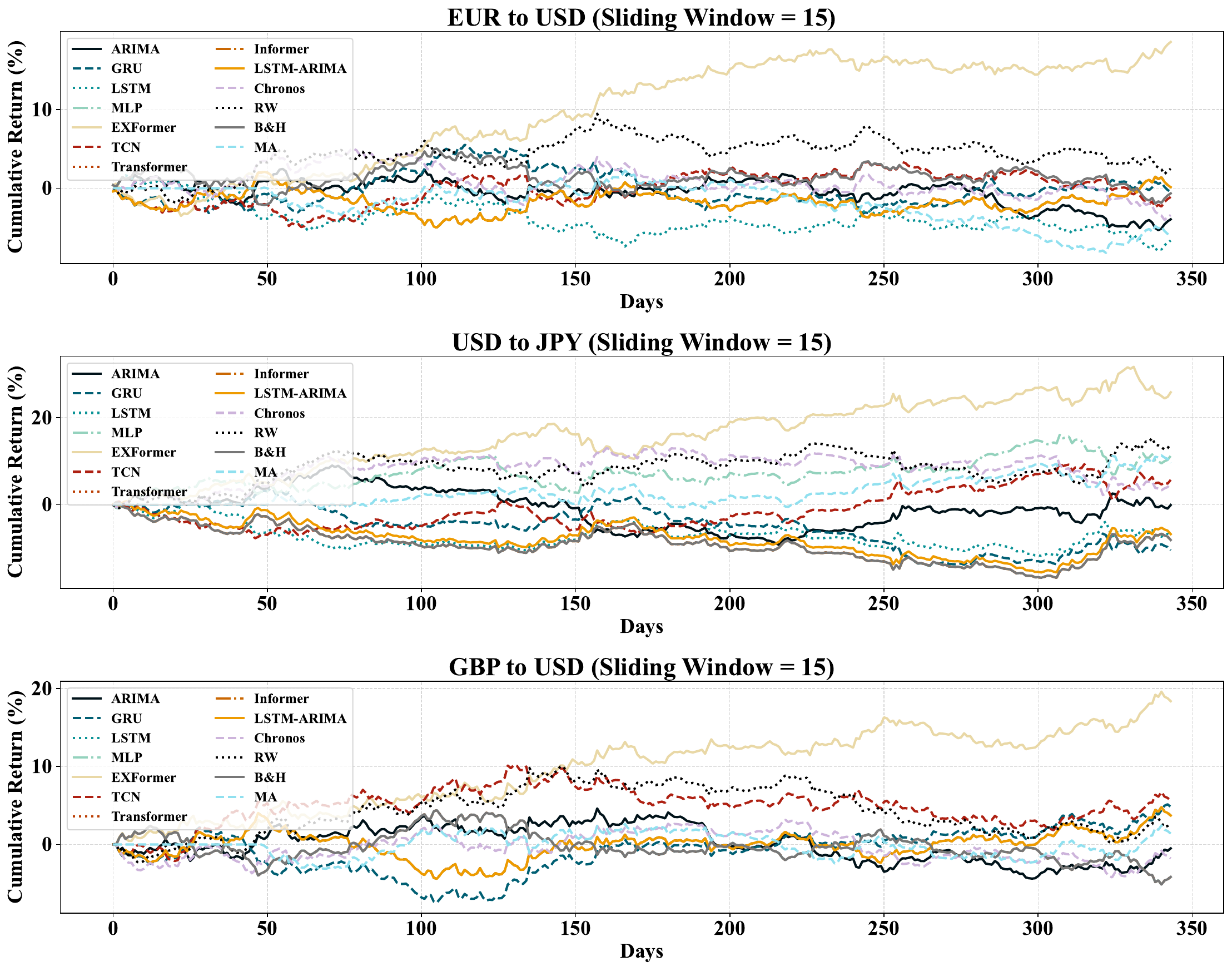}
  \caption{The cumulative returns of EXFormer and all benchmarks using a 15-day sliding window ($T = 15$) across EUR/USD, USD/JPY, and GBP/USD. The gold line represents EXFormer, while the black dotted line is the random walk benchmark. Other models are shown in various colors.}
  \label{fig:5}
\end{figure}

\begin{figure}[htbp]
  \centering
  \setlength{\leftskip}{-120pt}
  \includegraphics[width=210mm]{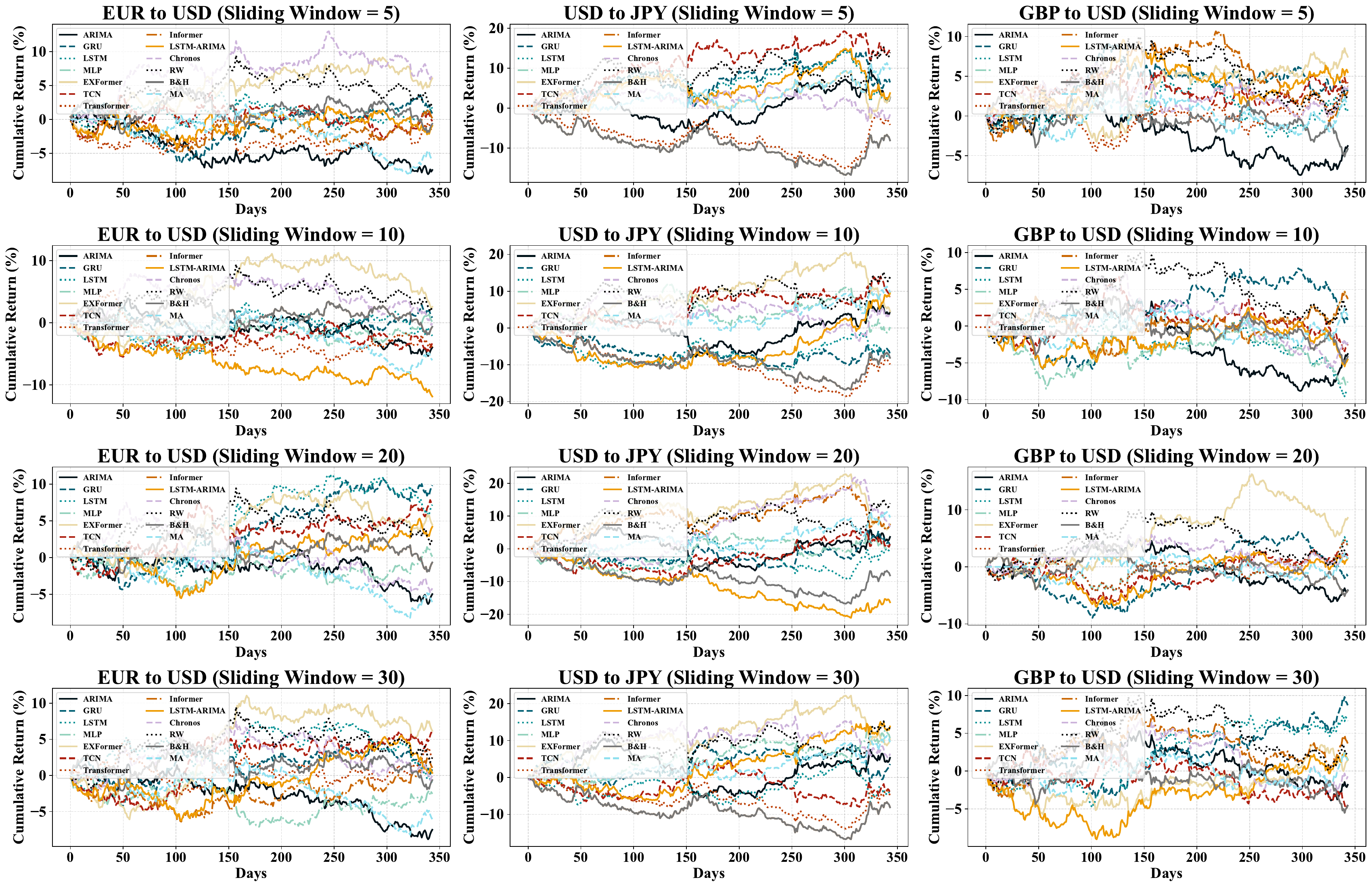}
  \caption{The cumulative returns of EXFormer and all benchmarks using sliding window sizes $T\in\{5,10,20,30\}$ across EUR/USD, USD/JPY, and GBP/USD. The gold line represents EXFormer, while the black dotted line is the random walk benchmark. Other models are shown in various colors.}
  \label{fig:6}
\end{figure}

Furthermore, Figure \ref{fig:5} illustrates the near 1 year cumulative return trajectories in the test set for EXFormer and all benchmarks when using a 15-day estimation window. Across EUR/USD, USD/JPY, and GBP/USD, EXFormer's strategy (gold line) embarks on a steady, upward path almost from day 0, culminating in roughly 19\%, 26\%, and 18 \% cumulative returns, respectively. In contrast, the random-walk (black dotted), Buy \& Hold (B\&H, solid gray), and Moving-Average (MA, cyan dashed) benchmarks hover near zero or decline over the test horizon, while other deep learning and econometric models deliver inferior end-values. The persistent outperformance enhances that EXFormer's forecasts translate into robust economic gains under a 15-day window.
Figure \ref{fig:6} presents the analogous cumulative return plots for alternative sliding windows \(T\in\{5,10,20,30\}\). For very short look-backs (5 and 10 days), EXFormer often leads early in the year, but its edge erodes in later months, with final returns converging toward zero. At longer windows (20 and 30 days), initial gains occasionally materialize but fail to persist, as the strategy becomes less responsive and more susceptible to regime shifts. These patterns highlight the importance of window length. Only the 15-day configuration yields both immediate and sustained trading profits, balancing adaptability to fresh information with resilience against noise.

To place the backtests in a more realistic trading context, we recompute all strategy returns under trading frictions. 
We model frictions as a fixed per-trade deduction equal to the sum of a 5\,bps transaction cost and a 2\,bps slippage cost, applied whenever the position changes sign or moves between \{-1,0,1\}.
This follows common practice in daily exchange rate returns backtesting. The three G10 pairs considered are among the most liquid instruments globally, so fixed bps costs provide a conservative proxy for bid-ask and execution uncertainty at the daily frequency.
Table~\ref{table:9} reports performance for the optimal 15-day window. 
EXFormer remains profitable across all three currency pairs after costs, with cumulative returns of 7.655\% (EUR/USD), 19.937\% (USD/JPY), and 9.216\% (GBP/USD), and Sharpe ratios of 1.419, 2.017, and 1.876, respectively, alongside comparatively low maximum drawdowns (5.208\%, 7.105\%, and 4.333\%). 
By contrast, several benchmarks turn negative once frictions are included (e.g., random walk and ARIMA across the three pairs, and Chronos on EUR/USD and GBP/USD), while others deliver modest gains in isolated cases (e.g., Informer on GBP/USD, MA on USD/JPY). 
The results indicate that EXFormer's statistical edge translates into economically meaningful, friction-aware profitability and remains competitive against recent transformer-type and hybrid baselines under realistic implementation assumptions.

\begin{table}[htbp]
 \centering
 \vspace*{-20mm} 
 \setlength{\leftskip}{-100pt}
 \caption{The backtesting results with transaction cost and slippage cost for EXFormer and all benchmarks using a 15-day sliding window ($T = 15$) across EUR/USD, USD/JPY, and GBP/USD.}
 \resizebox{200mm}{!}{%
\begin{tabular}{ccccccccc}
    \toprule
Type & Model & Mean Return (\%) & Minimum Return (\%) & Maximum Return (\%) & Max Drawdown (\%) & Sharpe Ratio & Sortino Ratio & Cumulative Return (\%) \\
\midrule
EUR/USD & Random Walk & -0.027 & -1.171 & 1.678 & 13.174 & -1.128 & -1.839 & -9.220 \\ \cline{2-9}
 & \textbf{EXFormer} & \textbf{0.022} & \textbf{-1.101} & \textbf{1.678} & \textbf{5.208} & \textbf{0.918} & \textbf{1.419} & \textbf{7.655} \\
 & Chronos & -0.039 & -1.171 & 1.678 & 15.935 & -1.610 & -2.576 & -12.851 \\
 & Informer & 0.001 & -1.061 & 1.678 & 6.906 & 0.047 & 0.076 & 0.140 \\
 & Transformer & 0.001 & -1.061 & 1.678 & 6.906 & 0.047 & 0.076 & 0.140 \\
 & TCN & -0.022 & -1.748 & 1.061 & 8.621 & -0.913 & -1.389 & -7.595 \\
 & GRU & -0.033 & -1.748 & 1.101 & 12.358 & -1.322 & -1.974 & -10.851 \\
 & LSTM & -0.021 & -1.678 & 1.061 & 9.904 & -0.877 & -1.282 & -7.291 \\
 & MLP & 0.001 & -1.061 & 1.678 & 6.906 & 0.047 & 0.076 & 0.140 \\
 & ARIMA & -0.039 & -1.171 & 1.608 & 15.198 & -1.588 & -2.621 & -12.668 \\
 & LSTM-ARIMA & 0.001 & -1.061 & 1.678 & 6.906 & 0.047 & 0.076 & 0.140 \\
 & Buy \& Hold & -0.001 & -1.678 & 1.061 & 6.645 & -0.047 & -0.070 & -0.648 \\
 & Moving-Average & -0.019 & -1.061 & 1.678 & 10.610 & -0.881 & -1.390 & -6.638 \\
 \midrule
USD/JPY & Random Walk & 0.004 & -2.257 & 1.916 & 14.847 & 0.097 & 0.148 & 0.648 \\ \cline{2-9}
 & \textbf{EXFormer} & \textbf{0.055} & \textbf{-2.187} & \textbf{1.777} & \textbf{7.105} & \textbf{1.507} & \textbf{2.017} & \textbf{19.937} \\
 & Chronos & -0.004 & -2.090 & 2.187 & 12.644 & -0.120 & -0.170 & -2.045 \\
 & Informer & -0.023 & -1.752 & 2.187 & 17.246 & -0.630 & -1.084 & -8.108 \\
 & Transformer & -0.023 & -1.752 & 2.187 & 17.246 & -0.630 & -1.084 & -8.108 \\
 & TCN & -0.007 & -2.090 & 2.117 & 12.718 & -0.200 & -0.276 & -3.026 \\
 & GRU & -0.050 & -2.160 & 2.187 & 20.680 & -1.364 & -2.129 & -16.182 \\
 & LSTM & -0.031 & -2.090 & 2.187 & 14.628 & -0.845 & -1.292 & -10.596 \\
 & MLP & 0.010 & -2.090 & 2.187 & 8.539 & 0.284 & 0.389 & 3.029 \\
 & ARIMA & -0.026 & -2.187 & 2.020 & 20.620 & -0.711 & -1.089 & -9.035 \\
 & LSTM-ARIMA & -0.019 & -1.752 & 2.187 & 16.224 & -0.519 & -0.893 & -6.827 \\
 & Buy \& Hold & -0.023 & -1.752 & 2.187 & 17.246 & -0.630 & -1.084 & -8.108 \\
 & Moving-Average & 0.027 & -2.090 & 2.187 & 5.289 & 0.789 & 1.106 & 9.111 \\
 \midrule
GBP/USD & Random Walk & -0.026 & -1.152 & 1.787 & 15.086 & -0.980 & -1.587 & -8.826 \\ \cline{2-9}
 & \textbf{EXFormer} & \textbf{0.027} & \textbf{-1.167} & \textbf{1.787} & \textbf{4.333} & \textbf{1.003} & \textbf{1.876} & \textbf{9.216} \\
 & Chronos & -0.034 & -1.787 & 1.362 & 13.323 & -1.298 & -1.979 & -11.334 \\
 & Informer & 0.011 & -1.152 & 1.787 & 8.120 & 0.435 & 0.680 & 3.716 \\
 & Transformer & -0.011 & -1.787 & 1.152 & 9.133 & -0.435 & -0.660 & -4.163 \\
 & TCN & -0.013 & -1.857 & 1.432 & 12.693 & -0.500 & -0.789 & -4.771 \\
 & GRU & 0.007 & -1.152 & 1.787 & 9.809 & 0.282 & 0.446 & 2.282 \\
 & LSTM & 0.011 & -1.152 & 1.787 & 8.120 & 0.435 & 0.680 & 3.716 \\
 & MLP & -0.011 & -1.787 & 1.152 & 9.133 & -0.435 & -0.660 & -4.163 \\
 & ARIMA & -0.028 & -1.152 & 1.717 & 13.135 & -1.062 & -1.710 & -9.475 \\
 & LSTM-ARIMA & 0.011 & -1.152 & 1.787 & 8.120 & 0.435 & 0.680 & 3.716 \\
 & Buy \& Hold & -0.011 & -1.787 & 1.152 & 9.133 & -0.435 & -0.660 & -4.163 \\
 & Moving-Average & 0.003 & -1.787 & 1.172 & 5.214 & 0.140 & 0.184 & 0.903 \\
    \bottomrule
\end{tabular}%
}
 \label{table:9}
\end{table}

Beyond the statistical and economic performance metrics, the profitability of EXFormer's trading strategies can be linked to established exchange rate theories. Persistent deviations from UIP imply that exchange rate returns embed risk premia not fully arbitraged away. EXFormer appears to capture these deviations, as reflected in consistently positive risk-adjusted returns. Moreover, the importance of commodity indices and global equity benchmarks in driving profitable trades is in line with portfolio balance theory, which emphasizes the role of cross-asset capital flows and investor risk preferences in determining exchange rates. These connections provide an economic rationale for the model's outperformance, suggesting that EXFormer does not merely exploit statistical patterns but identifies systematic drivers of currency risk premia consistent with theoretical exchange rate dynamics.

\subsection{Interpretability analysis}

\begin{figure}[htbp]
  \centering
  \setlength{\leftskip}{-110pt}
  \includegraphics[width=200mm]{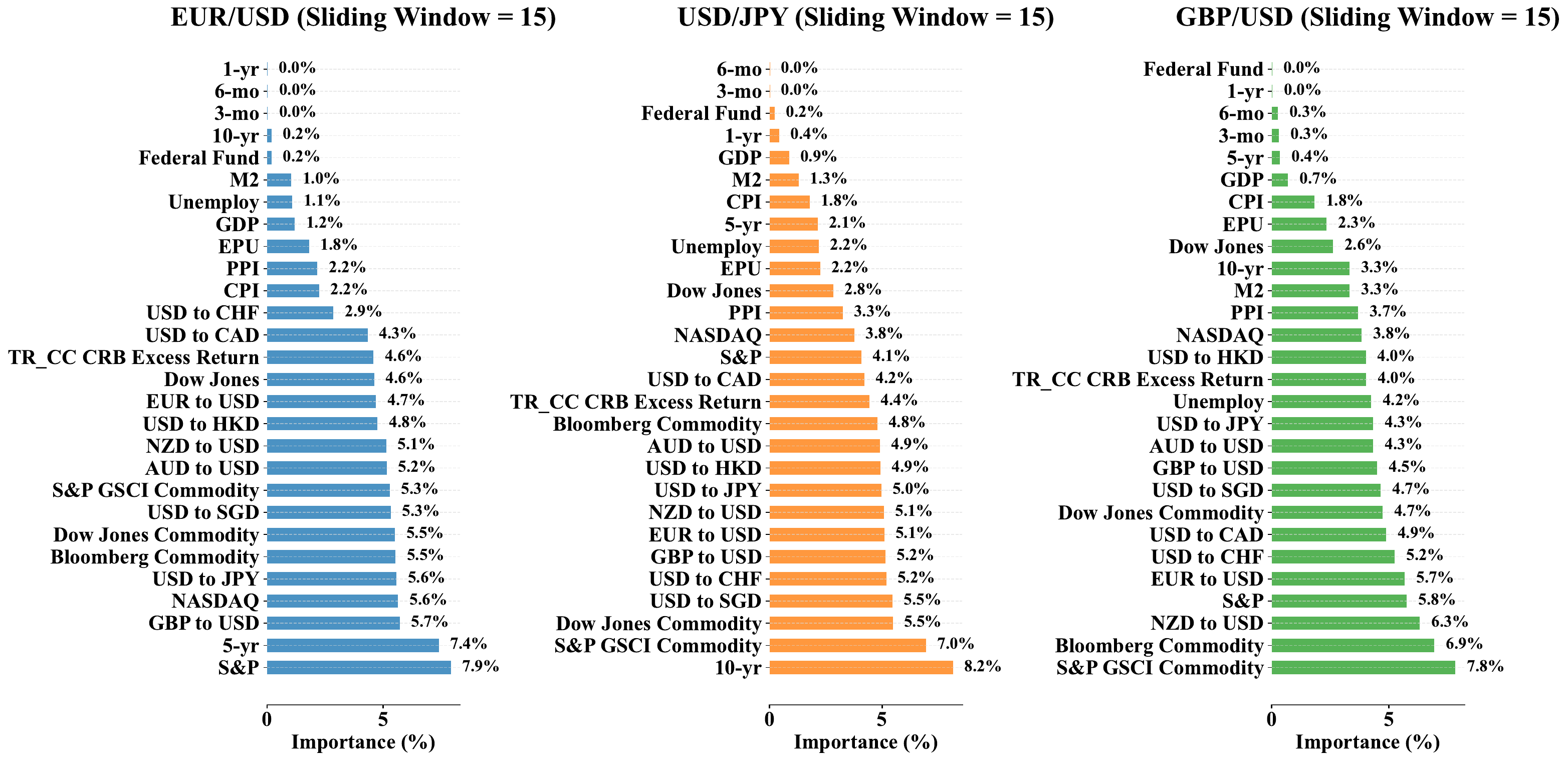}
  \caption{The global variables importance for EXFormer using a 15-day sliding window ($T = 15$) across EUR/USD, USD/JPY, and GBP/USD. The bars represent the percentage contribution of each predictor to the model's forecasting.}
  \label{fig:7}
\end{figure}

In the interpretability analysis, we first examine the global covariates importance learned by the dynamic variable selector when using the optimal 15-day estimation window (Figure~\ref{fig:7}). For EUR/USD, the three most influential predictors are the S\&P 500 index (7.9\,\%), the US 5-year treasury yield (7.4\,\%), and the GBP/USD returns (5.7\,\%), followed closely by NASDAQ (5.6\,\%) and USD/JPY (5.6\,\%). 
The commodity indices (Bloomberg and DJ Commodity at 5.5\,\%), cross rates such as USD/SGD (5.3\,\%) and AUD/USD (5.2\,\%), and the S\&P GSCI commodity index (5.3\,\%) also carry substantial weight. In contrast, short- and medium- term interest rates (1-year through 6-month and Federal Funds) register near-zero importance, reflecting their limited predictive value at this horizon.

The USD/JPY selector highlights a different constellation: the 10-year yield leads at 8.2\,\%, the S\&P GSCI commodity at 7.0\,\%, and the Dow Jones commodity at 5.5\,\%, with cross rates USD/SGD (5.5\,\%), USD/CHF (5.2\,\%), and GBP/USD (5.2\,\%) also prominent. The stock indices (S\&P\,500 at 4.1\,\%, NASDAQ at 3.8\,\%) and the CRB index (4.4\,\%) further underscore the importance of broad risk-asset and real-asset drivers for yen returns. Again, short-term yields and macro releases (e.g., CPI, GDP, unemployment) occupy lower ranks ($\leqslant 3$\,\%).

The GBP/USD's variable ranking parallels USD/JPY: S\&P GSCI Commodity (7.8\,\%), Bloomberg Commodity (6.9\,\%), NZD/USD (6.3\,\%), and the S\&P 500 index (5.8\,\%) top the list, followed by EUR/USD (5.7\,\%) and USD/CHF (5.2\,\%). The stock markets (NASDAQ 3.8\,\%, Dow Jones 2.6\,\%), CPI (1.8\,\%) and EPU (2.3\,\%) enter at intermediate levels, while interest rate covariates again contribute negligibly.

\begin{figure}[htbp]
  \centering
  \setlength{\leftskip}{-120pt}
  \vspace*{-25mm} 
  \includegraphics[width=210mm]{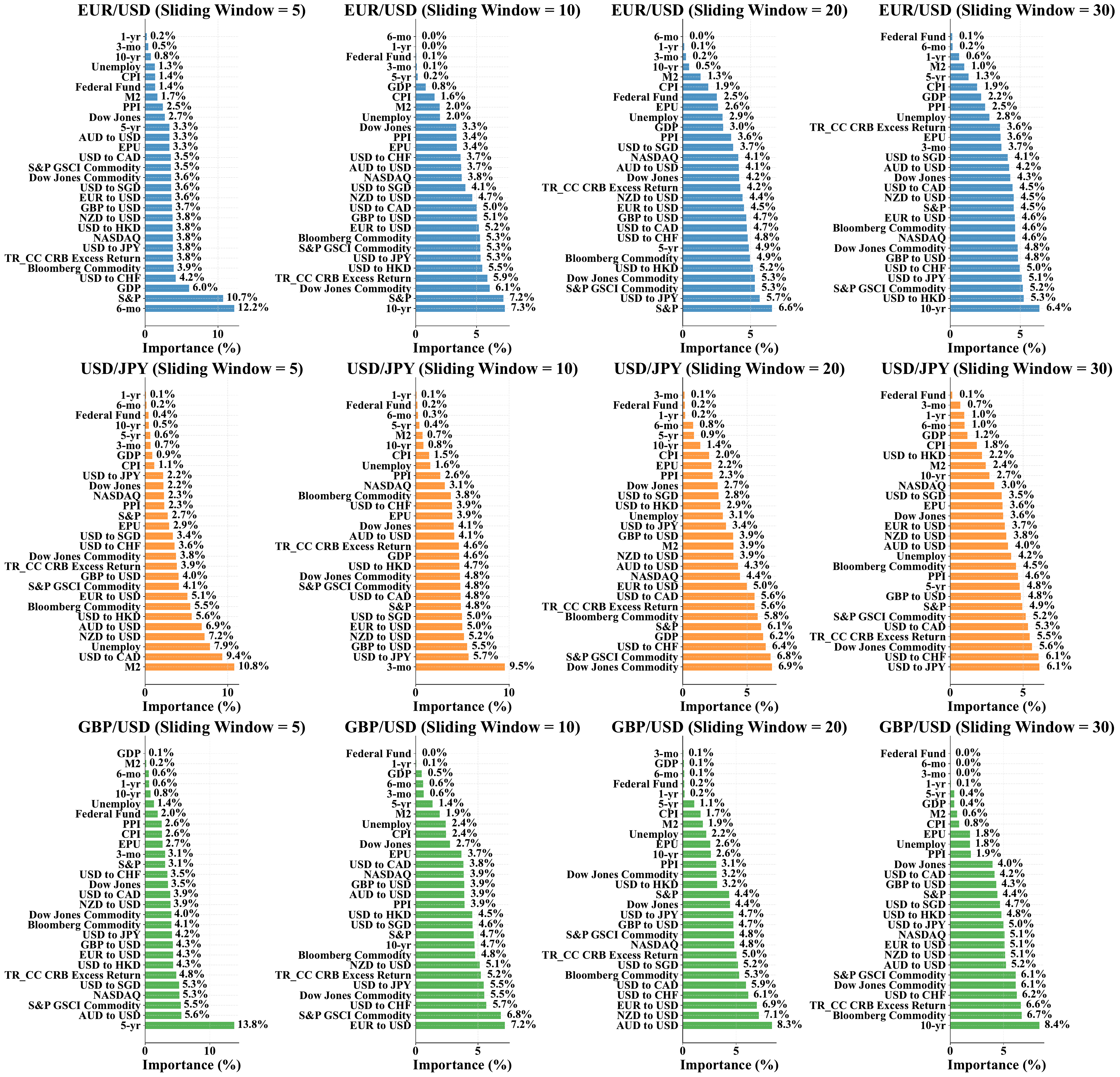}
  \caption{The global variables importance for EXFormer using sliding window sizes $T\in\{5,10,20,30\}$ across EUR/USD, USD/JPY, and GBP/USD. The bars represent the percentage contribution of each predictor to the model's forecasting.}
  \label{fig:8}
\end{figure}

Furthermore, Figure~\ref{fig:8} generalizes these patterns across alternative sliding windows. At very short sliding windows (5 and 10 days), stock indices and the 6-month yield sometimes dominate (e.g., the 6-month yield reaches 12.2\,\% for EUR/USD at \(T=5\)), reflecting the stronger role of near-term monetary policy signals. As \(T\) increases to 20 and 30 days, commodity indices and long-term yields regain prominence, while the importance of short-term macro data uniformly minimizes. The cross rates such as EUR/USD, USD/JPY, and GBP/USD maintain stable mid-rank importance regardless of window, confirming the value of exchange rate interdependencies. The global importance profiles demonstrate that EXFormer automatically allocates its attention to the most informative covariates in a horizon-dependent manner, coherently balancing risk-asset, fixed income, and exchange rate cross signals to optimize daily returns forecasting.

\begin{figure}[htbp]
  \centering
  \setlength{\leftskip}{-80pt}
  \vspace*{-25mm} 
  \includegraphics[width=180mm]{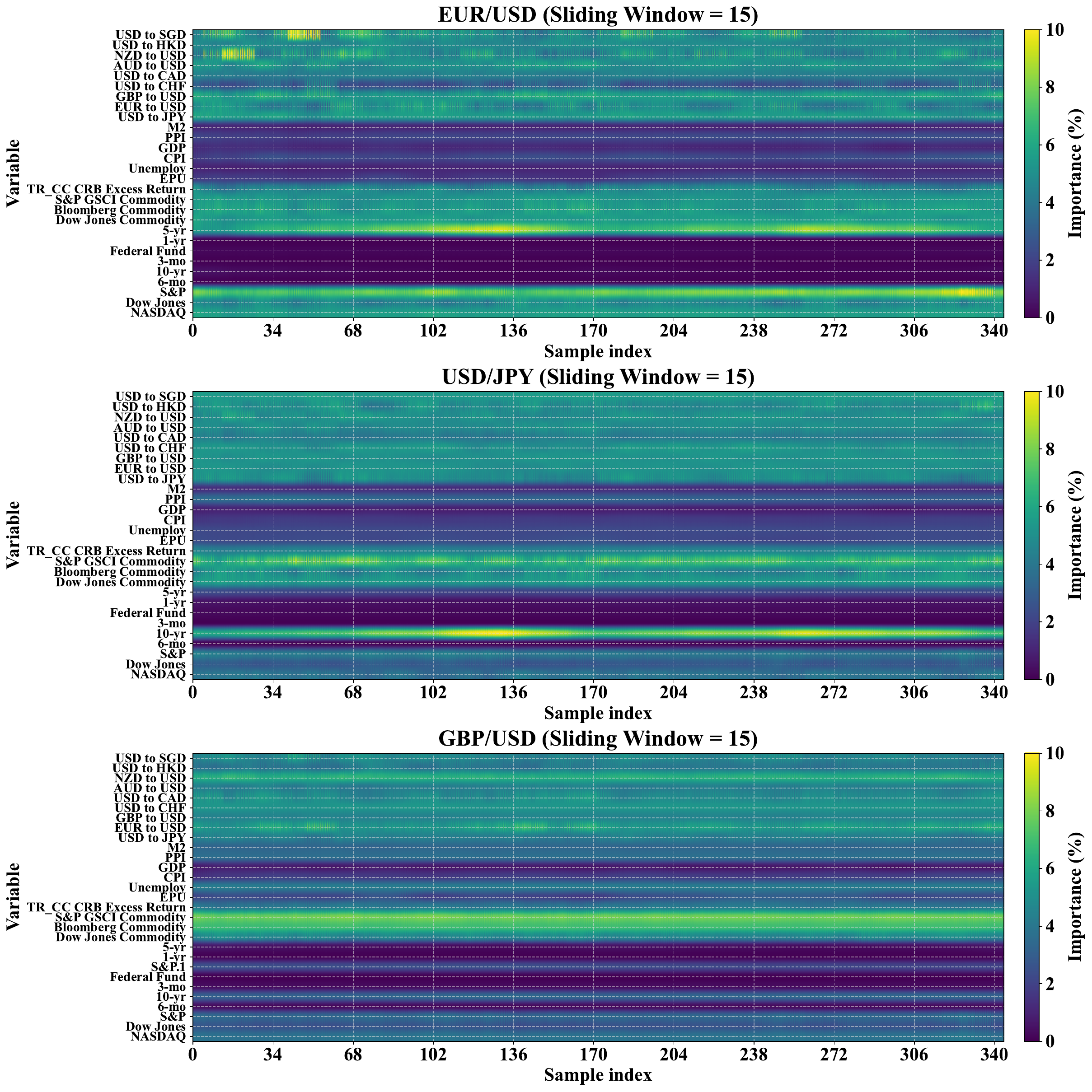}
  \caption{The time-varying variables importance for EXFormer using a 15-day sliding window ($T = 15$) across EUR/USD, USD/JPY, and GBP/USD. The heatmap shows the percentage contribution of each predictor to the model's forecasting over time.}
  \label{fig:9}
\end{figure}

Figure~\ref{fig:9} visualizes the time-varying importance for each covariate under the optimal 15-day window, revealing how EXFormer dynamically reallocates attention over the course of the out-of-sample set. 
For EUR/USD, the S\&P\,500 and the 5-year treasury yield persistently occupy the mid-to-top bands but exhibit pronounced spikes around sample indices corresponding to the 4th season of 2023, when stock market risk and medium-term yields suddenly surged in predictive value. The cross-rates such as NZD/USD and USD/SGD display brief bursts of importance in May and June 2023, likely reflecting commodity-driven dollar flows, while macro releases (CPI, GDP) remain in the lowest band with only ephemeral uplifts around major announcement dates.

USD/JPY's attention heatmap similarly highlights the 10-year yield and the S\&P\,GSCI commodity index as core drivers, with localized intensity peaks at sample indices coinciding with late 2023. The yen crosses (USD/CHF, GBP/USD) maintain steadier, moderate weights, whereas stock market indices (NASDAQ, Dow Jones) enter sporadically around months with heightened global risk aversion. 
In GBP/USD, the Bloomberg Commodity index dominates most of the year, lightening only when broad stock market gains momentarily outshine commodity trends in late 2023.
Across all three pairs, the dynamic variable selector endows EXFormer with the ability to ``turn on'' and ``off'' predictors in response to shifting market regimes, delivering the fine-grained, pre-hoc interpretability that post-hoc methods cannot.

\begin{figure}[htbp]
  \centering
  \setlength{\leftskip}{-100pt}
  \includegraphics[width=200mm]{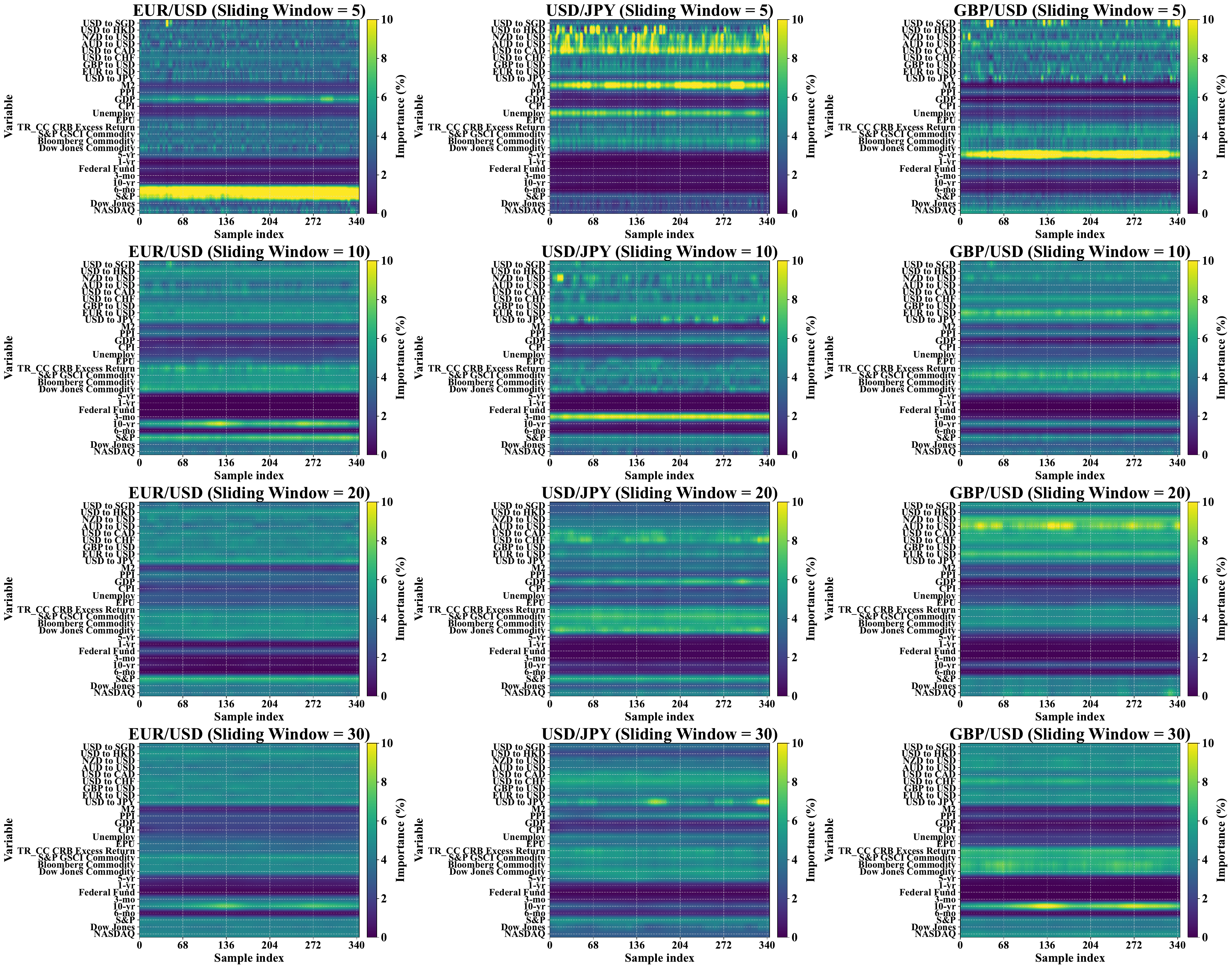}
  \caption{The time-varying variables importance for EXFormer using sliding window sizes $T\in\{5,10,20,30\}$ across EUR/USD, USD/JPY, and GBP/USD. The heatmap shows the percentage contribution of each predictor to the model's forecasting over time.}
  \label{fig:10}
\end{figure}

Figure~\ref{fig:10} extends this analysis to alternative sliding windows (\(T\in\{5,10,20,30\}\)). At \(T=5\), importance scores are highly erratic. The short-term yields and cross-rates flash bright bands nearly every week, reflecting over-responsiveness to noise in a very narrow look-back. With \(T=10\), the heatmaps are still noisy but begin to coalesce around commodity and stock market indices, though without the clear regime-driven spikes seen at \(T=15\). For \(T=20\) and \(T=30\), importance patterns smooth further. The long-term yields and broad commodity indices form the stable middle bands, while short-term macro data and cross-rates occupy low, nearly constant levels. These comparisons underscore that the 15-day configuration strikes the optimal balance between adaptability (capturing ephemeral but economically meaningful regime shifts) and stability (filtering out idiosyncratic noise), a balance that is visibly lost at both shorter and longer windows.

\subsection{Robustness check by market conditions}
To assess whether the performance gains documented above are robust to changing market environments, we stratify the out-of-sample period along two state dimensions and recompute directional accuracy for every model within each state.

The first is volatility conditions.
For each currency pair we calculate a 20-day rolling standard deviation of daily returns. The empirical distribution of this series is split at the 33$^{\text{rd}}$ and 66$^{\text{th}}$ percentiles, yielding three non-overlapping regimes: low, medium, and high volatility. 
The second is trend regimes.
Independently, we sum returns over a 20-day window to construct a cumulative trend indicator. Positive (negative) sums define a bull (bear) market sub-sample, capturing persistent directional swings in the exchange rate returns.

Within each regime we form the sign forecasts implied by (i) EXFormer, (ii) all benchmark models, and (iii) the random walk. The directional accuracy is the share of days on which the model predicts the correct sign of the returns. Because regime classification uses only past information, the test remains strictly out-of-sample. The full procedure is repeated for every estimation window (\(T\in\{5,10,15,20,30\}\)), generating the summary statistics reported in Tables~\ref{table:10}-\ref{table:13}.
The design allows verifying whether EXFormer's edge is confined to calm conditions or bull phases, or whether it persists when markets become turbulent or trend reversals dominate, the latter being situations in which many forecasting models typically break down.

Table~\ref{table:10} reports DA for the random walk and each competing model across five market states: high, medium, and low volatility, and bear versus bull trend regimes, using the 15-day sliding window that yielded the best overall performance. First, EXFormer consistently outperforms the random walk in every state for all three currency pairs: for EUR/USD, DA increases from 0.510 under the random walk to 0.568 in bear markets and 0.539 in bull markets. For USD/JPY, DA rises from 0.461 to 0.631 in bear markets. For GBP/USD from 0.500 to 0.543 in bear markets. Second, the largest gains occur in high volatility and bear market conditions, where traditional models often falter. For example, EUR/USD's DA climbs from 0.500 to 0.615 under high volatility. Third, no other benchmark delivers uniformly positive improvements. Transformer and MLP remain near 0.500, while ARIMA frequently underperforms the random walk in stressed regimes.

\begin{table}[htbp]
 \centering
  \vspace*{-20mm} 
 \setlength{\leftskip}{-40pt}
 \caption{The directional accuracy (DA) for the random walk and all models in five market conditions using a 15-day sliding window ($T = 15$) across EUR/USD, USD/JPY, and GBP/USD.}
 \resizebox{150mm}{!}{%
\begin{tabular}{ccccccc}
    \toprule
Type & Model & High Volatility & Medium Volatility & Low Volatility & Bear Market & Bull Market \\
\midrule
EUR/USD & Random Walk & 0.500 & 0.495 & 0.495 & 0.510 & 0.480 \\ \cline{2-7}
 & EXFormer & 0.615 & 0.509 & 0.527 & 0.568 & 0.539 \\
 & Chronos & 0.489 & 0.411 & 0.500 & 0.495 & 0.437 \\
 & Informer & 0.496 & 0.500 & 0.518 & 0.500 & 0.509 \\
 & Transformer & 0.496 & 0.500 & 0.518 & 0.500 & 0.509 \\
 & TCN & 0.504 & 0.500 & 0.464 & 0.469 & 0.515 \\
 & GRU & 0.437 & 0.464 & 0.527 & 0.448 & 0.503 \\
 & LSTM & 0.467 & 0.500 & 0.473 & 0.474 & 0.485 \\
 & MLP & 0.496 & 0.500 & 0.518 & 0.500 & 0.509 \\
 & ARIMA & 0.452 & 0.464 & 0.482 & 0.443 & 0.491 \\
 & LSTM-ARIMA & 0.496 & 0.500 & 0.518 & 0.500 & 0.509 \\
 \midrule
USD/JPY & Random Walk & 0.531 & 0.430 & 0.486 & 0.461 & 0.546 \\ \cline{2-7}
 & EXFormer & 0.548 & 0.589 & 0.625 & 0.631 & 0.465 \\
 & Chronos & 0.519 & 0.500 & 0.554 & 0.531 & 0.505 \\
 & Informer & 0.467 & 0.393 & 0.375 & 0.377 & 0.515 \\
 & Transformer & 0.467 & 0.393 & 0.375 & 0.377 & 0.515 \\
 & TCN & 0.548 & 0.509 & 0.616 & 0.569 & 0.525 \\
 & GRU & 0.474 & 0.402 & 0.384 & 0.392 & 0.505 \\
 & LSTM & 0.504 & 0.464 & 0.393 & 0.427 & 0.535 \\
 & MLP & 0.548 & 0.545 & 0.634 & 0.592 & 0.525 \\
 & ARIMA & 0.474 & 0.455 & 0.509 & 0.477 & 0.485 \\
 & LSTM-ARIMA & 0.504 & 0.393 & 0.375 & 0.396 & 0.515 \\
 \midrule
GBP/USD & Random Walk & 0.485 & 0.533 & 0.505 & 0.500 & 0.513 \\ \cline{2-7}
 & EXFormer & 0.556 & 0.571 & 0.527 & 0.543 & 0.563 \\
 & Chronos & 0.482 & 0.429 & 0.554 & 0.498 & 0.475 \\
 & Informer & 0.504 & 0.563 & 0.554 & 0.578 & 0.488 \\
 & TCN & 0.482 & 0.509 & 0.563 & 0.543 & 0.481 \\
 & Transformer & 0.496 & 0.438 & 0.446 & 0.422 & 0.513 \\
 & GRU & 0.504 & 0.500 & 0.563 & 0.538 & 0.500 \\
 & LSTM & 0.504 & 0.563 & 0.554 & 0.578 & 0.488 \\
 & MLP & 0.496 & 0.438 & 0.446 & 0.422 & 0.513 \\
 & ARIMA & 0.533 & 0.464 & 0.491 & 0.503 & 0.494 \\
 & LSTM-ARIMA & 0.504 & 0.563 & 0.554 & 0.488 & 0.558 \\
 \bottomrule
\end{tabular}%
}
 \label{table:10}
\end{table}

Tables~\ref{table:11}, \ref{table:12}, and \ref{table:13} extend the regime-level DA results across the other four sliding windows (5, 10, 20, and 30 days) for each currency pair. EXFormer's robustness persists across nearly every regime and window choice, with high volatility DA exceeding 0.54 in all configurations and surpassing 0.60 for USD/JPY at the medium volatility. Other architectures display greater sensitivity to window length and market state. While TCN and GRU occasionally rival EXFormer in isolated buckets, they never sustain across all states, and ARIMA's DA often falls below 0.50 in bear markets. The consistency of EXFormer's advantage, especially in turbulent or downward-trending environments, highlights its ability to dynamically extract multi-scale trends and select salient covariates under changing market conditions.

\begin{table}[htbp]
 \centering
 \setlength{\leftskip}{-20pt}
 \vspace*{-40mm} 
 \caption{The directional accuracy (DA) for the random walk and all models in five market conditions using sliding window size $T\in\{5,10,20,30\}$ across EUR/USD.}
 \resizebox{135mm}{!}{%
\begin{tabular}{ccccccc}
    \toprule
Sliding   Window & Model & High Volatility & Medium Volatility & Low Volatility & Bear Market & Bull Market \\
\midrule
5 & Random Walk & 0.500 & 0.495 & 0.495 & 0.510 & 0.480 \\ \cline{2-7}
 & EXFormer & 0.601 & 0.474 & 0.496 & 0.563 & 0.488 \\
 & Chronos & 0.493 & 0.452 & 0.491 & 0.472 & 0.488 \\
 & Informer & 0.500 & 0.496 & 0.526 & 0.508 & 0.506 \\
 & Transformer & 0.486 & 0.517 & 0.496 & 0.487 & 0.512 \\
 & TCN & 0.529 & 0.517 & 0.444 & 0.513 & 0.482 \\
 & GRU & 0.522 & 0.569 & 0.452 & 0.508 & 0.524 \\
 & LSTM & 0.515 & 0.517 & 0.461 & 0.513 & 0.482 \\
 & MLP & 0.500 & 0.517 & 0.496 & 0.503 & 0.506 \\
 & ARIMA & 0.435 & 0.483 & 0.496 & 0.457 & 0.482 \\
 & LSTM-ARIMA & 0.521 & 0.461 & 0.474 & 0.523 & 0.465 \\
 \midrule
10 & Random Walk & 0.500 & 0.495 & 0.495 & 0.510 & 0.480 \\ \cline{2-7}
 & EXFormer & 0.544 & 0.439 & 0.544 & 0.505 & 0.518 \\
 & Chronos & 0.522 & 0.412 & 0.500 & 0.516 & 0.441 \\
 & Informer & 0.500 & 0.509 & 0.474 & 0.495 & 0.494 \\
 & Transformer & 0.441 & 0.561 & 0.500 & 0.464 & 0.535 \\
 & TCN & 0.529 & 0.491 & 0.430 & 0.526 & 0.441 \\
 & GRU & 0.493 & 0.544 & 0.483 & 0.495 & 0.518 \\
 & LSTM & 0.544 & 0.500 & 0.483 & 0.505 & 0.518 \\
 & MLP & 0.500 & 0.526 & 0.491 & 0.505 & 0.506 \\
 & ARIMA & 0.449 & 0.447 & 0.500 & 0.428 & 0.506 \\
 & LSTM-ARIMA & 0.463 & 0.447 & 0.483 & 0.459 & 0.471 \\
 \midrule
20 & Random Walk & 0.500 & 0.495 & 0.495 & 0.510 & 0.480 \\ \cline{2-7}
 & EXFormer & 0.541 & 0.441 & 0.582 & 0.510 & 0.537 \\
 & Chronos & 0.481 & 0.427 & 0.469 & 0.490 & 0.426 \\
 & Informer & 0.519 & 0.500 & 0.478 & 0.510 & 0.488 \\
 & Transformer & 0.519 & 0.478 & 0.500 & 0.510 & 0.488 \\
 & TCN & 0.526 & 0.496 & 0.546 & 0.490 & 0.562 \\
 & GRU & 0.571 & 0.451 & 0.518 & 0.526 & 0.506 \\
 & LSTM & 0.556 & 0.532 & 0.446 & 0.500 & 0.531 \\
 & MLP & 0.481 & 0.523 & 0.500 & 0.490 & 0.512 \\
 & ARIMA & 0.459 & 0.478 & 0.446 & 0.448 & 0.475 \\
 & LSTM-ARIMA & 0.521 & 0.500 & 0.527 & 0.511 & 0.525 \\
 \midrule
30 & Random Walk & 0.500 & 0.495 & 0.495 & 0.510 & 0.480 \\ \cline{2-7}
 & EXFormer & 0.546 & 0.542 & 0.486 & 0.547 & 0.500 \\
 & Chronos & 0.515 & 0.477 & 0.439 & 0.531 & 0.415 \\
 & Informer & 0.462 & 0.505 & 0.561 & 0.500 & 0.513 \\
 & Transformer & 0.500 & 0.542 & 0.495 & 0.500 & 0.526 \\
 & TCN & 0.515 & 0.505 & 0.486 & 0.469 & 0.546 \\
 & GRU & 0.492 & 0.486 & 0.514 & 0.500 & 0.493 \\
 & LSTM & 0.539 & 0.514 & 0.477 & 0.516 & 0.507 \\
 & MLP & 0.531 & 0.486 & 0.514 & 0.516 & 0.507 \\
 & ARIMA & 0.454 & 0.458 & 0.467 & 0.453 & 0.467 \\
 & LSTM-ARIMA & 0.515 & 0.430 & 0.598 & 0.505 & 0.526 \\
\bottomrule
\end{tabular}%
}
 \label{table:11}
\end{table}

\begin{table}[htbp]
 \centering
 \setlength{\leftskip}{-20pt}
 \vspace*{-40mm} 
 \caption{The directional accuracy (DA) for the random walk and all models in five market conditions using sliding window size $T\in\{5,10,20,30\}$ across USD/JPY.}
 \resizebox{135mm}{!}{%
\begin{tabular}{ccccccc}
    \toprule
Sliding   Window & Model & High Volatility & Medium Volatility & Low Volatility & Bear Market & Bull Market \\
\midrule
5 & Random Walk & 0.531 & 0.430 & 0.486 & 0.461 & 0.546 \\ \cline{2-7}
 & EXFormer & 0.536 & 0.621 & 0.609 & 0.622 & 0.485 \\
 & Chronos & 0.507 & 0.522 & 0.440 & 0.493 & 0.485 \\
 & Informer & 0.464 & 0.391 & 0.379 & 0.378 & 0.515 \\
 & Transformer & 0.493 & 0.371 & 0.400 & 0.382 & 0.546 \\
 & TCN & 0.544 & 0.621 & 0.522 & 0.563 & 0.556 \\
 & GRU & 0.500 & 0.612 & 0.591 & 0.596 & 0.475 \\
 & LSTM & 0.486 & 0.612 & 0.644 & 0.600 & 0.505 \\
 & MLP & 0.464 & 0.379 & 0.391 & 0.378 & 0.515 \\
 & ARIMA & 0.515 & 0.535 & 0.435 & 0.485 & 0.525 \\
 & LSTM-ARIMA & 0.515 & 0.565 & 0.604 & 0.607 & 0.485 \\
 \midrule
10 & Random Walk & 0.531 & 0.430 & 0.486 & 0.461 & 0.546 \\ \cline{2-7}
 & EXFormer & 0.529 & 0.667 & 0.588 & 0.619 & 0.515 \\
 & Chronos & 0.515 & 0.439 & 0.439 & 0.453 & 0.505 \\
 & Informer & 0.471 & 0.386 & 0.377 & 0.377 & 0.515 \\
 & Transformer & 0.463 & 0.351 & 0.395 & 0.370 & 0.505 \\
 & TCN & 0.581 & 0.395 & 0.404 & 0.419 & 0.596 \\
 & GRU & 0.471 & 0.412 & 0.412 & 0.411 & 0.495 \\
 & LSTM & 0.493 & 0.509 & 0.439 & 0.468 & 0.515 \\
 & MLP & 0.507 & 0.605 & 0.614 & 0.615 & 0.455 \\
 & ARIMA & 0.485 & 0.553 & 0.474 & 0.502 & 0.505 \\
 & LSTM-ARIMA & 0.581 & 0.456 & 0.570 & 0.509 & 0.576 \\
 \midrule
20 & Random Walk & 0.531 & 0.430 & 0.486 & 0.461 & 0.546 \\ \cline{2-7}
 & EXFormer & 0.541 & 0.631 & 0.600 & 0.628 & 0.485 \\
 & Chronos & 0.462 & 0.604 & 0.469 & 0.553 & 0.459 \\
 & Informer & 0.541 & 0.600 & 0.631 & 0.628 & 0.485 \\
 & Transformer & 0.541 & 0.631 & 0.600 & 0.628 & 0.485 \\
 & TCN & 0.594 & 0.505 & 0.400 & 0.478 & 0.576 \\
 & GRU & 0.526 & 0.396 & 0.455 & 0.428 & 0.556 \\
 & LSTM & 0.526 & 0.387 & 0.436 & 0.412 & 0.566 \\
 & MLP & 0.466 & 0.505 & 0.482 & 0.455 & 0.556 \\
 & ARIMA & 0.534 & 0.523 & 0.473 & 0.502 & 0.535 \\
 & LSTM-ARIMA & 0.481 & 0.391 & 0.378 & 0.392 & 0.495 \\
 \midrule
30 & Random Walk & 0.531 & 0.430 & 0.486 & 0.461 & 0.546 \\ \cline{2-7}
 & EXFormer & 0.546 & 0.636 & 0.598 & 0.604 & 0.556 \\
 & Chronos & 0.508 & 0.523 & 0.570 & 0.551 & 0.485 \\
 & Informer & 0.454 & 0.393 & 0.374 & 0.367 & 0.515 \\
 & Transformer & 0.485 & 0.374 & 0.393 & 0.384 & 0.515 \\
 & TCN & 0.508 & 0.411 & 0.355 & 0.396 & 0.515 \\
 & GRU & 0.546 & 0.551 & 0.486 & 0.518 & 0.556 \\
 & LSTM & 0.492 & 0.617 & 0.523 & 0.551 & 0.515 \\
 & MLP & 0.515 & 0.542 & 0.579 & 0.555 & 0.515 \\
 & ARIMA & 0.523 & 0.551 & 0.421 & 0.494 & 0.515 \\
 & LSTM-ARIMA & 0.546 & 0.505 & 0.477 & 0.506 & 0.525 \\
 \bottomrule
\end{tabular}%
}
 \label{table:12}
\end{table}

\begin{table}[htbp]
 \centering
 \setlength{\leftskip}{-20pt}
 \vspace*{-40mm} 
 \caption{The directional accuracy (DA) for the random walk and all models in five market conditions using sliding window size $T\in\{5,10,20,30\}$ across GBP/USD.}
 \resizebox{135mm}{!}{%
\begin{tabular}{ccccccc}
    \toprule
Sliding   Window & Model & High Volatility & Medium Volatility & Low Volatility & Bear Market & Bull Market \\
\midrule
5 & Random Walk & 0.485 & 0.533 & 0.505 & 0.500 & 0.513 \\ \cline{2-7}
 & EXFormer & 0.558 & 0.535 & 0.591 & 0.604 & 0.506 \\
 & Chronos & 0.413 & 0.496 & 0.552 & 0.478 & 0.488 \\
 & Informer & 0.486 & 0.496 & 0.491 & 0.498 & 0.482 \\
 & Transformer & 0.500 & 0.543 & 0.565 & 0.575 & 0.482 \\
 & TCN & 0.507 & 0.535 & 0.496 & 0.536 & 0.482 \\
 & GRU & 0.536 & 0.457 & 0.548 & 0.498 & 0.537 \\
 & LSTM & 0.507 & 0.543 & 0.478 & 0.512 & 0.506 \\
 & MLP & 0.500 & 0.457 & 0.435 & 0.425 & 0.519 \\
 & ARIMA & 0.493 & 0.500 & 0.470 & 0.478 & 0.500 \\
 & LSTM-ARIMA & 0.515 & 0.470 & 0.569 & 0.517 & 0.519 \\
 \midrule
10 & Random Walk & 0.485 & 0.533 & 0.505 & 0.500 & 0.513 \\ \cline{2-7}
 & EXFormer & 0.515 & 0.544 & 0.561 & 0.581 & 0.485 \\
 & Chronos & 0.449 & 0.483 & 0.491 & 0.478 & 0.466 \\
 & Informer & 0.515 & 0.561 & 0.544 & 0.581 & 0.485 \\
 & Transformer & 0.485 & 0.456 & 0.439 & 0.419 & 0.516 \\
 & TCN & 0.471 & 0.491 & 0.456 & 0.468 & 0.478 \\
 & GRU & 0.515 & 0.447 & 0.456 & 0.483 & 0.466 \\
 & LSTM & 0.515 & 0.395 & 0.474 & 0.468 & 0.460 \\
 & MLP & 0.478 & 0.491 & 0.465 & 0.429 & 0.540 \\
 & ARIMA & 0.485 & 0.483 & 0.483 & 0.488 & 0.478 \\
 & LSTM-ARIMA & 0.500 & 0.483 & 0.421 & 0.473 & 0.466 \\
 \midrule
20 & Random Walk & 0.485 & 0.533 & 0.505 & 0.500 & 0.513 \\ \cline{2-7}
 & EXFormer & 0.504 & 0.559 & 0.555 & 0.574 & 0.491 \\
 & Chronos & 0.466 & 0.427 & 0.622 & 0.513 & 0.491 \\
 & Informer & 0.504 & 0.446 & 0.441 & 0.431 & 0.509 \\
 & Transformer & 0.496 & 0.559 & 0.555 & 0.569 & 0.491 \\
 & TCN & 0.489 & 0.487 & 0.509 & 0.492 & 0.497 \\
 & GRU & 0.511 & 0.514 & 0.509 & 0.533 & 0.484 \\
 & LSTM & 0.534 & 0.478 & 0.564 & 0.513 & 0.541 \\
 & MLP & 0.496 & 0.559 & 0.555 & 0.569 & 0.491 \\
 & ARIMA & 0.489 & 0.478 & 0.473 & 0.477 & 0.484 \\
 & LSTM-ARIMA & 0.534 & 0.509 & 0.505 & 0.518 & 0.516 \\
 \midrule
30 & Random Walk & 0.485 & 0.533 & 0.505 & 0.500 & 0.513 \\ \cline{2-7}
 & EXFormer & 0.515 & 0.598 & 0.514 & 0.552 & 0.527 \\
 & Chronos & 0.469 & 0.449 & 0.617 & 0.546 & 0.460 \\
 & Informer & 0.523 & 0.458 & 0.617 & 0.562 & 0.493 \\
 & Transformer & 0.515 & 0.439 & 0.439 & 0.433 & 0.513 \\
 & TCN & 0.423 & 0.570 & 0.598 & 0.495 & 0.560 \\
 & GRU & 0.492 & 0.579 & 0.551 & 0.572 & 0.493 \\
 & LSTM & 0.500 & 0.523 & 0.514 & 0.516 & 0.507 \\
 & MLP & 0.515 & 0.439 & 0.439 & 0.433 & 0.513 \\
 & ARIMA & 0.477 & 0.505 & 0.486 & 0.490 & 0.487 \\
 & LSTM-ARIMA & 0.477 & 0.533 & 0.589 & 0.526 & 0.533 \\
 \bottomrule
\end{tabular}%
}
 \label{table:13}
\end{table}

\subsection{Ablation experiments}
To quantify the individual contributions of EXFormer's architectural components, we conduct four ablations:
1) No MSC (multi-scale convolution). Remove the multi-scale convolutional extractor.
2) No SE (squeeze-and-excitation). Omit the squeeze-excitation block.
3) No DVS (dynamic variable selector). Disable the dynamic variable selector, using uniform feature weights.
4) Standard Attention. Replace our multi-scale trend-aware self-attention with traditional Transformer attention.
All variants are re-trained under identical settings and evaluated on 1 day-ahead directional accuracy.

Table~\ref{table:14} reports the DA for the full EXFormer and four ablated variants, as well as the random walk benchmark. Each entry shows DA along with the the Clark--West $t$-statistic in parentheses. Across all currency pairs and sliding windows, the full EXFormer delivers the highest DA, often significantly outperforming the random walk (e.g.\ USD/JPY at $T=15$: DA$=0.598$, $t=3.206^{***}$). Removal of multi-scale convolution (No MSC) collapses performance nearly to random walk levels. 
Excluding the squeeze-excitation block (No SE) reduces DA at certain degree, indicating the importance of channel-wise recalibration. 
Turning off the dynamic variable selector (No DVS) yields the largest drop up,highlighting the necessity of pre-hoc, time-varying feature weighting. 
Finally, replacing the multi-scale trend-aware attention with standard self-attention also costs model ability, validating the value of local trend incorporation. The ablation impacts peak at $T=15$, where the full model's architectural synergies most strongly enhance directional accuracy.

\begin{table}[htbp]
 \centering
 \setlength{\leftskip}{-60pt}
 \caption{The ablation results for EXFormer and all variants in 1 day-ahead. The table shows the directional accuracy (DA) with the Blaskowitz--Herwartz $t$-statistics for each model and sliding window size $T\in\{5,10,15,20,30\}$.}
 \resizebox{170mm}{!}{%
\begin{tabular}{lllllll}
    \toprule
 &  &  &  & Sliding Window &  &  \\ \cline{3-7}
Type & Model & 5 & 10 & 15 & 20 & 30 \\
\midrule
EUR/USD & Random Walk & 0.499 & 0.499 & 0.499 & 0.499 & 0.499 \\ \cline{2-7}
 & Full & 0.530 (0.974) & 0.510 (0.445) & 0.556 (1.647*) & 0.521 (0.632) & 0.525 (0.560) \\
 & No MSC & 0.493 (0.000) & 0.489 (-0.07) & 0.499 (0.068) & 0.497 (0.005) & 0.509 (0.202) \\
 & No SE & 0.472 (0.327) & 0.462 (-0.823) & 0.499 (0.010) & 0.449 (-1.358) & 0.486 (-0.406) \\
 & No DVS & 0.458 (-0.970) & 0.489 (-0.151) & 0.462 (-1.038) & 0.455 (-1.148) & 0.494 (-0.155) \\
 & Standard Attention & 0.491 (-0.068) & 0.451 (-1.050) & 0.496 (-0.001) & 0.492 (-0.147) & 0.497 (-0.075) \\
 \midrule
USD/JPY & Random Walk & 0.487 & 0.487 & 0.487 & 0.487 & 0.487 \\ \cline{2-7}
 & Full & 0.582 (2.633***) & 0.570 (2.365**) & 0.598 (3.206***) & 0.586 (2.763***) & 0.589 (2.788***) \\
 & No MSC & 0.550 (1.587) & 0.514 (0.622) & 0.527 (0.900) & 0.559 (1.723*) & 0.558 (1.735*) \\
 & No SE & 0.523 (0.940) & 0.486 (-0.072) & 0.524 (0.807) & 0.463 (-0.792) & 0.448 (-0.964) \\
 & No DVS & 0.474 (-0.369) & 0.464 (-0.626) & 0.471 (-0.463) & 0.506 (0.387) & 0.491 (0.059) \\
 & Standard Attention & 0.523 (0.963) & 0.442 (-1.131) & 0.426 (-1.597) & 0.525 (0.921) & 0.433 (-1.33) \\
 \midrule
GBP/USD & Random Walk & 0.507 & 0.507 & 0.507 & 0.507 & 0.507 \\ \cline{2-7}
 & Full & 0.560 (1.441) & 0.537 (0.902) & 0.550 (1.256) & 0.516 (0.167) & 0.542 (1.090) \\
 & No MSC & 0.531 (0.564) & 0.525 (0.479) & 0.51 (0.142) & 0.486 (-0.563) & 0.512 (-0.849) \\
 & No SE & 0.496 (-0.354) & 0.473 (-1.055) & 0.49 (-0.435) & 0.483 (-0.741) & 0.468 (-0.849) \\
 & No DVS & 0.453 (-1.451) & 0.440 (-1.661) & 0.446 (-1.439) & 0.455 (-1.406) & 0.465 (-1.009) \\
 & Standard Attention & 0.466 (-1.126) & 0.506 (-0.075) & 0.462 (-0.984) & 0.466 (-1.059) & 0.462 (-1.031) \\
 \bottomrule
\end{tabular}%
}
 \begin{tablenotes}
  \small
  \item \textbf{Notes:} The Blaskowitz--Herwartz $t$-statistics (in parentheses) test whether DA exceeds that of the random walk. Significance levels: $^{*}p<0.10$, $^{**}p<0.05$, $^{***}p<0.01$.
 \end{tablenotes}
 \label{table:14}
\end{table}

\begin{figure}[htbp]
  \centering
  \setlength{\leftskip}{-30pt}
  \includegraphics[width=140mm]{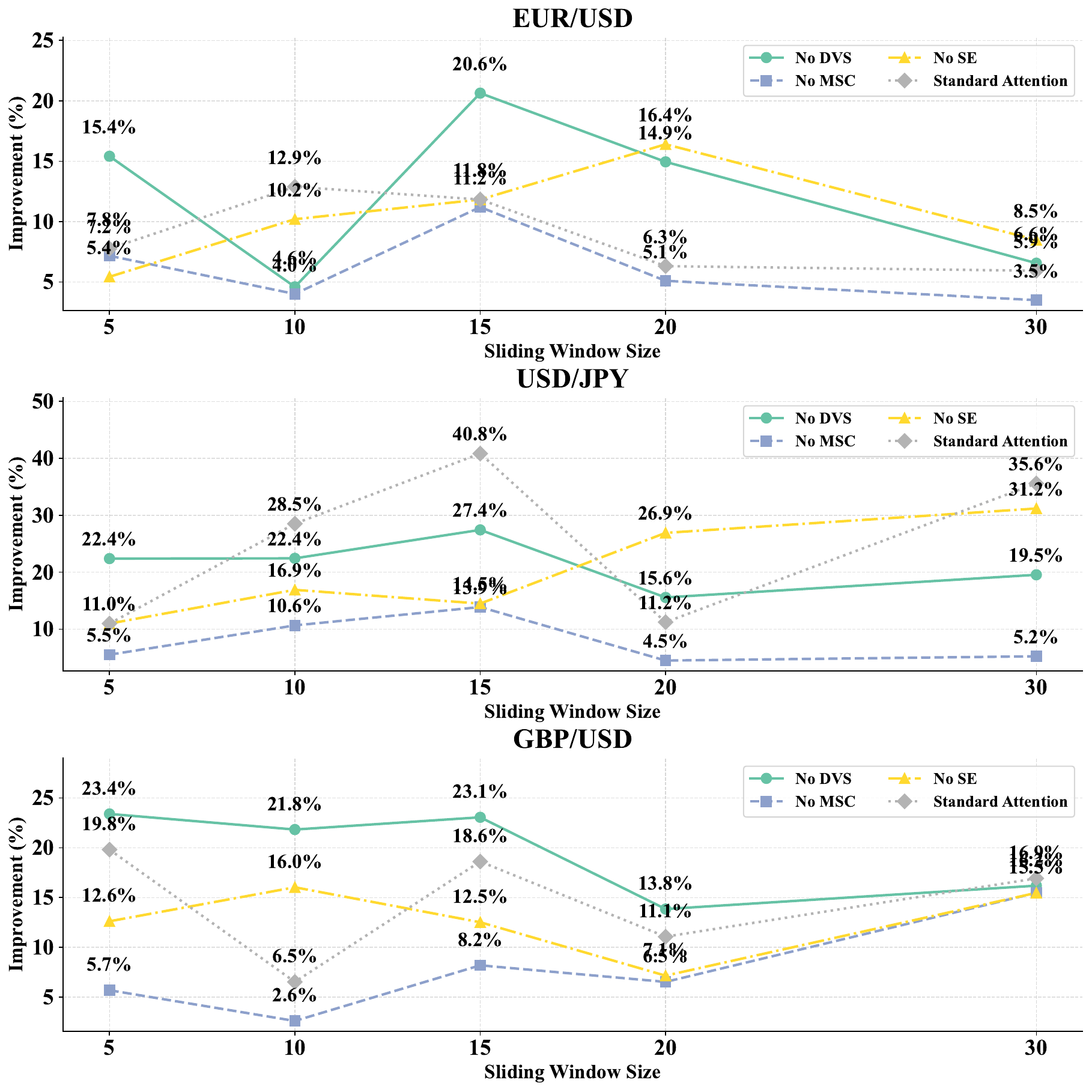}
  \caption{The directional accuracy (DA) improvement of EXFormer over each ablated variant for EUR/USD, USD/JPY, and GBP/USD in 1 day-ahead. Lines show the percentage point difference in DA between the full model and each ablated version across different sliding window $T\in\{5,10,15,20,30\}$. The larger values indicate greater importance of the removed component.}
  \label{fig:11}
\end{figure}

Figure~\ref{fig:11} summarizes how much the full EXFormer gains in 1 day-ahead directional accuracy over four ablated variants across different sliding windows. Two clear patterns emerge. First, the dynamic variable selector is indispensable. Removing it (``No DVS'', green) produces the largest and most persistent shortfall in all three pairs, with the full model exceeding the ablation by 20.6\% at $T=15$ for EUR/USD, 27.4\% at $T=15$ for USD/JPY, and 23.4\% at $T=5$ (and $23.1\%$ at $T=15$) for GBP/USD. Second, replacing the multi-scale trend-aware attention with traditional self-attention (gray diamonds) is detrimental for USD/JPY, where the full model's advantage peaks at a 40.8\% improvement at $T=15$ and remains large even at $T=30$ ($35.6\%$). By contrast, the penalties from dropping the squeeze-and-excitation (``No SE'', yellow) and multi-scale convolution (``No MSC'', blue) are moderate but consistent, typically $10$--$17\%$ and $5$--$12\%$, respectively, with maxima around the mid-range windows (e.g., EUR/USD $T=20$: 14.9\% for No SE; $T=15$: 11.8\% for No MSC). Across the currencies, the gaps generally enlarge from $T=10$ to $T=15$, signaling strong architectural synergies at this horizon, and then compress at $T=30$, especially for EUR/USD, where improvements fall to single digits. Overall, the figure shows that EXFormer's edge is driven primarily by pre-hoc, time-varying feature weighting (DVS) and multi-scale trend-aware attention; multi-scale convolutions and SE recalibration further contribute, but with smaller, steadier gains.

\section{Discussion}
\label{sec:sample6}
Despite a growing number of studies applying machine learning and neural networks to daily exchange rate returns forecasting, most have only compared their novel architectures against conventional baselines, such as simple neural networks or single factor models, and stopped short of benchmarking against the random walk or demonstrating statistical superiority over it \citep{windsor2022improving,mao2024unveiling}. For example, prior works by \citet{fischer2018deep} report backtest returns for their proposed networks but do not conduct formal statistical tests to confirm outperformance of the random walk. In contrast, EXFormer not only registers statistically significant gains in out-of-sample MSFE and directional accuracy, but also translates those gains into economically meaningful trading profits, outpacing buy and hold, moving average, and random walk strategies across multiple currency pairs.
Our built-in interpretability mechanism yields additional economic insights. Whereas studies such as \citet{afshan2024fintech} and \citet{singhal2019return} highlight the predictive power of crude oil and gold markets, our global importance analysis also shows that broad commodity indices (e.g. S\&P GSCI and Bloomberg Commodity) exert the largest influence on day-ahead exchange rate returns \citep{beckmann2020relationship}. 
Moreover, the dynamic variable weights reveal that EUR/USD returns are particularly sensitive to the stock markets and U.S. Treasury yields, corroborating findings by \citet{ben2024shorting} on the close coupling between the euro and global financial conditions.

Furthermore, our sliding window analysis points to an empirical optimum at 15 trading days for 1 day-ahead exchange rate returns prediction. While some econometric studies (e.g. using AIC or cross-validation) recommend longer or varying look-back lengths\citep{darvas2025forecasting}, and others claim robustness across any reasonably sized window \citep{byrne2018sources}, our results suggest that information beyond three weeks begins to weaken the model's ability to adapt to rapid market regime shifts. The finding has practical implications that careful tuning of the sliding window is as critical as neural network architecture when deploying deep learning for high-frequency financial forecasting.
Moreover, foreign exchange is one of the most liquid asset classes globally, with deep order books and minimal transaction frictions at daily horizons \citep{engel2023liquidity}. Therefore, liquidity constraints are unlikely to bias our predictive results or trading simulations, distinguishing this setting from equity or commodity markets where limited depth or slippage can materially affect forecast evaluation \citep{erdogan2015recession}.

Our findings contribute to the longstanding debate on the validity of classical exchange rate theories. The consistent profitability of EXFormer in daily forecasting underscores the well-documented short-horizon failure of UIP \citep{engel2019uncovered}, often referred to as the forward premium puzzle, whereby exchange rates do not fully adjust to eliminate predictable excess returns. This result aligns with the extensive empirical evidence that UIP is systematically broke at daily horizons \citep{candian2025imperfect,du2018deviations,gabaix2015international}.
At the same time, the interpretability analysis highlights the role of global commodities, equity indices, and long-term interest rates in shaping currency dynamics. Such evidence corresponds with the portfolio balance perspective \citep{ben2024shorting}, which emphasizes the importance of cross-asset capital flows and shifts in investor risk preferences in determining exchange rates \citep{verdelhan2018share,mueller2017exchange}. By jointly demonstrating UIP's empirical breakdown and providing support for portfolio balance mechanisms, our study situates the predictive and economic value of EXFormer within established theoretical frameworks and offers a bridge between modern machine-learning methods and classical international finance theories.

\section{Conclusion}
\label{sec:sample7}
In this paper, we have introduced EXFormer, a novel Transformer-based architecture tailored to the unique challenges of daily exchange rate returns forecasting. By integrating multi-scale convolutional feature extraction, squeeze-and-excitation recalibration, and a multi-scale trend-aware self-attention mechanism, alongside a dynamic variable selector that provides pre-hoc interpretability, EXFormer simultaneously captures high-frequency noise, medium-term cyclical patterns, and long-run trends, while adaptively weighting a rich set of 28 exogenous covariates. In an out-of-sample evaluation on EUR/USD, USD/JPY, and GBP/USD, EXFormer outperformed 9 baselines, including deep learning, hybrid models, and econometric approaches, in mean squared forecast error and directional accuracy, with superiority confirmed by the Clark--West test and dynamic regression tests. Furthermore, EXFormer's forecasts translated into economically trading profits, beating random walk, buy and hold, and moving-average strategies across multiple sliding windows, with a 15-day sliding window yielding the highest cumulative returns, Sharpe, and Sortino ratios.
When realistic frictions of 7 basis points per trade (5 bps transaction cost and 2 bps slippage) are incorporated, EXFormer still delivers positive and competitive performance, with cumulative returns of 7.6\% (EUR/USD), 19.9\% (USD/JPY), and 9.2\% (GBP/USD), while most benchmarks turn unprofitable or less than random walk. This demonstrates that the model's economic value is not an artifact of frictionless assumptions but persists in practice.

Methodologically, this study demonstrates the efficacy of embedding local-trend information directly into the self-attention computation via convolutional kernels of varying size, overcoming the pointwise matching limitation of standard Transformers in financial time series. The squeeze-and-excitation blocks proved vital for amplifying informative channels and suppressing redundant noise, while the dynamic variable selector endowed the model with a transparent, time-varying view of covariate importance, offering a pre-hoc explanation that uncovers shifting market drivers.
Practically, EXFormer offers a ready-to-deploy framework for risk managers, algorithmic traders, and policymakers. Its ability to generate statistically and economically robust forecasts, coupled with built-in interpretability, can inform more precise hedging strategies, systematic trading signals, and timely interventions by central banks. Our finding that a 15-day historical window strikes the optimal balance between adaptability and information sufficiency highlights the importance of carefully tuning look-back lengths to market regime dynamics, contradicting the common assumption that ``more data is always better'' for high-frequency exchange rate returns forecasting models.

Beyond predictive accuracy and trading profitability, this study also situates EXFormer's contributions within established exchange rate theories. The model's ability to generate persistent excess returns indicates that it captures deviations from UIP, which are often interpreted as manifestations of risk premia in exchange rate markets. Furthermore, the strong predictive role of commodity indices and global equity benchmarks uncovered by the dynamic variable selector aligns with portfolio balance theory, whereby cross-asset capital flows and shifts in investor risk preferences influence currency dynamics. Moreover, the pre-hoc interpretability of EXFormer reveals how these drivers vary over time. By linking empirical forecasting gains to theoretical constructs and event-driven variable importance shifts, EXFormer bridges the gap between statistical modeling and economic intuition, providing both robust predictions and a transparent lens into the mechanisms underlying currency risk premia.

The future work might extend EXFormer to multi-horizon and multi-currency portfolio forecasts, which may unlock new applications in cross-asset risk management. Adapting EXFormer's architecture to other high-volatility financial time series would test its generality and potentially uncover new structure in complex market dynamics.

\section*{CRediT authorship contribution statement}
\textbf{Dinggao Liu:} Conceptualization, Methodology, Writing - Original draft, Visualization.
\textbf{Robert Ślepaczuk:} Conceptualization, Investigation, Writing - Review \& Editing.
\textbf{Zhenpeng Tang:} Supervision, Funding acquisition, Writing - Review \& Editing.

\section*{Declaration of Competing Interest}
The authors declare that there is no conflict of interest regarding the publication of this paper.

\section*{Data availability}
Data will be made available on request.

\section*{Acknowledgements}
This research work was supported by the National Natural Science Foundation of China under Grant No. 72341030.


\begin{table}[htbp]
    \vspace{-20mm}
    \resizebox{130mm}{!}{%
    \begin{tabular}{ll}
        \hline
    \textbf{Abbreviation} &  \\
    ADF & Augmented Dickey-Fuller \\
    AIC & Akaike information criterion \\
    ARCH-LM & Autoregressive conditional heteroskedasticity-lagrange multiplier \\
    ARIMA & Autoregressive integrated moving average \\
    B\&H & Buy and hold \\
    CNN & Convolutional neural network \\
    DA & Directional accuracy \\
    DVS & Dynamic variable selector \\
    EPU & Economic policy uncertainty \\
    EUR & Euro \\
    GARCH & Generalized autoregressive conditional heteroskedasticity \\
    GBP & British pound \\
    GRU & Gated recurrent unit \\
    HAC & Heteroskedasticity and autocorrelation consistent \\
    JPY & Japanese yen \\
    LLM & Large language model \\
    LSTM & Long short-term memory \\
    MA & Moving average \\
    MAE & Mean absolute error \\
    MLP & Multilayer perceptron \\
    MSC & Multi-scale convolution \\
    MSFE & Mean squared forecast error \\
    RF & Random forest \\
    RMSE & Root mean square error \\
    RNN & Recurrent neural network \\
    RW & Random walk \\
    SE & Squeeze-and-excitation \\
    SHAP & SHapley Additive exPlanations \\
    SVM & Support vector machine \\
    TCN & Temporal convolutional network \\
    UIP & Uncovered interest parity \\
    USD & United States dollar \\
    \hline
    \end{tabular}%
    }
\end{table}

\clearpage
\appendix
\section{Appendix}
\label{sec:sample8}
\begin{table}[htbp]
    \centering
    \caption{ARCH-LM test for volatility clustering (returns).}
    \resizebox{\columnwidth}{!}{%
\begin{tabular}{llll}
    \toprule
\multicolumn{1}{l}{Type} & \multicolumn{1}{l}{Lag} & \multicolumn{1}{l}{ARCH-LM} & \multicolumn{1}{l}{} \\
\midrule
EUR to USD & Lags= 5 & LM-stat=161.334, p=0.000 & (F-stat=33.671,   p=0.000) \\
 & Lags=10 & LM-stat=210.208, p=0.000 & (F-stat=22.212, p=0.000) \\
 & Lags=20 & LM-stat=246.091, p=0.000 & (F-stat=13.103,   p=0.000) \\
 & Lags=30 & LM-stat=295.286, p=0.000 & (F-stat=10.606, p=0.000) \\
 \midrule
USD to JPY & Lags= 5 & LM-stat=137.300, p=0.000 & (F-stat=28.464, p=0.000) \\
 & Lags=10 & LM-stat=158.568, p=0.000 & (F-stat=16.513,   p=0.000) \\
 & Lags=20 & LM-stat=188.042, p=0.000 & (F-stat=9.847, p=0.000) \\
 & Lags=30 & LM-stat=236.035, p=0.000 & (F-stat=8.333, p=0.000) \\
 \midrule
GBP to USD & Lags= 5 & LM-stat=110.734, p=0.000 & (F-stat=22.788, p=0.000) \\
 & Lags=10 & LM-stat=121.737, p=0.000 & (F-stat=12.548, p=0.000) \\
 & Lags=20 & LM-stat=132.008, p=0.000 & (F-stat=6.805, p=0.000) \\
 & Lags=30 & LM-stat=137.323, p=0.000 & (F-stat=4.714, p=0.000) \\
 \bottomrule
\end{tabular}%
}
    \label{table:a0}
\end{table}

\begin{table}[htbp]
    \centering
     \vspace*{-40mm} 
    \setlength{\leftskip}{-20pt}
    \caption{Hyperparameter search space for all models.}
    \resizebox{140mm}{!}{%
\begin{tabular}{cccc}
    \toprule
Model & Hyperparameter & Type & Search Space \\
\midrule
EXFormer & Number of attention heads & Categorical & {1, 2, 4, 8} \\
 & Factor per head (dimension multiplier) & Integer (step=16) & {16, 32, 48, 64} \\
 & Hidden dimension size & Derived & Hidden dimension = heads $\times$ factor \\
 & Convolutional kernel size (branch 1) & Integer & [2--9] \\
 & Convolutional kernel size (branch 2) & Integer & [2--9], with constraint $k\_2 > k\_1$ \\
 & Convolutional kernel size (branch 3) & Integer & [2--9], with constraint $k\_3 > k\_2$ \\
 & Dropout rate & Continuous float & [0.00 -- 0.50] (step=0.1) \\
 & Learning rate & Continuous float (log-uniform) & [1e-4 -- 1e-2] \\
 & Batch size & Categorical & {32, 64, 128} \\
 \midrule
Informer & Number of encoder layers (e\_layers) & Integer & {1, 2, 3} \\
 & Number of decoder layers (d\_layers) & Integer & {1, 2} \\
 & Model dimension (d\_model) & Integer (step=32) & {32, 64, 96, 128} \\
 & Number of attention heads & Categorical & {2, 4} \\
 & Dropout rate & Continuous float & [0.00 -- 0.50] (step=0.1) \\
 & Learning rate & Continuous float (log-uniform) & [1e-4 -- 1e-2] \\
 & Batch size & Categorical & {32, 64, 128} \\
 \midrule
Transformer & Number of encoder layers & Integer & {1, 2, 3} \\
 & Model dimension (d\_model) & Integer (step=32) & {32, 64, 96, 128} \\
 & Number of attention heads & Categorical & {2, 4, 8} \\
 & Feed-forward network dimension & Integer (step=64) & {64, 128, 192, 256} \\
 & Dropout rate & Continuous float & [0.00 -- 0.50] (step=0.1) \\
 & Learning rate & Continuous float (log-uniform) & [1e-4 -- 1e-2] \\
 & Batch size & Categorical & {32, 64, 128} \\
 \midrule
TCN & Convolutional channel (num\_channels) & Categorical & {[16,16], [32,32]} \\
 & Kernel size & Categorical & {2, 3, 4, 5} \\
 & Dropout rate & Continuous float & [0.00 -- 0.50] (step=0.1) \\
 & Learning rate & Continuous float (log-uniform) & [1e-4 -- 1e-2] \\
 & Batch size & Categorical & {32, 64, 128} \\
 \midrule
GRU & Number of recurrent layers & Integer & {1, 2, 3} \\
 & Hidden state dimension & Integer (step=16) & {16, 32, 48, 64, 80, 96, 112, 128} \\
 & Dropout rate & Continuous float & [0.00 -- 0.50] (step=0.1) \\
 & Learning rate & Continuous float (log-uniform) & [1e-4 -- 1e-2] \\
 & Batch size & Categorical & {32, 64, 128} \\
 \midrule
LSTM & Number of recurrent layers & Integer & {1, 2, 3} \\
 & Hidden state dimension & Integer (step=16) & {16, 32, 48, 64, 80, 96, 112, 128} \\
 & Dropout rate & Continuous float & [0.00 -- 0.50] (step=0.1) \\
 & Learning rate & Continuous float (log-uniform) & [1e-4 -- 1e-2] \\
 & Batch size & Categorical & {32, 64, 128} \\
 \midrule
MLP & Hidden layer (hidden\_sizes) & Categorical & {[64,32], [128,64]} \\
 & Dropout rate & Continuous float & [0.00 -- 0.50] (step=0.1) \\
 & Learning rate & Continuous float (log-uniform) & [1e-4 -- 1e-2] \\
 & Batch size & Categorical & {32, 64, 128} \\
 \midrule
LSTM--ARIMA & ARIMA order parameters (p,d,q) & Fixed & (1,1,0) \\
 & LSTM number of recurrent layers (residual model) & Integer & {1, 2, 3} \\
 & LSTM hidden state dimension (residual model) & Integer (step=16) & {16, 32, 48, 64, 80, 96, 112, 128} \\
 & LSTM dropout rate (residual model) & Continuous float & [0.00 -- 0.50] (step=0.1) \\
 & Learning rate & Continuous float (log-uniform) & [1e-4 -- 1e-2] \\
 & Batch size & Categorical & {32, 64, 128} \\
 \midrule
Chronos & Pretrained model identifier & Fixed & amazon/chronos-t5-small \\
 & Batch size & Not applicable & Inference-only model \\
 & Learning rate & Not applicable & Inference-only model \\
 \bottomrule
\end{tabular}%
}
    \label{table:a1}
\end{table}

\begin{table}[htbp]
    \centering
    \setlength{\leftskip}{-117pt}
    \vspace*{-20mm} 
    \caption{The optimal hyperparameters for EUR/USD returns forecasting.}
    \resizebox{210mm}{!}{%
\begin{tabular}{ccccccccccc}
    \toprule
Type & Model & Sliding Window & Batch size & n\_heads & factor & hidden\_dim / hidden\_size /   d\_model & num\_layers & dropout\_rate & learning\_rate & Other Tuned Parameters \\
\midrule
EUR/USD & EXFormer & 5 & 64 & 2 & 16 & 32 & -- & 0.1 & 0.0002 & kernel\_sizes=(3,5,7) \\
 &  & 10 & 64 & 8 & 48 & 384 & -- & 0.2 & 0.0008 & kernel\_sizes=(3,5,7) \\
 &  & 15 & 64 & 1 & 16 & 16 & -- & 0.3 & 0.002 & kernel\_sizes=(3,5,7) \\
 &  & 20 & 64 & 1 & 32 & 32 & -- & 0.1 & 0.002 & kernel\_sizes=(3,5,7) \\
 &  & 30 & 64 & 8 & 16 & 128 & -- & 0.3 & 0.0001 & kernel\_sizes=(3,5,7) \\
\cline{2-11}
 & Informer & 5 & 64 & 4 & -- & 128 & -- & 0.1 & 0.001 & encoder\_layers=2, decoder\_layers=1 \\
 &  & 10 & 64 & 2 & -- & 64 & -- & 0.2 & 0.0005 & encoder\_layers=1, decoder\_layers=1 \\
 &  & 15 & 64 & 4 & -- & 128 & -- & 0.3 & 0.002 & encoder\_layers=2, decoder\_layers=2 \\
 &  & 20 & 32 & 2 & -- & 96 & -- & 0.1 & 0.0001 & encoder\_layers=1, decoder\_layers=1 \\
 &  & 30 & 64 & 4 & -- & 128 & -- & 0.3 & 0.0008 & encoder\_layers=3, decoder\_layers=1 \\
\cline{2-11}
 & Transformer & 5 & 64 & 8 & -- & 128 & 3 & 0.1 & 0.001 & dim\_feedforward=256 \\
 &  & 10 & 64 & 4 & -- & 96 & 2 & 0.2 & 0.0005 & dim\_feedforward=192 \\
 &  & 15 & 64 & 8 & -- & 128 & 3 & 0.3 & 0.002 & dim\_feedforward=256 \\
 &  & 20 & 32 & 2 & -- & 96 & 1 & 0.1 & 0.0001 & dim\_feedforward=128 \\
 &  & 30 & 64 & 4 & -- & 64 & 2 & 0.3 & 0.0008 & dim\_feedforward=192 \\
\cline{2-11}
 & TCN & 5 & 64 & -- & -- & -- & -- & 0.1 & 0.001 & num\_channels=[32,32], kernel\_size=5 \\
 &  & 10 & 64 & -- & -- & -- & -- & 0.2 & 0.0005 & num\_channels=[32,32], kernel\_size=3 \\
 &  & 15 & 64 & -- & -- & -- & -- & 0.3 & 0.002 & num\_channels=[32,32], kernel\_size=5 \\
 &  & 20 & 32 & -- & -- & -- & -- & 0.1 & 0.0001 & num\_channels=[16,16], kernel\_size=3 \\
 &  & 30 & 64 & -- & -- & -- & -- & 0.3 & 0.0008 & num\_channels=[32,32], kernel\_size=4 \\
\cline{2-11}
 & GRU & 5 & 64 & -- & -- & 128 & 2 & 0.1 & 0.001 & -- \\
 &  & 10 & 64 & -- & -- & 112 & 3 & 0.2 & 0.0005 & -- \\
 &  & 15 & 64 & -- & -- & 96 & 1 & 0.3 & 0.002 & -- \\
 &  & 20 & 32 & -- & -- & 64 & 2 & 0.1 & 0.0001 & -- \\
 &  & 30 & 64 & -- & -- & 128 & 3 & 0.3 & 0.0008 & -- \\
\cline{2-11}
 & LSTM & 5 & 64 & -- & -- & 128 & 2 & 0.1 & 0.001 & -- \\
 &  & 10 & 64 & -- & -- & 112 & 3 & 0.2 & 0.0005 & -- \\
 &  & 15 & 64 & -- & -- & 64 & 1 & 0.3 & 0.002 & -- \\
 &  & 20 & 32 & -- & -- & 96 & 2 & 0.1 & 0.0001 & -- \\
 &  & 30 & 64 & -- & -- & 128 & 3 & 0.3 & 0.0008 & -- \\
\cline{2-11}
 & MLP & 5 & 64 & -- & -- & -- & 2 & 0.1 & 0.001 & hidden\_units=[128,64] \\
 &  & 10 & 64 & -- & -- & -- & 2 & 0.2 & 0.0005 & hidden\_units=[64,32] \\
 &  & 15 & 64 & -- & -- & -- & 2 & 0.3 & 0.002 & hidden\_units=[128,64] \\
 &  & 20 & 32 & -- & -- & -- & 2 & 0.1 & 0.0001 & hidden\_units=[64,32] \\
 &  & 30 & 64 & -- & -- & -- & 2 & 0.3 & 0.0008 & hidden\_units=[128,64] \\
\cline{2-11}
 & LSTM--ARIMA & 5 & 64 & -- & -- & 128 (residual LSTM) & 2 & 0.1 & 0.001 & ARIMA\_order=(1,1,0) \\
 &  & 10 & 64 & -- & -- & 96 (residual LSTM) & 2 & 0.2 & 0.0005 & ARIMA\_order=(1,1,0) \\
 &  & 15 & 64 & -- & -- & 80 (residual LSTM) & 1 & 0.3 & 0.002 & ARIMA\_order=(1,1,0) \\
 &  & 20 & 32 & -- & -- & 64 (residual LSTM) & 2 & 0.1 & 0.0001 & ARIMA\_order=(1,1,0) \\
 &  & 30 & 64 & -- & -- & 112 (residual LSTM) & 3 & 0.3 & 0.0008 & ARIMA\_order=(1,1,0) \\
 \bottomrule
\end{tabular}%
}
    \label{table:a2}
\end{table}

\begin{table}[htbp]
    \centering
    \setlength{\leftskip}{-117pt}
    \caption{The optimal hyperparameters for USD/JPY returns forecasting.}
    \resizebox{210mm}{!}{%
\begin{tabular}{ccccccccccc}
    \toprule
Type & Model & Sliding Window & Batch size & n\_heads & factor & hidden\_dim / hidden\_size /   d\_model & num\_layers & dropout\_rate & learning\_rate & Other Tuned Parameters \\
\midrule
USD/JPY & EXFormer & 5 & 64 & 2 & 16 & 32 & -- & 0.2 & 0.005 & kernel\_sizes=(3,5,7) \\
 &  & 10 & 64 & 1 & 16 & 16 & -- & 0.2 & 0.0005 & kernel\_sizes=(3,5,7) \\
 &  & 15 & 64 & 2 & 48 & 96 & -- & 0.5 & 0.002 & kernel\_sizes=(3,5,7) \\
 &  & 20 & 64 & 4 & 48 & 192 & -- & 0.2 & 0.002 & kernel\_sizes=(3,5,7) \\
 &  & 30 & 64 & 8 & 48 & 384 & -- & 0.2 & 0.001 & kernel\_sizes=(3,5,7) \\
 \cline{2-11}
  & Informer & 5 & 64 & 4 & -- & 128 & -- & 0.1 & 0.001 & encoder\_layers=2, decoder\_layers=1 \\
 &  & 10 & 64 & 2 & -- & 64 & -- & 0.2 & 0.0006 & encoder\_layers=1, decoder\_layers=1 \\
 &  & 15 & 64 & 4 & -- & 128 & -- & 0.3 & 0.002 & encoder\_layers=2, decoder\_layers=2 \\
 &  & 20 & 32 & 2 & -- & 96 & -- & 0.1 & 0.0003 & encoder\_layers=1, decoder\_layers=1 \\
 &  & 30 & 64 & 4 & -- & 128 & -- & 0.3 & 0.0008 & encoder\_layers=3, decoder\_layers=1 \\
  \cline{2-11}
 & Transformer & 5 & 64 & 8 & -- & 128 & 3 & 0.1 & 0.001 & dim\_feedforward=256 \\
 &  & 10 & 64 & 4 & -- & 96 & 2 & 0.2 & 0.0008 & dim\_feedforward=192 \\
 &  & 15 & 64 & 8 & -- & 128 & 3 & 0.3 & 0.002 & dim\_feedforward=256 \\
 &  & 20 & 32 & 2 & -- & 96 & 2 & 0.1 & 0.0003 & dim\_feedforward=128 \\
 &  & 30 & 64 & 4 & -- & 64 & 2 & 0.3 & 0.0008 & dim\_feedforward=192 \\
  \cline{2-11}
 & TCN & 5 & 64 & -- & -- & -- & -- & 0.1 & 0.001 & num\_channels=[32,32], kernel\_size=5 \\
 &  & 10 & 64 & -- & -- & -- & -- & 0.2 & 0.0006 & num\_channels=[32,32], kernel\_size=3 \\
 &  & 15 & 64 & -- & -- & -- & -- & 0.3 & 0.002 & num\_channels=[32,32], kernel\_size=5 \\
 &  & 20 & 32 & -- & -- & -- & -- & 0.1 & 0.0003 & num\_channels=[16,16], kernel\_size=3 \\
 &  & 30 & 64 & -- & -- & -- & -- & 0.3 & 0.0008 & num\_channels=[32,32], kernel\_size=5 \\
  \cline{2-11}
 & GRU & 5 & 64 & -- & -- & 128 & 2 & 0.1 & 0.001 & -- \\
 &  & 10 & 64 & -- & -- & 112 & 2 & 0.2 & 0.0006 & -- \\
 &  & 15 & 64 & -- & -- & 96 & 1 & 0.3 & 0.002 & -- \\
 &  & 20 & 32 & -- & -- & 128 & 2 & 0.1 & 0.0003 & -- \\
 &  & 30 & 64 & -- & -- & 128 & 3 & 0.3 & 0.0008 & -- \\
  \cline{2-11}
 & LSTM & 5 & 64 & -- & -- & 128 & 2 & 0.1 & 0.001 & -- \\
 &  & 10 & 64 & -- & -- & 112 & 2 & 0.2 & 0.0006 & -- \\
 &  & 15 & 64 & -- & -- & 96 & 1 & 0.3 & 0.002 & -- \\
 &  & 20 & 32 & -- & -- & 128 & 2 & 0.1 & 0.0003 & -- \\
 &  & 30 & 64 & -- & -- & 128 & 3 & 0.3 & 0.0008 & -- \\
  \cline{2-11}
 & MLP & 5 & 64 & -- & -- & -- & 2 & 0.1 & 0.001 & hidden\_units=[128,64] \\
 &  & 10 & 64 & -- & -- & -- & 2 & 0.2 & 0.0005 & hidden\_units=[64,32] \\
 &  & 15 & 64 & -- & -- & -- & 2 & 0.3 & 0.002 & hidden\_units=[128,64] \\
 &  & 20 & 32 & -- & -- & -- & 2 & 0.1 & 0.0003 & hidden\_units=[64,32] \\
 &  & 30 & 64 & -- & -- & -- & 2 & 0.3 & 0.0008 & hidden\_units=[128,64] \\
  \cline{2-11}
 & LSTM--ARIMA & 5 & 64 & -- & -- & 128 (residual LSTM) & 2 & 0.1 & 0.001 & ARIMA\_order=(1,1,0) \\
 &  & 10 & 64 & -- & -- & 96 (residual LSTM) & 2 & 0.2 & 0.0006 & ARIMA\_order=(1,1,0) \\
 &  & 15 & 64 & -- & -- & 80 (residual LSTM) & 1 & 0.3 & 0.002 & ARIMA\_order=(1,1,0) \\
 &  & 20 & 32 & -- & -- & 64 (residual LSTM) & 2 & 0.1 & 0.0003 & ARIMA\_order=(1,1,0) \\
 &  & 30 & 64 & -- & -- & 112 (residual LSTM) & 3 & 0.3 & 0.0008 & ARIMA\_order=(1,1,0) \\
 \bottomrule
\end{tabular}%
}
    \label{table:a3}
\end{table}

\begin{table}[htbp]
    \centering
    \setlength{\leftskip}{-117pt}
    \caption{The optimal hyperparameters for GBP/USD returns forecasting.}
    \resizebox{210mm}{!}{%
\begin{tabular}{ccccccccccc}
    \toprule
Type & Model & Sliding Window & Batch size & n\_heads & factor & hidden\_dim / hidden\_size /   d\_model & num\_layers & dropout\_rate & learning\_rate & Other Tuned Parameters \\
\midrule
GBP/USD & EXFormer & 5 & 64 & 1 & 16 & 16 & -- & 0.2 & 0.001 & kernel\_sizes=(3,5,7) \\
 &  & 10 & 64 & 1 & 48 & 48 & -- & 0.5 & 0.002 & kernel\_sizes=(3,5,7) \\
 &  & 15 & 64 & 2 & 48 & 96 & -- & 0.3 & 0.002 & kernel\_sizes=(3,5,7) \\
 &  & 20 & 64 & 2 & 64 & 128 & -- & 0.3 & 0.004 & kernel\_sizes=(3,5,7) \\
 &  & 30 & 64 & 8 & 32 & 256 & -- & 0.3 & 0.003 & kernel\_sizes=(3,5,7) \\
   \cline{2-11}
 & Informer & 5 & 64 & 4 & -- & 128 & -- & 0.1 & 0.001 & encoder\_layers=2, decoder\_layers=1 \\
 &  & 10 & 64 & 2 & -- & 64 & -- & 0.2 & 0.0005 & encoder\_layers=1, decoder\_layers=1 \\
 &  & 15 & 64 & 4 & -- & 128 & -- & 0.3 & 0.002 & encoder\_layers=2, decoder\_layers=2 \\
 &  & 20 & 32 & 2 & -- & 96 & -- & 0.1 & 0.0001 & encoder\_layers=1, decoder\_layers=1 \\
 &  & 30 & 64 & 4 & -- & 128 & -- & 0.3 & 0.0008 & encoder\_layers=3, decoder\_layers=1 \\
    \cline{2-11}
 & Transformer & 5 & 64 & 8 & -- & 128 & 3 & 0.1 & 0.001 & dim\_feedforward=256 \\
 &  & 10 & 64 & 4 & -- & 128 & 2 & 0.2 & 0.0005 & dim\_feedforward=256 \\
 &  & 15 & 64 & 8 & -- & 128 & 3 & 0.3 & 0.002 & dim\_feedforward=256 \\
 &  & 20 & 32 & 2 & -- & 96 & 1 & 0.1 & 0.0001 & dim\_feedforward=128 \\
 &  & 30 & 64 & 4 & -- & 64 & 2 & 0.3 & 0.0008 & dim\_feedforward=192 \\
    \cline{2-11}
 & TCN & 5 & 64 & -- & -- & -- & -- & 0.1 & 0.001 & num\_channels=[32,32], kernel\_size=5 \\
 &  & 10 & 64 & -- & -- & -- & -- & 0.2 & 0.0005 & num\_channels=[32,32], kernel\_size=3 \\
 &  & 15 & 64 & -- & -- & -- & -- & 0.3 & 0.002 & num\_channels=[32,32], kernel\_size=5 \\
 &  & 20 & 32 & -- & -- & -- & -- & 0.1 & 0.0001 & num\_channels=[16,16], kernel\_size=3 \\
 &  & 30 & 64 & -- & -- & -- & -- & 0.3 & 0.0008 & num\_channels=[32,32], kernel\_size=5 \\
    \cline{2-11}
 & GRU & 5 & 64 & -- & -- & 128 & 2 & 0.1 & 0.001 & -- \\
 &  & 10 & 64 & -- & -- & 112 & 3 & 0.2 & 0.0005 & -- \\
 &  & 15 & 64 & -- & -- & 128 & 1 & 0.3 & 0.002 & -- \\
 &  & 20 & 32 & -- & -- & 96 & 2 & 0.1 & 0.0001 & -- \\
 &  & 30 & 64 & -- & -- & 128 & 3 & 0.3 & 0.0008 & -- \\
    \cline{2-11}
 & LSTM & 5 & 64 & -- & -- & 128 & 2 & 0.1 & 0.001 & -- \\
 &  & 10 & 64 & -- & -- & 112 & 3 & 0.2 & 0.0005 & -- \\
 &  & 15 & 64 & -- & -- & 128 & 1 & 0.3 & 0.002 & -- \\
 &  & 20 & 32 & -- & -- & 96 & 2 & 0.1 & 0.0001 & -- \\
 &  & 30 & 64 & -- & -- & 128 & 3 & 0.3 & 0.0008 & -- \\
    \cline{2-11}
 & MLP & 5 & 64 & -- & -- & -- & 2 & 0.1 & 0.001 & hidden\_units=[128,64] \\
 &  & 10 & 64 & -- & -- & -- & 2 & 0.2 & 0.0005 & hidden\_units=[64,32] \\
 &  & 15 & 64 & -- & -- & -- & 2 & 0.3 & 0.002 & hidden\_units=[128,64] \\
 &  & 20 & 32 & -- & -- & -- & 2 & 0.1 & 0.0001 & hidden\_units=[64,32] \\
 &  & 30 & 64 & -- & -- & -- & 2 & 0.3 & 0.0008 & hidden\_units=[128,64] \\
    \cline{2-11}
 & LSTM--ARIMA & 5 & 64 & -- & -- & 128 (residual LSTM) & 2 & 0.1 & 0.001 & ARIMA\_order=(1,1,0) \\
 &  & 10 & 64 & -- & -- & 96 (residual LSTM) & 2 & 0.2 & 0.0005 & ARIMA\_order=(1,1,0) \\
 &  & 15 & 64 & -- & -- & 80 (residual LSTM) & 1 & 0.3 & 0.002 & ARIMA\_order=(1,1,0) \\
 &  & 20 & 32 & -- & -- & 64 (residual LSTM) & 2 & 0.1 & 0.0001 & ARIMA\_order=(1,1,0) \\
 &  & 30 & 64 & -- & -- & 112 (residual LSTM) & 3 & 0.3 & 0.0008 & ARIMA\_order=(1,1,0) \\
 \bottomrule
\end{tabular}%
}
    \label{table:a4}
\end{table}

\clearpage
\bibliographystyle{model1-num-names}
\bibliography{EXcas-refs}

\end{document}